\documentclass[aps,prd,onecolumn,superscriptaddress,showpacs,floatfix,preprintnumbersm,nofootinbib]{revtex4-2}
\usepackage[dvipsnames]{xcolor}
\usepackage{dsfont}
\usepackage{bm}
\usepackage[english]{babel}
\usepackage[T1]{fontenc}
\usepackage{physics}
\usepackage{lmodern}
\usepackage{mathtools}
\usepackage{cancel}
\usepackage{mathtools}
\usepackage{relsize}
\usepackage{amssymb}
\usepackage{slashed}
\usepackage{enumerate}
\usepackage{appendix}
\usepackage{comment}
\usepackage{amsmath}
\makeatletter
\g@addto@macro\bfseries{\boldmath}
\makeatother
\usepackage{graphicx}
\usepackage{tikz}
\usetikzlibrary{shapes,arrows,shadows,automata,positioning}
\newdimen\nodeDist
\usepackage[
  colorlinks=true,
  linkcolor={red!50!black},
  citecolor={blue!50!black},
  filecolor=black,
  urlcolor=black,
  breaklinks=true
  ]{hyperref}

\definecolor{lcolor}{rgb}{0.5,0,0}
\definecolor{citcolor}{rgb}{0,0,1}

\DeclareMathOperator*{\SumInt}{%
\mathchoice%
  {\ooalign{$\displaystyle\sum$\cr\hidewidth$\displaystyle\int$\hidewidth\cr}}
  {\ooalign{\raisebox{.14\height}{\scalebox{.7}{$\textstyle\sum$}}\cr\hidewidth$\textstyle\int$\hidewidth\cr}}
  {\ooalign{\raisebox{.2\height}{\scalebox{.6}{$\scriptstyle\sum$}}\cr$\scriptstyle\int$\cr}}
  {\ooalign{\raisebox{.2\height}{\scalebox{.6}{$\scriptstyle\sum$}}\cr$\scriptstyle\int$\cr}}
}

\renewcommand{\tilde}{\widetilde}
\renewcommand{\Re}{\operatorname{Re}}

\newcommand{\rmii}[1]{{\mbox{\tiny\rm{#1}}}}

\newcommand\varpm{\mathbin{\vcenter{\hbox{%
  \oalign{\hfil{$+$}\hfil\cr
          \noalign{\kern-.3ex}
          \rmii{$(-)$}\cr}%
}}}}
\newcommand{\re}{\mathop{\mbox{Re}}}
\newcommand{\im}{\mathop{\mbox{Im}}}

\begin{document}

\title{Momentum expansions in finite-density perturbative calculations}
\begin{flushright}
HIP-2025-24/TH
\end{flushright}
\author{Mika Nurmela}
\email{mika.nurmela@helsinki.fi}
\affiliation{Department of Physics and Helsinki Institute of Physics,
P.O.~Box 64, FI-00014 University of Helsinki, Finland}
\author{Juuso Österman}
\email{juuso.s.osterman@helsinki.fi}
\affiliation{Department of Physics and Helsinki Institute of Physics,
P.O.~Box 64, FI-00014 University of Helsinki, Finland}

\begin{abstract}
\noindent
Complex-valued Feynman integrals in the imaginary time formalism and zero-temperature limit suffer from particular types of infrared divergences that can not be regulated by integration dimension alone. Related problems leading to integration order dependent results are even further pronounced in the presence of additional scales such as external momenta. This plays a noticeable role in systems featuring fermionic degrees of freedom such as cold Quantum Chromodynamics, where loop integrals are complexified by chemical potential(s). Working in the limit of vanishing temperature, we utilize novel complex-valued extensions to bubble Feynman integrals and study momentum expansions of fermionic loop integrals. The expansions are then used to illustrate the mechanisms of manifested discrepancies between orders of integration, associated with the residue theorem. Finally, we address the issues by introducing a representation avoiding the observed ambiguity and briefly overview classes of integrals insensitive to problems from external momenta.

\end{abstract}

\maketitle
\newpage
\tableofcontents
\newpage

\section{Introduction}
\label{sec:introduction}
\noindent
Despite their central role in (analytic) studies of quantum field theories, higher-order correlation functions often contain pathological infinities that need to be regulated (and renormalized) in practical computations \cite{Collins:1984xc, peskin, Weinberg:1995mt, Schwartz:2014sze}. The most popular such method in real-valued vacuum theories at vanishing temperature, $T=0$, involves using the integration dimension as the sole regulator ($d$ or $D = d+1$ in literature)\footnote{The utility of dimensional regularization can be largely attributed to the particularly attractive properties of Euler functions in analytic continuation \cite{hooft}.} \cite{tHooft:1972tcz,Bollini,Cicuta:1972jf}. In the context of thermal field theory, where Lorentz symmetry is broken already in the path integral description via both temperature and chemical potential, $\mu$, dimension acts as the definitive regulator present in the spatial sector of the momentum integrations \cite{Kapusta:1989tk, lainebook, Ghiglieri:2020dpq}. In this work, we specifically focus on the imaginary-time description of thermal field theory at vanishing temperature, characterized by the Euclidean metric. In this context, the temporal sector of integrations does not formally introduce an additional regulator to the explicitly one-dimensional integration. However, in \cite{Gorda:2022yex, Osterman:2023tnt, Ostermanthesis}  infinitesimal temperature has been observed to work as an effective (and necessary) regulator pertaining to computations involving a non-vanishing chemical potential, contributing to a complex-valued scale ($i \mu$).

The hierarchy $|\mu| \gg T \simeq 0$ \textcolor{black}{is characteristic of the complex-valued system of integrals contributing to} the \textcolor{black}{perturbative regime of the} Equation of State of cold quark matter \cite{Vuorinen:2003fs,Kurkela:2009gj,Vuorinen:2016pwk}. \textcolor{black}{While structural similarities extend to scattering amplitude computations involving complex-valued momenta \cite{Britto:2005fq} and masses \cite{Denner:1999gp,Armadillo:2022ugh}, this work is categorically formulated towards problems arising in the low-temperature description of Quantum Chromodynamics (QCD). With the intent to determine the viability of quark matter occurring in neutron star cores, the perturbative regime of QCD pressure is utilized to constrain possible values of EoS in more reasonable densities \cite{Annala:2019puf,Vuorinen:2024qws}.} Given the slow convergence of the weak-coupling expansions associated with the QCD pressure \cite{Blaizot:2003iq,Ipp:2003yz, Brambilla:2014jmp}, the on-going research extends \textcolor{black}{to isolating the coefficients in the order of $\mathcal{O}(\alpha_s^3)$ in terms of the strong coupling \cite{Gorda:2018gpy,Gorda:2023mkk,Gorda:2023zwy,Gorda:2023zwy, Navarrete:2024zgz, Karkkainen:2025nkz}. The remaining unresolved contributions to the coefficient of $\alpha_s^3$ consist of non-factorizing four-loop Feynman integrals.}

In these exceedingly challenging computations, efficient analytic computation strategies are required, with there being a particular interest towards approaches involving the residue theorem \cite{Freedman:1976ub, Kurkela:2009gj,ghisou, Sappi:2020yuj,Gorda:2022yex}. Residue-driven methods are characterized by the strict $T=0$ temporal limit\footnote{The strict limit can be understood as the naive interpretation of the vanishing temperature, where the Matsubara summation becomes a temporal integral with a range of solid real line $\mathbb{R}$.} and the corresponding integrations preceding spatial ones in order. In general these methods have been observed to greatly simplify the structures appearing in loop computations \cite{ghisou, Gorda:2022yex, Ostermanthesis}, and their use is further underscored by their prominent role in deriving multi-loop results, extending to three loops and beyond \cite{Kurkela:2009gj, Karkkainen:2025nkz}. \textcolor{black}{However, the intricacy and scale of the related computation motivate the use of sophisticated methods for evaluating the integrals, which provide valuable cross-checks \cite{Osterman:2023tnt}.}

Here we continue studying the properties and utility of residue-driven approaches, and provide a careful comparison by working in parallel using leading low-temperature limits arising from Matsubara sums and Euler-Maclaurin formula \cite{Stegun,Vuorinen:2003fs,Gorda:2022yex,Ostermanthesis}. We choose to predominantly work at the one-loop level, as dictated by the earliest and therefore simplest occurrence of an observed discrepancy between different orders of integration in the presence of external momenta. Such one-loop expressions, in particular polarization functions, have contributed significantly to extracting results in multi-loop computations \cite{Gorda:2022fci, Gorda:2022zyc,Gorda:2023zwy, Gorda:2023mkk,Karkkainen:2025nkz}. Working specifically using hierarchies assigned by Hard Thermal Loop (HTL) effective theories \cite{Pisarski:1988vd,Braaten:1989mz,Frenkel:1989br,Taylor:1990ia,Braaten:1991gm,Frenkel:1991ts,Blaizot:1999ap,Blaizot:2001nr} to external momenta in the context of finite density \cite{Kurkela:2016was,lainebook,Gorda:2018gpy,Sappi:2020yuj,Ghiglieri:2020dpq}, we illustrate a novel type of discrepancy by studying the simplest self-energy (or polarization) function with internal fermionic lines\footnote{\textcolor{black}{The four-loop integral contributions to QCD pressure are the first to involve these discrepancies in the associated calculation (pertaining for example to isolating divergent behaviour). Notably computations in the three-loop order do not require such care, given that the crucial integral results are explicitly known at non-vanishing temperature \cite{Vuorinen:2003fs,Karkkainen:2025nkz}.}}.
 
In particular, we emphasize that the mechanism studied in this work leading to results dependent on integration order is distinct from the one discussed in \cite{Gorda:2022yex}, which is associated with fermionic poles with chemical potential(s) as the exclusive scale(s). To further illustrate, let us first express the set of Feynman bubble integrals discussed in \cite{Gorda:2022yex} at nonvanishing temperature using Matsubara sum formalism:
 \begin{equation}
\begin{split}
\label{eq:MatsubarainitialS} 
\SumInt_{P}^f \frac{1}{P^{2\alpha} }\equiv T \sum_{n = -\infty }^\infty \int_p \frac{1}{\{[\pi T ( 2 n +1)+ i \mu]^2+p^2 \}^\alpha},
    \end{split}
\end{equation}
where we utilize minimal subtraction scheme and regulate the spatial dimension $d$ such that
\begin{equation}
    \int_p \equiv \left( \frac{e^\gamma \Lambda^2}{4 \pi} \right)^{\frac{3-d}{2}} \frac{1}{{(2 \pi)^d}}\oint \text{d} \Omega_d \int_0^\infty \text{d} p\ p^{d-1}.
\end{equation} 
Here fermionic (characterized by $f$) loop momenta are written as Fourier modes in Matsubara formalism such that
\begin{equation}
    p_{0,n} \equiv  (2n+1)\pi T +i\mu.
\end{equation}
In the case of  $T \rightarrow 0$ (temporal) integral computations, we opt to use a convention $p_{0,n}\rightarrow p_0+i\mu$, where $p_0$ is a continuous real-valued integration variable.

In the context of this work and \cite{Gorda:2022yex}, the primary interest lies with vanishing temperature limit, which can be studied using Euler-Maclaurin formalism (for more details see appendix \ref{sec:EM-subsec}) applied in the integration order shown above (associated with the sought-after physical results \cite{Vuorinen:2003fs, Gorda:2022yex}). To reverse the order of integration in the same limit, the Matsubara summation is re-written using complex contours and the residues of the Fermi-Dirac distribution function $n_\text{F} [i \beta (p_0-i\mu)]$, where $\beta = \frac{1}{T}$ (see also eq.~\eqref{eq:FermiDiracDef}). Notably, via the strict $T=0$ limit the Fermi-Dirac distribution function simplifies to a Heaviside step function $\theta[\text{Im}p_0 - \mu]$ for $\mu > 0$. Utilizing this limit and assuming $\alpha \in \mathbb{Z}_+$\footnote{We emphasize that the explicit residue prescription is well-defined only for integer values of the exponent parameter $\alpha$. However, as discussed in \cite{Gorda:2022yex, Ostermanthesis}, the strict residue results can be associated with specific contour lines in the complex plane, yielding an interpretation to further real-valued parameters.}, we can interpret eq.~\eqref{eq:MatsubarainitialS} as
\begin{equation}
\SumInt_{P}^f \frac{1}{P^{2\alpha} } \overset{T \to 0^+}{\longmapsto}   \int_p  i \text{Res} \left\{ \frac{1}{[(p_0 + i \mu)^2 +p^2]^\alpha}\right\}_{p_0= i(p-\mu)},
\end{equation}
where the residue is computed using semicircular complex contour along the upper complex half-plane (for more details see appendix \ref{sec:temporalintegrationandresidue}). However, this strict implementation of a limit does not necessarily include all physically relevant contributions, and can even lead to breakdown of integration order independent results, and hence consistency of loop computations\cite{Collins:1984xc} if no remedy is applied.

Specifically in \cite{Gorda:2022yex}, it was observed that 
\begin{eqnarray}
 \underset{T \to 0^+}{\text{lim}} \SumInt_{P}^f \frac{1}{P^{2} } &=&  \int_p  i \text{Res} \left\{ \frac{1}{(p_0 + i \mu)^2 +p^2}\right\}_{p_0= i(p-\mu)},\\
     \underset{T \to 0^+}{\text{lim}} \SumInt_{P}^f \frac{1}{P^{2\alpha} } &\overset{\alpha \neq 1}{\neq}&  \int_p  i \text{Res} \left\{ \frac{1}{[(p_0 + i \mu)^2 +p^2]^\alpha}\right\}_{p_0= i(p-\mu)},
\end{eqnarray}
with the parametric agreement occurring serendipitously through the convergent left-sided limit $\alpha \rightarrow 1^{-}$. The corresponding discrepancy surfaces through the unaccounted effects from the regime of an exceedingly small but nonvanishing temperature. These are again pertinent to the integration order on the right-hand side and the strict $T=0$ limit applied carelessly onto the temporal distribution function.

In contrast, this work explicitly studies and observes a discrepancy occurring even concerning first order fermionic poles in the presence of real-valued external momenta $S^\nu = (s_0, \vec{s} )$. The simplest example of this is 
\begin{equation}
\label{eq:introresidueinequalityfull}
     \underset{T \to 0^+}{\text{lim}} \SumInt_{P}^f \frac{1}{P^{2} (P-S)^2 } \neq  \sum_{n=0}^1\int_p  i \text{Res} \left\{ \frac{1}{(p_0 + i \mu)^2 +p^2} \frac{1}{(p_0 + i \mu-s_0)^2 +q^2}\right\}_{p_0= in (p-\mu)+(1-n)(iq-i\mu+s_0)},
\end{equation}
and is studied through momentum expansions in the hierarchy \textcolor{black}{$|s_0|,|\Vec{s}| \ll \mu$} and $s_0 \neq 0$ with the shorthand $q = |\vec{p}-\vec{s}|$. Of particular note here is that the discrepancy extends to even the leading part of the expansion, implying that for arbitrary hierarchy between components of $S^\nu$
\begin{equation}
\label{eq:introresidueinequalityleading}
    \sum_{n=0}^1\int_p i \underset{|S| \rightarrow 0^+}{\text{lim}}  \text{Res} \left\{ \frac{1}{(p_0 + i \mu)^2 +p^2} \frac{1}{(p_0 + i \mu-s_0)^2 +q^2}\right\}_{p_0= in (p-\mu)+(1-n)(iq-i\mu+s_0)} \neq \underset{T \to 0^+}{\text{lim}} \SumInt_{P}^f \frac{1}{P^{4}  }.
\end{equation}
However, fixing $s_0 = 0$, the two sides of \eqref{eq:introresidueinequalityfull} can be shown to agree, given sufficient regularization (for details see sections  \ref{sec:spatialexpansionresidue} and \ref{sec:Fpspatialdirect}). The leading case of this equality --- agreement between the sides of eq.~\eqref{eq:introresidueinequalityleading} --- was previously discussed in \cite{Gorda:2022yex}.  The other extreme case (with fixed $\vec{s} = \vec{0}$) modifies the sides of eq.~\eqref{eq:introresidueinequalityleading} to explicitly resemble the distinct solutions of bubble Feynman integrals presented (also) in \cite{Gorda:2022yex}. For a detailed description of the differences between corresponding momentum expansions, see sections \ref{sec:temporalexpansionresidueasymps0} and \ref{sec:spatialexpansasympts0}.

We further extend the results of \cite{Gorda:2022yex} by generalizing the one-loop bubble integral (master) formulae of --- the notation following that of \cite{Osterman:2023tnt,Ostermanthesis} ---
\begin{eqnarray}
    \mathcal{I}_\alpha (\mu) &\equiv & \underset{T \to 0^+}{\text{lim}} \SumInt_{P}^f \frac{1}{P^{2 \alpha}  },\\
   \text{Res} \left[ \mathcal{I}_\alpha \right] (\mu) &\equiv &  \int_p  i \text{Res} \left\{ \frac{1}{[(p_0 + i \mu)^2 +p^2]^\alpha}\right\}_{p_0= i(p-\mu)}
\end{eqnarray}
to permit complex-valued chemical potentials (see section \ref{sec:extensiontocomplex}). The generalizations are then utilized to streamline the momentum expansions involving first simpler subsets in section \ref{sec:momentumasymptotes} and the most general case in section \ref{sec:generalfermionicproblem}. For their distinct values, see eqs.~\eqref{eq:masterformualIa} and \eqref{eq:residuebubblemaster} for a real-valued chemical potential, and eqs.~\eqref{eq:complexresiduebubble} and \eqref{eq:complexmumasterbubble} for their complex-valued generalizations.

To remove the discrepancy associated with the strict $T=0$ residue prescription, we construct a modified temporal integrand by utilizing the Fermi-Dirac distribution function and Feynman parametrization (see section \ref{sec:comparison}). This representation is then used to highlight the structural problems of the naive application of standard residue theorem in terms of the resulting (post-temporal integration) spatial integrands (for more details see sections \ref{sec:comparison} and \ref{sec:generalcomparison} as well as appendix \ref{app:powerseries}).

Many related results are further contextualized in the appendices. Specifically, in appendices \ref{sec:bosons} and \ref{app:distinctchemicalpotentials}, we observe cases leading to identical results in either integration order, involving bosonic version of the left-hand side of eq.~\eqref{eq:introresidueinequalityfull} -- characterized by $\text{Im} s_0 =  \mu$ -- as well as a case with an imaginary-valued external temporal momentum. Lastly in appendix \ref{app:sunset}, we study the characteristics of different temporal integration orders arising in the simplest non-trivial two-loop Feynman integral.

\section{Mathematical preface}
\label{sec:preface}
\noindent
In this section, we provide further details to our notation and conventions. With these we briefly review a closely related set of problems of bubble loop integrals (lacking external momenta), which has had a significant impact on computation methods and definitions utilized in this work.
\subsection{Contour formalism}
\label{sec:contourformalism}
\begin{figure}[h!]
\centering
\begin{tikzpicture}[scale=0.8]
\def\blockh{1.2}
\def\blockl{3.0}
  \draw[->,thick] (-\blockl-1,-2*\blockh+1) -- (\blockl+1,-2*\blockh+1) node[above] {$\re(p_0)$};
  \draw[->,thick] (0,-2*\blockh) -- (0,2.5*\blockh) node[right] {$\im(p_0)$};
  \draw[-, loosely dotted, line width = 0.4mm,black!75] (-\blockl+0.1,\blockh-0.75) -- (-0.1,\blockh-0.75) node[black, above left] {};
   \draw[-, loosely dotted, line width = 0.4mm,black!75] (0.1,\blockh-0.75) -- (\blockl,\blockh-0.75) node[black, above left] {};
 %\draw[-, thick, loosely dotted,red!75] (-\blockl,\blockh-0.75) -- (\blockl,\blockh-0.75) node[black, above left] {};
 \draw[-latex,thick,blue!75] (-\blockl,\blockh+0.5) -- (0,\blockh+0.5) node[black, above left] {};
  \draw[-latex,thick,blue!75] (0,\blockh-2) -- (-\blockl,\blockh-2) node[midway,below] {};
  \draw[-latex,thick,blue!75] (0,\blockh+0.5) -- (\blockl,\blockh+0.5) node[black, above left] {};
 % \draw[-latex,thick,blue!75] (0,+\blockh-0.5) -- (0,+\blockh+0.5) node[black,midway,left] {};
  \draw[-latex,thick,blue!75] (\blockl,\blockh-2) -- (0,\blockh-2) node[black, above left] {};
%\draw[-latex,densely dashed,blue!50] (\blockl,+\blockh+0.5) -- (\blockl,+\blockh-2) node[right] {};
%\draw[-latex,densely dashed,blue!50] (-\blockl,+\blockh-2) -- (-\blockl,+\blockh+0.5) node[right] {};
  %\draw[-latex] (\blockl/2,+\blockh) arc (0:-270:\blockh/4) node[black,below] {};
  \draw[red,fill] (0,+\blockh+0.5) circle [radius=1.5pt] node[black,above right] {$+i\mu+i\eta$};
  \draw[red,fill] (0,+\blockh-2) circle [radius=1.5pt] node[black,below right] {$+i\mu-i\eta$};

\end{tikzpicture}
\begin{tikzpicture}[scale=0.8]
\def\blockh{1.2}
\def\blockl{3.0}
  \draw[->,thick] (-\blockl-1,-2*\blockh+1) -- (\blockl+1,-2*\blockh+1) node[above] {$\re(p_0)$};
  \draw[->,thick] (0,-2*\blockh) -- (0,2.5*\blockh) node[right] {$\im(p_0)$};

 \draw[-latex,thick,blue!75] (-\blockl,\blockh+0.5) -- (0,\blockh+0.5) node[black, above left] {};
  \draw[-latex,thick,blue!75] (0,\blockh-2) -- (-\blockl,\blockh-2) node[midway,below] {};
  \draw[-latex,thick,blue!75] (0,\blockh+0.5) -- (\blockl,\blockh+0.5) node[black, above left] {};
  \draw[-latex,dashed,purple!75] (0.1,+\blockh-2) -- (0.1,+\blockh+0.5) node[black,midway,left] {};
  \draw[-latex,dashed,purple!75] (-0.1,+\blockh+.5) -- (-0.1,+\blockh-2) node[black,midway,left] {};
  \draw[-latex,thick,blue!75] (\blockl,\blockh-2) -- (0,\blockh-2) node[black, above left] {};
\draw[-latex,densely dashed,blue!75] (\blockl,+\blockh+0.5) -- (\blockl,+\blockh-2) node[right] {};
\draw[-latex,densely dashed,blue!75] (-\blockl,+\blockh-2) -- (-\blockl,+\blockh+0.5) node[right] {};
  \draw[-latex] (\blockl/2+0.6,+\blockh-0.6) arc (0:-270:\blockh/2) node[black,below] {};
   \draw[-latex] (-\blockl/2+0.6,+\blockh-0.6) arc (0:-270:\blockh/2) node[black,below] {};
  \draw[red,fill] (0,+\blockh+0.5) circle [radius=1.5pt] node[black,above right] {$+i\mu+i\eta$};
  \draw[red,fill] (0,+\blockh-2) circle [radius=1.5pt] node[black,below right] {$+i\mu-i\eta$};

\end{tikzpicture}
\caption[]{%
  Left: The line integral representation of the contour $\gamma$ is described with blue solid lines. It resembles two solid lines parallel to the real axis apart from the two removed points signified by the red dots along the imaginary axis. The thinly dotted black line indicates the positions of the poles arising from the Fermi-Dirac distributions function. Note in particular that $p_0=i\mu$ is not included in these poles.
  Right: The addition of the two pairs of dashed line integrals does not change the result of the integration. The dashed blue lines are placed at infinitely far along the real axis where the integrand is assumed to systematically vanish. The dashed purple lines are running in opposite directions along the imaginary axis, and therefore canceling out, provided that the integrated function is holomorphic in the integration contours. This facilitates the contour to be used for summation of the residues corresponding to the Fermi-Dirac distribution function. The image is adapted from \cite{Osterman:2023tnt, Ostermanthesis}. 
  }
  \label{fig:contouroverlap}
\end{figure}
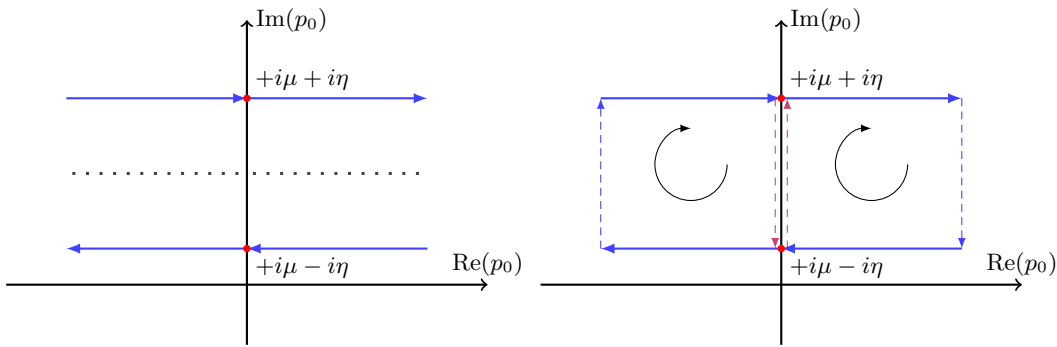

\noindent
This work predominantly focuses on the properties of finite density integrals occurring at vanishing (infinitesimally small) limit of temperature, where the chemical potentials (and external momenta) dominate as scale(s). Accordingly, most of our computations to follow are performed utilizing carefully chosen integral limits that are discussed in detail in \cite{lainebook,Gorda:2022yex, Osterman:2023tnt, Ostermanthesis}. In order to avoid excessively lengthy expressions, we use the following shorthands for complex contour expressions, relevant to the vanishing temperature limits in our computations

\begin{eqnarray}
\oint_{P}^f &=& \oint_{p_0}^f \int_p,\\
\label{eq:gammacontour}
\oint_{p_0}^f &=& \oint_\gamma \frac{\text{d} p_0}{2 \pi},
\end{eqnarray}
where the contour $\gamma$ consists of four lines parallel to the real axis such that $\gamma = (i \mu + i \eta - \infty, i \mu + i \eta) \times(i \mu+i\eta, i \mu+i\eta + \infty) \times (i \mu - i \eta +  \infty, i\mu-i\eta) \times(i\mu-i\eta, i \mu - i \eta - \infty)\}$ or as seen on the left-hand side of Fig. \ref{fig:contouroverlap}.

These divisions occurring at the imaginary axis correspond to the natural split of the odd valued Matsubara summation of eq.~\eqref{eq:MatsubarainitialS}. The contour $\gamma$ represents the minimal rectangular shape enveloping all the first order poles of the Fermi-Dirac distribution function
\begin{equation}
\label{eq:FermiDiracDef}
    \tilde{n}_\text{F} (p_0) \equiv n_\text{F}[i\beta (p_0-i\mu)] = \frac{1}{\exp\left[i\beta(p_0-i\mu) \right]+1},
\end{equation}
given as an equivalent set to that of the Matsubara indices being summed over in the fermionic sum of eq.~\eqref{eq:MatsubarainitialS}: $\{(2n+1) \pi T + i \mu\}_{n\in \mathbb{Z}}$. As discussed in both \cite{Gorda:2022yex, Osterman:2023tnt}, the strict low-temperature limit of the Fermi-Dirac distribution function is that of a step function such that 
\begin{equation}
     \tilde{n}_\text{F} (p_0) \overset{T \to 0}{\longrightarrow} \theta\left[\text{Im} p_0-\mu \right],
\end{equation}
which facilitates for a regular enough function $f(p_0)$\footnote{Lack of divergences occurring inside the contour is sufficient condition. However, as discussed in \cite{Ostermanthesis, Osterman:2023tnt, Gorda:2022yex} corresponding exclusion of specific boundary values may lead to wrong results. Notably, for the one-loop master integral of eq.~\eqref{eq:MatsubarainitialS} performing spatial integration first leads to an well-behaved temporal integrand entity (without any additional conditions), i.e. a desired $f(p_0)$.}:
\begin{equation}
\begin{split}
    \oint_{p_0}^f \tilde{n}_\text{F} (p_0) f(p_0) &\overset{T \to 0}{\longrightarrow} \left[\int_{-\infty+ i\eta}^{i\eta} + \int_{i\eta}^{\infty+i\eta} \right] \frac{\text{d} p_0}{2 \pi} f(p_0+i\mu)=\left[\int_{-\infty}^{0} + \int_{0}^{\infty} \right] \frac{\text{d} p_0}{2 \pi} f(p_0+i\mu)\\
    &\equiv \text{P.V.} \int_{-\infty}^\infty  \frac{\text{d} p_0}{2 \pi} f(p_0+i\mu)
    \end{split}
\end{equation}
To match the discussion of eq.~\eqref{eq:MatsubarainitialS} with the steps of \cite{Osterman:2023tnt}, we denote the class of master integrals under scrutiny -- corresponding to the leading term in the low-temperature expansion -- as
\begin{equation}
\label{eq:leadingIalimit}
    \mathcal{I}_\alpha (\mu) = \underset{T \to 0}{\text{lim}} \oint_P^f \frac{\tilde{n}_\text{F}(p_0)}{(P^2)^\alpha},
\end{equation}
or 
\begin{equation}
\label{eq:leadingRIalimit}
    \text{Res} \left[\mathcal{I}_\alpha \right] (\mu) =  \int_p \oint_{p_0}^f \frac{1}{(P^2)^\alpha}\underset{T \to 0}{\text{lim}}\tilde{n}_\text{F}(p_0),
\end{equation}
with the challenges arising for corresponding integrand $f(p_0) = (p_0^2+p^2)^{-\alpha}$ and discrepancies between integration orders deferred to appendices \ref{sec:EM-subsec} and \ref{sec:temporalintegrationandresidue} concerning a real-valued chemical potential $\mu \in \mathbb{R}$ (originally presented somewhat similarly in \cite{Gorda:2022yex}). The results of the two master integrals are extended to general complex-valued chemical potential, $\mu \in \mathbb{C}$, in the following subsection.

\subsection{Bubble integrals with complex-valued chemical potential}
\label{sec:extensiontocomplex}
\noindent
Pertaining to our actual target integral containing external momenta $S^\nu = (s_0, \vec{s})$, we find it exceedingly useful to study first a generalization of the bubble integrals $\mathcal{I}_\alpha (\mu)$ and $\text{Res}[\mathcal{I}_\alpha ](\mu)$ discussed in \cite{Gorda:2022yex} and appendices \ref{sec:EM-subsec}---\ref{sec:temporalintegrationandresidue}. Specifically,  let $\mu \in \mathbb{C}$ and consider the strict residue integral at $T=0$. Then we apply a suitable (as dictated by the real part of chemical potential) semicircular contour covering a complex half-plane to find 
\begin{equation}
\begin{split}
\label{eq:complexresiduebubble}
& \left.\text{Res} \left[\mathcal{I}_\alpha \right] (\mu) \right|_{\mu \in \mathbb{C}}\\
    =&\sum_{n=0}^1 \theta\left[(-1)^n \text{Re} \mu\right]\int_p i \text{Res} \left[ \frac{1}{[(p_0-\text{Im}\mu+i \text{Re}\mu-ip)(p_0-\text{Im} \mu+i\text{Re}\mu+ip)]^\alpha}\right]_{p_0=\text{Im}\mu-i \text{Re} \mu+(1-2n)ip}\\
    &= -\left(\frac{e^\gamma \Lambda^2}{4\pi} \right)^\frac{3-d}{2} \frac{\Gamma \left(\alpha-\frac{1}{2}\right)}{\sqrt{\pi} (4\pi)^\frac{d}{2} \Gamma (\alpha) \Gamma \left( \frac{d}{2} \right)} \frac{\left|\text{Re}\mu\right|^{d+1-2\alpha}}{d+1-2\alpha}\\
    &= \text{Res}\left[ \mathcal{I}_\alpha \right] (|\text{Re} \mu|).
    \end{split}
\end{equation}
In the context of this work, the subset case of $\alpha = 1$
\begin{equation}
\text{Res}\left[ \mathcal{I}_1 \right] (|\text{Re} \mu|) = \mathcal{I}_1 (|\text{Re}\mu|)
\end{equation}
is of particular interest, yielding a straightforward but degenerate (in the complex sense) solution in terms of the closed forms of real-valued master integrals given in eqs.~\eqref{eq:leadingIalimit} and \eqref{eq:leadingRIalimit}. In the context of this parametric subset case --- extending to complex-valued chemical potentials --- we can show that integration in either order yields identical results (or even with the formal leading term in the low-temperature expansion) by utilizing similar steps to those shown in appendix \ref{sec:EM-subsec}.

To demonstrate the agreement, let us begin by considering
\begin{equation}
 \left.\mathcal{I}_\alpha (\mu)\right|_{\mu \in \mathbb{ C}}  =\underset{T \to 0^+}{\text{lim}} T \sum_{n=-\infty}^\infty \int_p \frac{1}{\{[(2n+1)\pi T-\text{Im} \mu+i \text{Re}\mu]^2+p^2\}^\alpha},
\end{equation}
the innermost spatial integral of which can be analytically continued (for all $\alpha \in \mathbb{R}_+$) to a scaled version of the Euler beta function if $(2n+1)\pi T \neq \text{Im} \mu$ \footnote{ This condition onto temporal loop momentum manifests unavoidably in the final results, even should one regulate the expression using the exponent parameter $\alpha$.}. Continuing with this condition, we find a structurally similar result to that of the $\mu > 0$ case:
\begin{equation}
\begin{split}
   \left.\mathcal{I}_\alpha (\mu)\right|_{\mu \in \mathbb{ C}}=\left( \frac{e^\gamma \Lambda^2}{4 \pi} \right)^\frac{3-d}{2} \frac{\Gamma \left(\alpha-\frac{d}{2} \right)}{(4\pi)^\frac{d}{2}\Gamma(\alpha)} \underset{T \to 0^+}{\text{lim}} T \sum_{n=-\infty}^\infty \left\{[(2n+1)\pi T+i\mu]^2\right\}^{\frac{d}{2}-\alpha}.
   \end{split}
\end{equation}
Using next Euler-Maclaurin formulae (see appendix \ref{sec:EM-subsec} and particularly eqs.~\eqref{eq:simplestMatsubarcase}---\eqref{eq:quadraticbubblematsubara} ), we can extract the leading contribution from the temporal sum as an integral equivalent with the structure arising from the strict vanishing-temperature limit, begun with the spatial integration. Specifically, let $p_0 \neq \text{Im} \mu$ and write \footnote{For convenience and clarity, we have written the decomposition here best aligned with $\text{Im} \mu > 0$, but the other hierarchy would lead to the same result, albeit with the second and first integral being combined instead of the last two.}
\begin{equation}
\begin{split}
\label{eq:PVcomplexsplit}
 &\left.\mathcal{I}_\alpha (\mu)\right|_{\mu \in \mathbb{ C}}
  =\text{P.V.}\int_{-\infty}^\infty \frac{\text{d} p_0 }{2 \pi}  \int_p \frac{1}{[(p_0-\text{Im} \mu+i \text{Re}\mu)^2+p^2]^\alpha}\\
  &=  \left( \frac{e^\gamma \Lambda^2}{4 \pi} \right)^\frac{3-d}{2} \frac{\Gamma \left(\alpha-\frac{d}{2} \right)}{2 \pi (4\pi)^\frac{d}{2}\Gamma(\alpha)} \left[ \int_0^\infty \text{d} p_0 \left[(p_0 + i \text{Re}\mu)^2 \right]^\frac{d-2\alpha}{2} + \int_0^{\text{Im} \mu} \text{d} p_0 \left[ (p_0-i \text{Re}\mu)^2 \right]^\frac{d-2\alpha}{2}\right.\\
  &\left.+\int_{\text{Im} \mu}^\infty \text{d} p_0 \left[ (p_0-i\text{Re}\mu)^2 \right]^\frac{d-2\alpha}{2}\right], 
  \end{split}
\end{equation}
where we have performed a change of integration variables such that
\begin{equation}
\label{eq:minusp0oneloopintegral}
    \int_{-\text{Im} \mu}^0 \text{d} p_0 \left[(p_0 + i \text{Re}\mu)^2 \right]^\frac{d-2\alpha}{2} =  \int_0^{\text{Im} \mu} \text{d} p_0 \left[ (p_0-i \text{Re}\mu)^2 \right]^\frac{d-2\alpha}{2}.
\end{equation}
By piece-wise utilizing the partial integration result of
\begin{equation}
\label{eq:quadratichelpintegral2}
    \int_a^b \text{d} x [(x+A)^2]^{k} = \frac{1}{2k+1} \left\{(x+A)[(x+A)^2]^{k} \right\}_{x=a}^{x=b}.
\end{equation}
and (or) by noting that we are allowed to combine the latter two integrals of eq.~\eqref{eq:PVcomplexsplit}\footnote{The integrand receives a finite value at $p_0= \text{Im} \mu$, specifically $(\text{Im}\mu-i \text{Re} \mu)^{\frac{d-2\alpha}{2}}$ and the function is continuous over this point, enabling a zero-width addition to the integration interval without the result changing.}, we find the full result as 
\begin{equation}
\begin{split}
  \left.\mathcal{I}_\alpha (\mu)\right|_{\mu \in \mathbb{ C}}
  &=  \left( \frac{e^\gamma \Lambda^2}{4 \pi} \right)^\frac{3-d}{2} \frac{\Gamma \left(\alpha-\frac{d}{2} \right)}{2 \pi (4\pi)^\frac{d}{2}\Gamma(\alpha)} \left[ \int_0^\infty \text{d} p_0 \left[(p_0 + i \text{Re}\mu)^2 \right]^\frac{d-2\alpha}{2} + \int_0^{\infty} \text{d} p_0 \left[ (p_0-i \text{Re}\mu)^2 \right]^\frac{d-2\alpha}{2}\right]\\
  &= -\left( \frac{e^\gamma \Lambda^2}{4 \pi} \right)^\frac{3-d}{2} \frac{\Gamma \left(\alpha-\frac{d}{2} \right)}{2 \pi (4\pi)^\frac{d}{2}\Gamma(\alpha)} \frac{i \text{Re} \mu}{d+1-2\alpha} \left\{ \left[\left(i \text{Re} \mu \right)^2 \right]^{\frac{d-2\alpha}{2}}-\left[\left(-i \text{Re} \mu \right)^2 \right]^{\frac{d-2\alpha}{2}}  \right\},
  \label{eq:Ialpha_correct}
  \end{split}
\end{equation}
which can be straightforwardly simplified using steps of appendix \ref{sec:EM-subsec} to be given in terms of the master integral $\mathcal{I}_\alpha (\mu)$ such that
\begin{equation}
\label{eq:complexmumasterbubble}
\begin{split}
 %&\underset{T \to 0^+}{\text{lim}} T \sum_{n=-\infty}^\infty \int_p \frac{1}{\{[(2n+1)\pi T-\text{Im} \mu+i \text{Re}\mu]^2+p^2\}^\alpha}\\
\left.\mathcal{I}_\alpha (\mu)\right|_{\mu \in \mathbb{ C}} &= -\left( \frac{e^\gamma \Lambda^2}{4 \pi} \right)^\frac{3-d}{2} \frac{1}{(4\pi)^\frac{d}{2}\Gamma (\alpha) \Gamma\left(\frac{d}{2}+1-\alpha\right)}  \frac{\left| \text{Re}\mu \right|^{d+1-2\alpha}}{d+1-2\alpha} \\
 &= \mathcal{I}_\alpha \left(|\text{Re} \mu| \right).
\end{split}
\end{equation}

It is worth noting  here that both approaches (orders of integration) retain their  master integral structure even when moving to complex-valued chemical potentials -- thus keeping all the properties therein (specifically the means of partial integration to derive the difference terms). Furthermore, the result maintains the equality of integration orders occurring for first order poles of the propagator integrand, notably with any placement of the vanishing-temperature limit. The specific structures of eqs. \eqref{eq:complexresiduebubble} and \eqref{eq:complexmumasterbubble} are exceedingly useful in studying the expansions involved in eq.~\eqref{eq:introresidueinequalityfull}, permitting us to compactly identify and isolate structures in a well-defined manner.

Finally, we stress that too careless steps in integrations or expansions can introduce spurious --- even complex-valued --- contributions when deriving eq.~\eqref{eq:complexmumasterbubble}. To help prevent such potential sources of mistakes, we have written plainly and thoroughly discussed related computations in appendix \ref{sec:FalseExpansions}. We acknowledge that such problems naturally extend to more intricate integral structures. Accordingly, appendix \ref{sec:FalseExpansions} discusses an additional example utilizing a structure from  section \ref{sec:momentumasymptotes}.

\section{Asymptotes of the external momentum}
\label{sec:momentumasymptotes}
\noindent
In this section we begin studying the novel problem associated with external momenta. As indicated earlier, we limit ourselves to discussing structures of the form 
\begin{equation}
\label{eq:problemintegral}
    \mathcal{J}_{\alpha_1 \alpha_2}(\mu, \vec{s}, s_0) \equiv \underset{T \to 0}{\text{lim}}\SumInt_P^f \frac{1}{P^{2\alpha_1} (P-S)^{2\alpha_2}} %\underset{T \to 0}{\text{lim}} \oint_P^f \frac{\tilde{n}_\text{F} (p_0)}{P^2 (P-S)^2},
\end{equation}
{\begin{equation}
\label{eq:problemintegralres}
    \text{Res}\left[\mathcal{J}_{\alpha_1\alpha_2} \right](\mu, \vec{s}, s_0) \equiv \int_p \oint_{p_0}^f\frac{1}{P^{2\alpha_1} (P-S)^{2\alpha_2}} \underset{T \to 0}{\text{lim}}\tilde{n}_\text{F} (p_0),
\end{equation}
for $\alpha_1 = \alpha_2 =1$ (for more details involving the limits see appendix \ref{sec:temporalintegrationandresidue} and \cite{Gorda:2022yex,Ostermanthesis}). The propagators herein superficially contain only first order poles, seen to agree in any order of limits and integrations for the case without external momenta. In section \ref{sec:extensiontocomplex}, this property was seen work organically even with complex-valued chemical potentials. However, in the following subsection we explicitly study two simplified cases:  i) $\vec{s} = \vec{0}$, ii) $s_0 = 0$ \footnote{We additionally discuss $s_0 = i\mu$ corresponding to bosonic poles in subsection \ref{sec:bosons}, and produce here the basic tools to discuss such results with.}. Before comparing with different order of integration, we start by observing the leading parts of the power series utilizing the residue theorem to recognize a systematic problem. In the subsequent subsection we compute the expressions in the reverse order of integration in alignment with the leading terms of thermal sums, to be compared against the results of the former subsection. In the final subsection, we present a strategy that yields consistent results in any order of integration, combining both Feynman parametrization and the presence of distribution functions (at infinitesimal temperature). Furthermore we remind the reader that the physical motivation of our computation revolves around such high-density systems that are characterized by the hierarchy \textcolor{black}{$|s_0|,|\Vec{s}| \ll \mu$} (or $|\mu|$ for general real-valued chemical potential). While the divergences appearing in \textcolor{black}{$|s_0|,|\Vec{s}| \gg \mu$} hierarchy can be seen to correspond to renormalization \cite{peskin,Collins:1984xc}, some divergences of the \textcolor{black}{$|s_0|,|\Vec{s}| \ll \mu$} hierarchy persist. These divergences are canceled against HTL, where the cancellation of the divergences coincide with the \textcolor{black}{$|s_0|,|\Vec{s}| \ll \mu$} hierarchy of the naive perturbation theory \cite{Gorda:2022zyc,Karkkainen:2025nkz}. Consequently, we do not compute results tracing the well-known vacuum results of the form (working in $D = d+1$ spatial dimensions)
\begin{equation}
    \int_P \frac{1}{P^2 (P-S)^2} = \left(\frac{e^\gamma \Lambda^2}{4\pi} \right)^\frac{3-d}{2} \frac{\Gamma \left(2-\frac{d+1}{2} \right)}{(4\pi)^\frac{d+1}{2} } \frac{\Gamma \left(\frac{d+1}{2}-1 \right)}{\Gamma \left(d-1 \right)} (S^2)^{\frac{d+1}{2}-3}
\end{equation}
and small corrections in the reverse hierarchy.  While not essential to the arguments here, we have further addressed the connection of the computation scheme to the standard vacuum theory results in appendix \ref{app:Slimit}.

We emphasize that any results we are looking for should be categorically in alignment with exchange of the order between the integrations and limits onto the external momenta. Most notably, we utilize the following set of (schematic) equalities
\begin{equation}
   \underset{S \to 0}{\text{lim}} \SumInt_P^f \frac{1}{P^2 (P-S)^2} = \SumInt_P^f  \underset{S \to 0}{\text{lim}} \frac{1}{P^2 (P-S)^2} = \SumInt_P^f \frac{1}{P^4} \label{eq:leadinglimit}
\end{equation}
as necessary (but not sufficient) condition for a valid integration result, specifically when $T \to 0^+$. Additionally, we want to remind the reader that while we do describe a subset of results to be in agreement between different methods pertaining to first order poles, any integral with second or higher order fermionic poles contains corrections not accessible with conventional residue theorem as discussed in \cite{Gorda:2022yex} and in the previous section.

\subsection{Residue results}
\label{sec:residueasymptoteS}
\noindent
Following the construction presented in section \ref{sec:residuenotation}, we perform the computation in the strict $T=0$ limit utilizing a semicircle shaped contour covering half of the complex plane and explicitly work with real-valued $\mu$ \footnote{For $\mu > 0$ the upper and $\mu < 0$ the lower. The $\mu=0$ case holds no interested for us, as it corresponds to the vacuum theory result discussed above.}. Denoting $q = |\vec{p}-\vec{s}|$, the integral given eq.~\eqref{eq:problemintegralres} becomes
\begin{equation}
\begin{split}
\label{eq:residueT0fulllist}
&\text{Res}\left[\mathcal{J}_{11} \right](\mu, \vec{s}, s_0)\\
    &=\theta(\mu )\int_p  \sum_{n=0}^1 i \text{Res} \left[ \frac{1}{(p_0+i\mu+ip)(p_0+i\mu-ip)(p_0+i\mu-s_0+iq)(p_0+i\mu-s_0-iq)} \right]_{p_0  + i \mu =(1-n) ip + n(iq+ s_0)}\\
    &+\theta(-\mu )\int_p  \sum_{n=0}^1 i \text{Res} \left[ \frac{1}{(p_0+i\mu+ip)(p_0+i\mu-ip)(p_0+i\mu-s_0+iq)(p_0+i\mu-s_0-iq)} \right]_{p_0  + i \mu =(n-1) ip - n(iq+ s_0)}\\
    &=\int_p \left[\frac{\theta(p-|\mu|)}{2p}\frac{1}{(ip-s_0)^2+q^2}+\frac{\theta(q-|\mu-\text{Im} s_0|)}{2q}\frac{1}{(iq+s_0)^2+p^2} \right],
    \end{split}
\end{equation}
where we initially work with $s_0 \in \mathbb{C}$ to enable the study of bosonic propagator structures characterized by $\text{Im} s_0 = \mu$ and structures involving distinct chemical potential when $\text{Im} s_0 \neq \mu$ (as opposed to standard fermionic structures with $\text{Im} s_0 = 0$). Additionally, we reiterate that the Heaviside step function structures are associated with the semicircular contour in either half-plane. In the main text we focus on expressions with exclusively fermionic pole structures, for which small-momentum expansions generate novel correction terms via delta functions, associated with infinitesimal temperature. For computations involving bosonic poles or two distinct chemical potentials, see appendices \ref{sec:bosons} and \ref{app:distinctchemicalpotentials} respectively.

For explicit computations, it is not only useful but \textcolor{black}{also} common practice to symmetrize the latter part of the integrand (separated by summation) with the former part. Given that the spatial integration covers all of the $\mathbb{R}^d$ space, we are entitled to write instead
\begin{equation}
\label{eq:residuecollection}
  \text{Res}\left[\mathcal{J}_{11} \right](\mu, \vec{s}, s_0) =\int_p \left[\frac{\theta(p-|\mu|)}{2p}\frac{1}{S^2-2ip s_0 - 2psz}+\frac{\theta(p-|\mu-\text{Im} s_0|)}{2p}\frac{1}{S^2+2ips_0+2psz} \right],
\end{equation}
where the spatial external momentum was completely removed from the Heaviside step functions via a linear shift $\vec{p}\rightarrow \vec{p}+\vec{s}$ and we utilized a compact expression for the squared external four-momentum $S^2 = s_0^2+ s^2$, as well as the shorthand $s \textcolor{black}{ \equiv } |\vec{s}|$. \textcolor{black}{For integrals involving one external momentum, the angular dependence ($z$) appears solely from the dot product $\Vec{p}\cdot \Vec{s}\equiv psz$.} In further sections, we utilize $\mu$ in place of $|\mu|$ to simplify the steps (and as it does not change any of the results). We additionally emphasize that this type of symmetrization removes access to the previously discussed delta function corrections from expansions involving spatial external momentum.

\subsubsection{Spatial expansion $\Vec{s} \neq \vec{0}$, $s_0 = 0$}\label{sec:spatialexpansionresidue}
\noindent
Let us begin with the case $s_0 = 0$ (applied to eq.~\eqref{eq:residueT0fulllist}), the leading order of which ($s \rightarrow 0$) was discussed in \cite{Gorda:2022yex} as an alternative means of deriving the result of $\mathcal{I}_2 (\mu)$:
\begin{equation}
\label{eq:limitingspatialI2}
\begin{split}
  \underset{s \rightarrow 0}{\text{lim}}\text{Res}\left[\mathcal{J}_{11} \right](\mu, \vec{s}, 0) &\equiv  \int_p\underset{q \to p}{\text{lim}} \frac{1}{q^2-p^2} \left[-\frac{\theta(q-\mu)}{2q}+\frac{\theta(p-\mu)}{2p}\right]\\&=\int_p\frac{1}{2p}\frac{\text{d}}{\text{d}p}\left[-\frac{\theta(p-\mu)}{2p}\right]= \mathcal{I}_2 (\mu)
  \end{split}
\end{equation}
As this result suggests, the leading order does agree with the leading temperature expansion (and therefore the opposite integration order). However, as discussed in \cite{Ostermanthesis, Gorda:2022yex, Osterman:2023tnt}, the contributions from non-vanishing temperature impact the distribution functions. Given that all the scales in this simple one-loop integral are (initially) contained within the step functions of eq.~\eqref{eq:residuecollection}, we can actually observe that with suitable regularization the full expansion over $\frac{s}{\mu}$ is observed to align with the opposite integration order (see section \ref{sec:FPasymptoteS} ).

To address the above, we continue from the last line of eq.~\eqref{eq:residuecollection}:
\begin{equation}
    \text{Res}\left[\mathcal{J}_{11} \right](\mu, \vec{s}, 0) = \frac{1}{s} \int_p \frac{\theta(p-\mu)}{2p} \left[\frac{1}{2pz +s}-\frac{1}{2pz-s} \right] = - \int_p\frac{\theta(p-\mu)}{p} \frac{1}{4p^2 z^2 - s^2} 
\end{equation}
and note further that the conventional method of extracting angular integrals \cite{Blumenson,Stillinger:1977mt,Somogyi:2011ir} conforms to
\begin{equation}
      \int_p = \left( \frac{e^\gamma \Lambda^2}{4 \pi} \right)^\frac{3-d}{2} \frac{1}{(2\pi)^d}\oint \text{d} \Omega_{d-1} \int_0^\infty \text{d} p\ p^{d-1} \int_{-1}^1 \text{d} z \left(1-z^2\right)^{\frac{d-3}{2}}.
\end{equation}
Specifically looking into the angular integration regulated solely by the dimension $d$, we note that the vanishing external momentum limit is problematic. The associated divergence can be seen by considering the angular integration in the neighbourhood of origin $z=0$ by fixing $0 < \eta \ll 1$ and writing 
\begin{equation}
    \int_{-1}^1 \text{d}z \frac{\left(1-z^2\right)^\frac{d-3}{2}}{4p^2 z^2-s^2} \overset{s = 0}{\longmapsto} \int_{-\eta}^\eta \text{d}z \frac{\left(1-z^2\right)^\frac{d-3}{2}}{4p^2 z^2} = \infty.
\end{equation}
This contradicts the convergent structure implied in eq.~\eqref{eq:limitingspatialI2}. The problem follows from the information conveyed by a Dirac delta function being transferred to the propagator via the introduced angular coordinate. The observable problem logically follows from divergences generated independent of the end points of radial integration, and hence falling outside the effects of dimensional regularization.  

To employ expressions with angular coordinates properly, and hence allowing the sought-after limits to be studied, we introduce a modified spatial integration measure with additional regulator exponent $b$ such that 
\begin{equation}
    \int_p^{\mathcal{R}} \equiv \left( \frac{e^\gamma \Lambda^2}{4 \pi} \right)^\frac{3-d}{2} \frac{1}{(2\pi)^d}\oint \text{d} \Omega_{d-1} \int_0^\infty \text{d} p\ p^{d-1} \lim_{b\rightarrow 0}\int_{-1}^1 \text{d} z \left(1-z^2\right)^{\frac{d-3}{2}}\left(z^2\right)^b.
    \label{eq:b_regulated_integral}
\end{equation}
Note that we introduce the limit for $b\rightarrow 0$ in the definition of the regulated integral. With this redefinition, we are able to compute any angular integral admitting negative monomial powers of the angular coordinate $z$, analytically continuing Euler gamma functions pertaining to values $b \rightarrow 0$. Thus, we can produce the following power series 
\begin{equation}
    \begin{split}
    \label{eq:spatialresidueintermediateexpansion}
        \text{Res}\left[\mathcal{J}_{11} \right](\mu, \vec{s}, 0) &=\frac{1}{s} \int_p^{\mathcal{R}}\frac{\theta(p-\mu)}{2p} \left[\frac{1}{2pz +s}-\frac{1}{2pz-s} \right]\\
        &=- \sum_{n=0}^\infty s^{2n} \int_p^{\mathcal{R}}  \frac{\theta (p-\mu)}{p(2 p z)^{2+2n}}. 
    \end{split}
\end{equation}
Each individual term of this expansion is straightforward to compute utilizing the following closed-form expression for angular integrals 
\begin{equation}
\label{eq:angularbetafunctionhelp}
    \int_{-1}^1 \text{d} z (1-z^2)^{\frac{d-3}{2}} (z^2)^{a} = \frac{\Gamma \left(\frac{1}{2}+a \right) \Gamma \left( \frac{d-1}{2} \right)}{\Gamma \left(a+\frac{d}{2} \right)},
\end{equation}
where $a>-1/2$, but the result can be extended to real values $a  \not\in \{-\mathbb{N}_{0}-\frac{1}{2}\} \cup \{-\mathbb{N}_{0}-\frac{d}{2}\}$ via analytic continuation. Then the corresponding spatial integrals --- on the right-hand side of eq.~\eqref{eq:spatialresidueintermediateexpansion} --- read
\begin{equation}
    \begin{split}
  - \int_p^{\mathcal{R}}  \frac{\theta (p-\mu)}{p(2 p z)^{2+2n}}%&= -\left( \frac{e^\gamma \Lambda^2}{4 \pi} \right)^\frac{3-d}{2} \frac{2 \pi^\frac{d-1}{2}}{\Gamma\left( \frac{d-1}{2} \right) (2 \pi)^d} \frac{1}{2^{2n+2}} \int_0^\infty \text{d} p \theta(p-\mu) p^{d-4-2n} \lim_{b\rightarrow 0}\int_{-1}^1 \text{d} z (1-z^2)^{\frac{d-3}2{}} (z^2)^{b-1-n} \\
    &=  \left( \frac{e^\gamma \Lambda^2}{4 \pi} \right)^\frac{3-d}{2} \frac{ 1}{(2 \pi)\Gamma\left( \frac{d-1}{2} \right) (4 \pi)^\frac{d-1}{2}} \frac{\mu^{d-3-2n}}{2^{2n+1} (d-3-2n)}   \frac{\Gamma \left(-\frac{1}{2}-n \right) \Gamma \left( \frac{d-1}{2} \right)}{\Gamma \left(-1-n + \frac{d}{2} \right)}\\
    %&= -\left(-1 \right)^{n} \frac{(n+1)!}{2^{2n+1}} \left( \frac{1}{2} \right)_{n+1}^{-1}  \left( \frac{e^\gamma \Lambda^2}{4 \pi} \right)^\frac{3-d}{2}  \frac{1}{(4 \pi)^\frac{d}{2} \Gamma (n+2) \Gamma \left( \frac{d}{2}+1-(n+2) \right)} \frac{\mu^{d+1-2(n+2)}}{d+1-2(n+2)}\\
    &= (-1)^n \left[ \binom{2n+1}{n} \right]^{-1} \mathcal{I}_{2+n}(\mu),
    \end{split}
\end{equation}
where we utilized the master integrals given in eq.~\eqref{eq:masterformualIa} to express the results in a compact manner and facilitate straightforward comparison between different methods (compare against results in section \ref{sec:FPasymptoteS}). Thus, the full power series expansion is given by
\begin{equation}
\label{eq:residuevecexpansion}
    \text{Res}\left[\mathcal{J}_{11} \right](\mu, \vec{s}, 0)  = \sum_{n=0}^\infty(-s^2)^n \left[ \binom{2n+1}{n} \right]^{-1} \mathcal{I}_{2+n}(\mu).
\end{equation}

\subsubsection{Temporal expansion $\Vec{s}=\vec{0}$, $s_0\neq0$ }
\label{sec:temporalexpansionresidueasymps0}
\noindent
Let us next consider the second subset of interest, without a spatial component of external momentum and with $s_0$ assumed to be real. Using the residue calculation result of \eqref{eq:residuecollection}, we straightforwardly obtain utilizing the assumed hierarchy $\mu \gg |s_0|$:
\begin{equation}
\begin{split}
\label{eq:naiveresidueexp}
    \text{Res}\left[\mathcal{J}_{11} \right](\mu, 0, s_0)  &= \int_p \frac{\theta(p-\mu)}{2p} \frac{1}{(ip-s_0)^2+p^2}+\int_p \frac{\theta(p-\mu)}{2p} \frac{1}{(ip+s_0)^2+p^2}\\
     &=  \int_p \frac{\theta(p-\mu)}{4p} \frac{1}{s_0} \left[ \frac{1}{-ip +\frac{s_0}{2} }-\frac{1}{-i p  -\frac{s_0}{2}} \right]\\
     &=  \int_p \frac{\theta(p-\mu)}{4p} \frac{i}{p s_0} \left\{ \sum_{n=0}^\infty \left( \frac{i s_0}{p} \right)^n \frac{(-1)^n-(1)^n}{2^n}  \right\}\\
     &= -\left( \frac{e^\gamma \Lambda^2}{4\pi}\right)^{\frac{3-d}{2}} \frac{ \mu^{d-2}}{2 (4\pi)^\frac{d}{2} \Gamma \left( \frac{d}{2}\right)} \frac{i}{s_0}\left\{ \sum_{n=0}^\infty \left( \frac{i s_0}{\mu} \right)^n \frac{(-1)^n-(1)^n}{2^n(d-2-n)}  \right\}\\
     &=\sum_{n=0}^\infty (-s_0^2)^n \left[ \binom{2n+1}{n} \right]^{-1} \text{Res}[\mathcal{I}_{2+n}] (\mu),
    \end{split}
\end{equation}
where the last identity utilizes the strict $T = 0$ master integral structure of eq.~\eqref{eq:residuebubblemaster}. We specifically emphasize that the leading order result in the expansion does not match the result presented in \eqref{eq:leadinglimit} but rather the corresponding naive integral variant. This exemplifies how the residue theorem breaks down even concerning specific set of first order pole problems. This discrepancy is addressed and remedied in section \ref{sec:comparison}.

\subsection{Spatial results}
\label{sec:FPasymptoteS}
\noindent
In this section we discuss computations utilizing the Feynman parametrization
\begin{equation}
    \frac{1}{A_1^{\alpha_1} A_2^{\alpha_2}} = \frac{\Gamma \left( \alpha_1+\alpha_2 \right)}{\Gamma (\alpha_1) \Gamma (\alpha_1)} \int_0^1 \text{d} x \frac{x^{\alpha_1-1} (1-x)^{\alpha_2-1}}{[xA_1+(1-x)A_2]^{\alpha_1+\alpha_2}},\, \forall A_i \neq 0, \text{Re} \alpha_i > 0,
\end{equation}
which facilitate us to map the dual propagator structure onto a complex-valued generalization of vacuum one-loop integrals for initial spatial integration, following the idea popularized (initially by Feynman \cite{Feynman:1949zx}) by widespread textbooks such as \cite{peskin,Schwartz:2014sze}\footnote{For complex valued parameters $\{A_k\}$ a short derivation utilizing the complex valued generalization of Euler beta function is discussed in for example \cite{Ostermanthesis}.}. To apply this, we need to study the conditions i) required by Feynman parametrization, ii) the new propagators structure sets onto the spatial integration and iii) how these conditions reflect onto the final results. We systematically address the ordering of the introduced parametric integration and the loop integrations to justify the steps we take in each relevant subsection. The organization of these computations follow that of the previous subsection \ref{sec:residueasymptoteS}. In addition to this, we emphasize that given the presence of dimensional regularization and the order of integration, we can and will systematically only consider the leading thermal effect with the distribution function taken to its step function limit. Accordingly all temporal integrals are as a rule (in this subsection) computed as the principal value integrals discussed in section \ref{sec:preface}. 

\subsubsection{Spatial expansion $\Vec{s} \neq \vec{0}$, $s_0 = 0$}
\label{sec:Fpspatialdirect}
\noindent
To be able to consider the spatial integration in terms of the common practices of dimensional regularization, pertaining to 
\begin{equation}
\label{eq:sFPinitint}
    \mathcal{J}_{11}(\mu, \vec{s}, 0)= \text{P.V.} \int_{-\infty}^\infty \frac{\text{d}p_0}{2\pi} \int_p \frac{1}{[(p_0+i\mu)^2+p^2][(p_0+i\mu)^2+|\vec{p}-\vec{s}|^2]},
\end{equation}
the convenience of analytic continuation from Euler beta functions is only achievable from Feynman parametrization in conjunction with a suitable change of integration variables enacted onto the two propagators. The Feynman parametrization -- in itself -- sets rather modest conditions onto the propagator elements (with already fixed exponents). Specifically $\forall (p_0,p) \in (\mathbb{R}\backslash\{0\}_, \mathbb{R}^d)$ these read 
\begin{eqnarray}
    (p_0+i\mu)^2 + p^2 &\neq&0,\\
    (p_0+i\mu)^2 + |\vec{p}-\vec{s}|^2 &\neq& 0.
\end{eqnarray}
A sufficient condition is therefore $p_0 \neq 0$, which arises from the leading thermal integral part of the fermionic Matsubara sum. Then given the presence of dimensional regularization, we can justify exchanging the order between the parametric integral and spatial integration (with the integration seen as explicitly finite and Fubini's theorem {-- or results independent of integration order -- therefore applicable). Thus, we seek to study
\begin{equation}
    \begin{split}
        \mathcal{J}_{11}(\mu, \vec{s}, 0)&= \text{P.V.} \int_{-\infty}^\infty \frac{\text{d}p_0}{2\pi} \int_p \int_0^1 \text{d} x  \frac{1}{[|\vec{p}-x \vec{s}|^2+(p_0+i\mu)^2+x(1-x)s^2]^2}\\
        &= \text{P.V.} \int_{-\infty}^\infty \frac{\text{d}p_0}{2\pi} \int_0^1 \text{d} x \int_p  \frac{1}{[p^2+(p_0+i\mu)^2+x(1-x)s^2]^2},
    \end{split}
\end{equation}
which we can interpret as a straightforward (scaled) complex-valued extension of the Euler beta function exactly when
\begin{equation}
    (p_0+i\mu)^2+x(1-x) s^2 \not \in \mathbb{R}_{-},
\end{equation}
which notably again is satisfied $\forall p_0 \neq 0$  \cite{Gorda:2022yex}. Thus, the spatial integral leads to the straightforward one-loop vacuum result albeit with a complex-valued power structure:
\begin{equation}
\begin{split}
    \mathcal{J}_{11}(\mu, \vec{s}, 0)&= \text{P.V.} \int_{-\infty}^\infty \frac{\text{d}p_0}{2\pi} \int_0^1 \text{d} x \int_p  \frac{1}{[|\vec{p}-x \vec{s}|^2+(p_0+i\mu)^2+x(1-x)s^2]^2}\\
    &= \left( \frac{e^\gamma \Lambda^2}{4 \pi} \right)^\frac{3-d}{2} \frac{\Gamma \left(2-\frac{d}{2} \right)}{(4 \pi)^\frac{d}{2}}\text{P.V.} \int_{-\infty}^\infty \frac{\text{d}p_0}{2\pi}  \int_0^1 \text{d} x  \left[ (p_0+i\mu)^2+x(1-x) s^2 \right]^\frac{d-4}{2}.
    \end{split}
\end{equation}
Given that the power function is regulated by the dimension\footnote{The absence of a need for the regulated integral, following from analytic continuation via Euler beta functions, provides a compelling example of the well-defined nature of this integration order.}, we can explicitly assume that either order of integration yields finite and equal values, and therefore change again the order of integration. Thus, the temporal integral at hand is a slight generalization of the cases studied in section \ref{sec:preface}, with an added positive (quadratic) scale. By noting that $|p_0+i\mu| = \sqrt{p_0^2+\mu^2}$, we can expand the power function utilizing $\frac{x(1-x)s^2}{(p_0+i\mu)^2}$ as the expansion parameter, given that we work with the hierarchy $\mu \gg s$. This leads to each term of the expansion being proportional to the temporal integral representation of $\{\mathcal{I}_n (\mu)\}_{n \geq 2}$, with $n \in \mathbb{N}$. Specifically, the expansion can be expressed such that
\begin{equation}
\label{eq:FPvecexpansion}
    \begin{split}
        \mathcal{J}_{11}(\mu, \vec{s}, 0)&=  \left( \frac{e^\gamma \Lambda^2}{4 \pi} \right)^\frac{3-d}{2} \frac{\Gamma \left(2-\frac{d}{2} \right)}{(4 \pi)^\frac{d}{2}} \int_0^1\text{d}x \, \text{P.V.} \int_{-\infty}^\infty \frac{\text{d}p_0}{2\pi}   \left[ (p_0+i\mu)^2+x(1-x) s^2 \right]^\frac{d-4}{2}\\
        %&= \left( \frac{e^\gamma \Lambda^2}{4 \pi} \right)^{\frac{3-d}{2}} \frac{\Gamma \left(2-\frac{d}{2}\right)}{(4 \pi)^\frac{d}{2}} \int_0^1 \text{d} x \int_0^\infty \frac{\text{d} p_0}{2\pi} \left\{\left[ (p_0+i\mu)^2 \right]^\frac{d-4}{2}+\left[ (p_0-i\mu)^2 \right]^\frac{d-4}{2} \right\}\\
      %&-s^2 \left[\int_0^1 \text{d} x \, x(1-x) \right] \left( \frac{e^\gamma \Lambda^2}{4 \pi} \right)^{\frac{4-d}{2}} \frac{\Gamma \left(3-\frac{d}{2}\right)}{(4 \pi)^\frac{d}{2}} \int_0^\infty \frac{\text{d} p_0}{2\pi} \left\{\left[ (p_0+i\mu)^2 \right]^\frac{d-6}{2}+\left[ (p_0-i\mu)^2 \right]^\frac{d-6}{2} \right\} \dots\\
      &=\sum_{n=0}^\infty \frac{(-s^2)^n \left( n! \right)^2 (n+1)!}{(2n+1)! n!} \left( \frac{e^\gamma \Lambda^2}{4 \pi} \right)^{\frac{4-d}{2}} \frac{\Gamma \left(2+n-\frac{d}{2} \right)}{(4\pi)^\frac{d}{2} \Gamma \left(2+n \right)}\\
      %&\times\int_0^\infty \frac{\text{d} p_0}{2\pi} \left\{\left[ (p_0+i\mu)^2 \right]^\frac{d-4-2n}{2}+\left[ (p_0-i\mu)^2 \right]^\frac{d-4-2n}{2} \right\} \\
      &= \sum_{n=0}^\infty \frac{(-s^2)^n n!  (n+1)!}{(2n+1)!} \mathcal{I}_{2+n} (\mu)\\
      &\equiv \sum_{n=0}^\infty \left[ \binom{2n+1}{n} \right]^{-1} (-s^2)^n \mathcal{I}_{2+n} (\mu).
    \end{split}
\end{equation}

\subsubsection{Temporal expansion $\Vec{s}=\vec{0}$, $s_0\neq0$ }
\label{sec:spatialexpansasympts0}
\noindent
Studying instead the case without spatial component of the external momentum, we note that while Feynman parametrization is required to identify the initial spatial integration as a complex-valued generalization of the Euler beta function, no additional change of integration variable is required. However, an additional condition is required to facilitate the extension towards the beta function, structurally in alignment with what we encountered in the computation of section \ref{sec:extensiontocomplex}.

Let us study the conditions explicitly, and start by noting that Feynman parametrization is well-defined for 
\begin{equation}
\label{eq:s0FPinitint}
    \mathcal{J}_{11}(\mu, 0, s_0)= \text{P.V.} \int_{-\infty}^\infty \frac{\text{d}p_0}{2\pi} \int_p \frac{1}{[(p_0+i\mu)^2+p^2][(p_0+i\mu-s_0)^2+p^2]},
\end{equation}
if 
\begin{eqnarray}
    (p_0+i\mu)^2+p^2 &\neq& 0,\\
    (p_0+i\mu-s_0)^2+p^2&\neq& 0.
\end{eqnarray}
Given that for the beta function integration, we fix $p \in \mathbb{R}_+$, the resulting conditions read  $p_0 \not \in \{0,s_0\}$. Assuming these conditions, we find a Feynman parametrization such that
\begin{equation}
\label{eq:s0FPinitint}
    \mathcal{J}_{11}(\mu, 0, s_0)= \text{P.V.} \int_{-\infty}^\infty \frac{\text{d}p_0}{2\pi} \int_p \int_0^1 \text{d} x \frac{1}{[p^2+(p_0+i\mu-x s_0)^2+x(1-x)s_0^2]^2}.
\end{equation}
We exchange the order of the spatial and parametric integration, then the spatial integral converges to the complex generalization of the Euler beta function for each $x$ if $p_0-x s_0 \neq 0$\footnote{Again the alternative method is to regulate the exponent of the propagator structure instead \cite{Osterman:2023tnt}.}. Thus, we find
\begin{equation}
\begin{split}
\label{eq:afterspatilprincipalvalue1}
   \mathcal{J}_{11}(\mu, 0, s_0)=  \left( \frac{e^\gamma \Lambda^2}{4 \pi} \right)^\frac{3-d}{2} \frac{\Gamma \left(2-\frac{d}{2} \right)}{(4 \pi)^\frac{d}{2}}\text{P.V.} \int_{-\infty}^\infty \frac{\text{d}p_0}{2\pi}  \int_0^1 \text{d} x  \left[ (p_0+i\mu-xs_0)^2+x(1-x) s_0^2 \right]^\frac{d-4}{2},
    \end{split}
\end{equation}
where the principal value temporal integral excludes the following values: $p_0 \not \in \{0,s_0,xs_0\}$ as per the conditions for each individual value of $x$. Again given the presence of dimensional regularization, we can one final time exchange the order of integration. 

Following a similar approach as with the purely spatial external momenta, we note immediately that $|p_0+i\mu-xs_0| = \sqrt{(p_0-xs_0)^2+\mu^2}\gg x(1-x)s_0^2$, which enables us to expand the temporal integrand such that
\begin{equation}
\begin{split}
    \mathcal{J}_{11}(\mu, 0, s_0)= \left( \frac{e^\gamma \Lambda^2}{4 \pi} \right)^\frac{3-d}{2} \frac{\Gamma \left(2-\frac{d}{2} \right)}{(4 \pi)^\frac{d}{2}} \int_0^1 \text{d} x \text{P.V.} \int_{-\infty}^\infty \frac{\text{d}p_0}{2\pi} \left[ (p_0+i\mu-xs_0)^2\right]^\frac{d-4}{2}  \left[ 1 +\frac{x(1-x) s_0^2}{(p_0+i\mu-xs_0)^2} \right]^\frac{d-4}{2}.
    \end{split}
\end{equation}
In a manner similar to the computation in section \ref{sec:extensiontocomplex} (see eq.~\eqref{eq:PVcomplexsplit}), we note that the previously excluded values $p_0 \in \{0, s_0\}$ can be re-inserted to the principal value integration without altering the result (owing to dimensional regularization), leading to the sought after expansion to be
\begin{equation}
\begin{split}
\label{eq:FPtempexpansion}
      \mathcal{J}_{11}(\mu, 0, s_0)&=\left( \frac{e^\gamma \Lambda^2}{4 \pi} \right)^\frac{3-d}{2} \frac{\Gamma \left(2-\frac{d}{2} \right)}{(4 \pi)^\frac{d}{2}} \int_0^1 \text{d} x \text{P.V.} \int_{-\infty}^\infty \frac{\text{d}p_0}{2\pi} \left[ (p_0+i\mu-xs_0)^2\right]^\frac{d-4}{2}  \left[ 1 +\frac{x(1-x) s_0^2}{(p_0+i\mu-xs_0)^2} \right]^\frac{d-4}{2}\\
      &=\sum_{n=0}^\infty (-s_0^2)^n \left\{\int_0^1 \text{d} x [x(1-x)]^n \right\} \left( \frac{e^\gamma \Lambda^2}{4 \pi} \right)^{\frac{4-d}{2}} \frac{\Gamma \left(2+n-\frac{d}{2} \right)}{(4\pi)^\frac{d}{2} \Gamma \left(1+n \right)} \\
      &\times\int_0^\infty\frac{\text{d} p_0}{2\pi} \left\{\left[ (p_0+i\mu)^2 \right]^\frac{d-4-2n}{2}+\left[ (p_0-i\mu)^2 \right]^\frac{d-4-2n}{2} \right\} \\
      &\equiv \sum_{n=0}^\infty \left[ \binom{2n+1}{n} \right]^{-1} (-s_0^2)^n \mathcal{I}_{2+n} (\mu),
      \end{split}
\end{equation}
leading to a result with equivalent coefficients to those of eq.~\eqref{eq:FPvecexpansion}, and even more notably in alignment with the sought-after $s_0 \rightarrow 0$ limit of eq.~\eqref{eq:leadinglimit}.

\subsection{Comparison and remedy}\label{sec:comparison}
\noindent
In the previous two subsections, we have explicitly shown the details of each integration order, and their corresponding results. As suggested already in sections \ref{sec:spatialexpansionresidue} and \ref{sec:Fpspatialdirect}, the $s_0 = 0$ scenarios lead to explicitly equal results, as a direct consequence of the Heaviside step functions serving fully the role of the Fermi-Dirac distribution functions in equipping and conveying all scales present of the computation (compare eqs.~\eqref{eq:residuevecexpansion} and \eqref{eq:FPvecexpansion}). However, as seen in eqs.~\eqref{eq:naiveresidueexp} and \eqref{eq:FPtempexpansion}, not only is each order of expansion different between the two results, but also the strict $T = 0$ residue result yields an undesired result for the leading expression. This explicitly suggests that the residue result is unable to convey the results from non-vanishing (infinitesimal) temperature unlike the opposite order of integration does (as conveyed by the complex-valued extension of the Euler beta integral).

To remedy this specific case and enable the use of the residue theorem (leading to the sought-after expansion of eq.~\eqref{eq:FPtempexpansion}), we suggest that two properties are explicitly required from the modified residue integrand i) inclusion of the Fermi-Dirac distribution following the statements of \cite{Gorda:2022yex, Ostermanthesis} and ii) manifestly second order pole in the leading order of expansion. To achieve this, we suggest a simultaneous Feynman parametrization and inclusion of Fermi-Dirac distribution (without strict $T = 0$) prior to performing the temporal integration:
\begin{equation}
\begin{split}
\label{eq:remedyconstrunction}
  &\text{P.V.} \int_{-\infty}^\infty \frac{\text{d}p_0}{2\pi}  \frac{1}{[(p_0+i\mu)^2+p^2][(p_0+i\mu-s_0)^2+p^2]}\\
  &\mapsto\underset{T \to 0}{\text{lim}}\text{P.V.} \int_{-\infty}^\infty \frac{\text{d} p_0}{2 \pi} \int_0^1 \text{d} x \frac{\tilde{n}_\text{F} (p_0+i\mu)}{[p^2+(1-x)(p_0+i\mu)^2+x(p_0+i\mu-s_0)^2]^2}\\
  &\mapsto \int_0^1 \text{d} x \underset{T \to 0}{\text{lim}} i \text{Res} \left\{\frac{\tilde{n}_\text{F} (p_0+i\mu)}{[(p_0+i\mu-xs_0)^2+p^2+x(1-x)s_0^2]^2} \right\}_{p_0 =i\sqrt{p^2+x(1-x) s_0^2} - i \mu + xs_0}\\
  &= \int_0^1 \text{d} x \underset{T \to 0}{\text{lim}} i \text{Res} \left\{\frac{\tilde{n}_\text{F} (p_0+i\mu)}{[(p_0+i\mu)^2+p^2+x(1-x)s_0^2]^2} \right\}_{p_0 =i\sqrt{p^2+x(1-x) s_0^2} - i \mu }
    \end{split}
\end{equation}
wherein we note that the low-temperature limits of the Fermi-Dirac distribution and its derivatives are insensitive to real-valued shifts of the temporal coordinate such that
\begin{equation}
    \underset{T \to 0}{\text{lim}}\tilde{n}_\text{F}^{(k)} (p_0+i\mu \pm |a|) = \underset{T \to 0}{\text{lim}}\tilde{n}_\text{F}^{(k)} (p_0+i\mu ) = \theta^{(k)} \left(\text{Im} p_0 -\mu \right), 
\end{equation}
which justifies the last equality. As discussed earlier, Feynman parametrization sets the conditions $p_0 \not \in \{0, s_0\}$ onto the temporal integration. Inclusion of these conditions -- both at non-vanishing and vanishing temperature -- is irrelevant to the convergence of the temporal integral studied within the semicircular contour covering the upper complex half-plane, which instead implies the condition
\begin{equation}
    \sqrt{p^2+x(1-x)s_0^2}-\mu > 0.
\end{equation}
Thus, we find the target integrand to be well-behaved (under dimensional regularization and the cut-off condition), and can continue with the computation, interchanging the order of the spatial and parametric integrations. Then adopting the following abbreviation
\begin{equation}
    \text{n-Res} \left[ \frac{1}{(P-xS)^4} \right] \equiv  i \text{Res} \left\{\frac{\tilde{n}_\text{F} (p_0+i\mu)}{[(p_0+i\mu)^2+p^2+x(1-x)S^2]^2} \right\}_{p_0 =i\sqrt{p^2+x(1-x) S^2} - i \mu },
\end{equation}
the full formula is obtained through the mapping
\begin{equation}
    \begin{split}
    \label{eq:mappedFPresidues0}
        &\int_p  i\sum_{n=0}^1\text{Res} \left[ \frac{1}{[(p_0+i\mu)^2+p^2][(p_0+i\mu-s_0)^2+p^2]} \right]_{p_0 = ip-i\mu+ns_0}\\
        &\longmapsto \int_0^1\text{d} x \int_p  \underset{T \to 0}{\text{lim}}  \text{n-Res} \left[ \frac{1}{(P-xS)^4} \right]_{s=0}
        %&\longmapsto \int_0^1\text{d} x \int_p  \underset{T \to 0}{\text{lim}} i \text{Res} \left\{\frac{\tilde{n}_\text{F} (p_0+i\mu)}{[(p_0+i\mu)^2+p^2+x(1-x)s_0^2]^2} \right\}_{p_0 =i\sqrt{p^2+x(1-x) s_0^2} - i \mu }.
    \end{split}
\end{equation}

Let us explicitly observe the most crucial aspects of this integral construct:
\begin{itemize}
    \item The first exchanges of integration order take place between the temporal principal value integration at non-vanishing temperature and the parametric integral, rather than between the naive residue integration and the parametric integration. After the temporal integration of eq.~\eqref{eq:mappedFPresidues0} has been carried out, the relevant small parameter expansion --- with respect to $\frac{s_0}{\mu}$ --- can be studied in any order of integration.
    \item Once the second order pole and the distribution function are present, the temporal integral can be interpreted faithfully through the semicircular contour and the corresponding residue theorem result. 
    \item The differences from the strict $T=0$ residue theorem result are in alignment with the non-vanishing temperature derivation of $\mathcal{I}_2 (\mu)$ in \cite{Gorda:2022yex}. However, here the systematic order-by-order corrections manifest in each term of the $\frac{s_0^2}{\mu^2}$ power series (as is demonstrated below). 
\end{itemize}

\subsubsection{Explicit expansions}
\noindent
Having introduced a suggested solution to the problematic limit, let us next i) establish that it produces the correct expansion (for the case of $s=0$) and ii) confirm that the same strategy can be utilized to produce the sought-after results also in the case of purely spatial external momenta with $s_0 =0$.

To study i), following the convention of \cite{Osterman:2023tnt} we denote
\begin{equation}
    \tilde{n}_\text{F}' (p_0) = \frac{\text{d}}{\text{d} p_0} n_\text{F} [i \beta (p_0-i\mu)] = i \beta n_\text{F}' [i\beta(p_0-i\mu)]
\end{equation}

and then identify innermost residue structure of eq.~\eqref{eq:mappedFPresidues0} as
\begin{equation}
\begin{split}
\label{eq:finTress0}
   % & i \text{Res} \left\{\frac{\tilde{n}_\text{F}(p_0+i\mu)}{[(p_0+i \mu)^2+x(1-x)s_0^2 + p^2]^2} \right\}_{p_0 =i\sqrt{p^2+x(1-x) s_0^2} - i \mu }\\
    \text{n-Res} \left[ \frac{1}{(P-xS)^4} \right]_{s=0}&=-\frac{  \beta n_\text{F}'(-\beta \sqrt{p^2+x(1-x)s_0^2} + \beta \mu )}{4 [p^2+ x(1-x)s_0^2]} +\frac{   n_\text{F}(-\beta \sqrt{p^2+x(1-x)s_0^2} + \beta \mu )}{4 [p^2+ x(1-x)s_0^2]^\frac{3}{2}}  \\
    &=\frac{1}{2p} \frac{\text{d}}{\text{d}p} \left\{-\frac{   n_\text{F}(-\beta \sqrt{p^2+x(1-x)s_0^2} + \beta \mu )}{2 \sqrt{p^2+x(1-x)s_0^2}}\right\}\\
    \end{split}
\end{equation}
Before performing any further steps, we Taylor expand this expression in terms of the small parameter $x(1-x) s_0^2$ ($\ll p$) and find
\begin{equation}
\label{eq:smallparameterexpansions02}
    \begin{split}
        %&\frac{1}{2p} \frac{\text{d}}{\text{d}p} \left\{-\frac{   n_\text{F}(-\beta \sqrt{p^2+x(1-x)s_0^2} + \beta \mu )}{2 \sqrt{p^2+x(1-x)s_0^2}}\right\}\\
         &\text{n-Res} \left[ \frac{1}{(P-xS)^4} \right]_{s=0}\\
         &=\left(-\frac{1}{2p}\frac{\text{d}}{\text{d}p} \right) \left\{ \sum_{n=0}^\infty \frac{1}{n!}\left[\left(\frac{\text{d}}{\text{d}\left[ x(1-x)s_0^2\right]} \right)^n\frac{  n_\text{F}(-\beta \sqrt{p^2+x(1-x)s_0^2} + \beta \mu )}{2 [p^2+x (1-x)s_0^2]^\frac{1}{2}}\right]_{s_0=0} \left[x(1-x)s_0^2 \right]^n \right\}\\
    &= \left(-\frac{1}{2p}\frac{\text{d}}{\text{d}p} \right) \left\{ \sum_{n=0}^\infty \frac{1}{n!}\left[\left(-\frac{1}{2p}\frac{\text{d}}{\text{d} p} \right)^n\frac{  n_\text{F}(-\beta \sqrt{p^2+x(1-x)s_0^2} + \beta \mu )}{2 [p^2+x (1-x)s_0^2]^\frac{1}{2}}\right]_{s_0=0} \left[-x(1-x)s_0^2 \right]^n \right\}\\
    &=\sum_{n=0}^\infty\frac{1}{n!} \left(-\frac{1}{2p}\frac{\text{d}}{\text{d} p} \right)^{n+1}\frac{ n_\text{F}[-\beta ( p- \mu)]}{2 p} \left[-x(1-x)s_0^2 \right]^n, 
    \end{split}
\end{equation}
where we noted that the $x(1-x)s_0^2$ and $p^2$ appear in an equal footing, so the derivative can be shifted to act on the $p^2$ term and consequently the condition $s_0=0$ can be set trivially. Studying the vanishing-temperature limit in conjunction with the two remaining integrations, we can straightforwardly write this expansion in terms of the master integrals $\mathcal{I}_\alpha (\mu)$ of eq.~\eqref{eq:masterformualIa}, by utilizing partial integration in alignment with eq.~\eqref{eq:partialintegrationbubble}. This yields
\begin{equation}
\begin{split}
\label{eq:Taylors0expansionpartdiff}
\int_0^1 \text{d} x \int_p \underset{T \to 0}{\text{lim}} \text{n-Res} \left[ \frac{1}{(P-xS)^4} \right]_{s=0}
    &=\int_0^1 \text{d} x \int_p\left\{ \sum_{n=0}^\infty\frac{1}{n!} \left(-\frac{1}{2p}\frac{\text{d}}{\text{d} p} \right)^{n+1}\frac{  \theta\left(p-\mu \right)}{2 p} \left[-x(1-x)s_0^2 \right]^n \right\}\\
    &=\sum_{n=0}^\infty \mathcal{I}_{2+n} (\mu) (-s_0^2)^n\left[ \binom{2n+1}{n} \right]^{-1}, 
    \end{split}
\end{equation}
with each term in the expansion in alignment with the reverse order of integration, seen in eq.~\eqref{eq:FPtempexpansion}. We note that the same result can be straightforwardly found by isolating the hypergeometric representation (\textcolor{black}{see footnote} \ref{fn:hypergeom}) corresponding to the low-temperature limit of the last line of eq.~\eqref{eq:finTress0}:
\begin{equation}
\begin{split}
\label{eq:hypergeomderivationtos0expansion}
    & \int_0^1 \text{d} x\int_p \left(-\frac{1}{2p}\frac{\text{d}}{\text{d}p} \right) \left\{ \frac{  \theta\left(\sqrt{p^2+x(1-x) s_0^2}-\mu \right)}{2 [p^2+x (1-x)s_0^2]^\frac{1}{2}} \right\}\\
    &=\left( \frac{e^\gamma\Lambda^2}{4\pi}\right)^\frac{3-d}{2} \frac{(d-2) }{2(4\pi)^\frac{d}{2}\Gamma\left( \frac{d}{2}\right)}  \int_0^1 \text{d} x \int_0^\infty \text{d} y y^{\frac{d}{2}-2}  \left\{ \frac{  \theta\left(y-\mu^2+x(1-x)s_0^2 \right)}{2 [y+x (1-x)s_0^2]^\frac{1}{2}} \right\}\\
    &=-\left( \frac{e^\gamma\Lambda^2}{4\pi}\right)^\frac{3-d}{2} \frac{(d-2) }{2(4\pi)^\frac{d}{2}\Gamma\left( \frac{d}{2}\right)}  \int_0^1 \text{d} x\frac{(\mu^2-x(1-x) s_0^2)^{\frac{d-3}{2}}}{d-3}{}_2 F_1 \left[\frac{1}{2}, \frac{3-d}{2},\frac{5-d}{2},-\frac{x(1-x) s_0^2}{\mu^2-x(1-x)s_0^2} \right]
    \end{split}
\end{equation}
and through a Pfaff transformation \cite{Stegun} and expansion in the hierarchy $\frac{x(1-x)s_0^2}{\mu^2} \ll 1$, we find
\begin{equation}
\begin{split}
  \int_0^1 \text{d} x \int_p \underset{T \to 0}{\text{lim}} \text{n-Res} \left[ \frac{1}{(P-xS)^4} \right]_{s=0} &=\mathcal{I}_2 (\mu)\int_0^1 \text{d} x{}_2 F_1 \left[\frac{4-d}{2}, \frac{3-d}{2},\frac{5-d}{2},\frac{x(1-x) s_0^2}{\mu^2} \right]\\
    &= \sum_{n=0}^\infty \mathcal{I}_{2+n} (\mu) (-s_0^2)^n\left[ \binom{2n+1}{n} \right]^{-1}. 
    \end{split}
\end{equation}

To study similarly the purely spatial external momenta, $\{s\neq0 ,\, s_0 = 0\}$,  we note that the conditions of the Feynman parametrization at the start of section \ref{sec:Fpspatialdirect} are identical in this instance. The exchange of integration order between the temporal and parametric integral is a well defined operation under the semi-circular contour of
\begin{equation}
    \begin{split}
    \label{eq:finiteTresiduespatials}
        & \int_0^1 \text{d} x \underset{T \to 0}{\text{lim}}  \int_{-\infty}^\infty \frac{\text{d}p_0}{2\pi} \frac{\tilde{n}_\text{F} (p_0+i\mu)}{[(p_0+i\mu)^2+|\vec{p}-x \vec{s}|^2+x(1-x)s^2]^2},
    \end{split}
\end{equation}
which sets the condition (conveyed in parallel by the Fermi-Dirac distribution function)
\begin{equation}
    |\vec{p}-x\vec{s}|^2+x(1-x)s^2-\mu^2 > 0.
\end{equation}
To simplify the following steps, let us first exchange the integration order between the parametric ($x$) integration and the spatial integration. Then we can utilize the integration range of the spatial integration and perform a change of variables such that $\vec{p}-x\vec{s} \mapsto \vec{p}$. Thus, we can write the result of the relevant temporal residue as 
\begin{equation}
 \label{eq:spatialvariantofnFtrepresentation}
    \begin{split}
        \text{n-Res} \left[ \frac{1}{(P-xS)^4} \right]_{s_0=0} 
       &=-\frac{  \beta n_\text{F}'(-\beta \sqrt{p^2+x(1-x)s^2} + \beta \mu )}{4 [p^2+ x(1-x)s^2]} +\frac{   n_\text{F}(-\beta \sqrt{p^2+x(1-x)s^2} + \beta \mu )}{4 [p^2+ x(1-x)s^2]^\frac{3}{2}}\\
       &= \frac{1}{2p} \frac{\text{d}}{\text{d}p} \left[-\frac{   n_\text{F}(-\beta \sqrt{p^2+x(1-x)s^2} + \beta \mu )}{2 [p^2+x(1-x)s^2]^\frac{1}{2}}\right].
    \end{split}
\end{equation}
By observing that this result is structurally identical to the temporal external momentum scenario apart from the exchange $s_0^2 \mapsto s^2$, we can immediately write the full solution following steps similar to those given above:
\begin{equation}
    \begin{split}
       %&\int_0^1 \text{d} x\int_p \underset{T \to 0}{\text{lim}} i \text{Res} \left[\frac{\tilde{n}_\text{F} (p_0+i\mu)}{[(p_0+i\mu)^2+|\vec{p}-x \vec{s}|^2+x(1-x)s^2]^2} \right]_{p_0=-i \mu+i\sqrt{|\vec{p}-x\vec{s}|^2+x(1-x)s^2}} \\
       &\int_0^1 \text{d} x\int_p \underset{T \to 0}{\text{lim}} \text{n-Res} \left[ \frac{1}{(P-xS)^4} \right]_{s_0=0} = \sum_{n=0}^\infty \left[ \binom{2n+1}{n} \right]^{-1} (-s^2)^n \mathcal{I}_{2+n}(\mu),
    \end{split}
\end{equation}
yielding again a result in alignment with both eq.~\eqref{eq:residuevecexpansion} and \eqref{eq:FPvecexpansion}.

\subsubsection{Differences in residue expansions}
\label{sec:differencetermasymptotesresidue}
\noindent
Before moving to consider more general structures of external momenta $\{S^\nu\}_\nu$, let us lastly study more explicitly the impact of Feynman parametrization and the Fermi-Dirac distribution on the power series solutions. In each scenario, we isolate first the strict $T = 0$ residue theorem contributions, then consider the strict $T= 0$ residue theorem of the Feynman parametrized variant and lastly add also the Fermi-Dirac contribution to fully include non-vanishing temperature to the leading order results. 

Let us consider the $s = 0$ scenario first. Neglecting both Feynman parametrization and the distribution function, we find via the residue theorem interpretation of the temporal integral (once again using the semi-circular contour on top half of the complex plane)  
\begin{equation}
    \begin{split}
    \label{eq:naiveresiduesection4}
    %&\textcolor{red}{\int_{-\infty}^\infty \frac{\text{d} p_0}{2\pi}\frac{1}{[(p_0+i\mu)^2+p^2][(p_0+i\mu-s_0)^2+p^2]}}\\
         &i \sum_{n=0}^1 \text{Res} \left\{ \frac{1}{[(p_0+i\mu)^2+p^2][(p_0+i\mu-s_0)^2+p^2]}\right\}_{p_0 = -i\mu+ip +ns_0}\\
    &=\frac{\theta(p-\mu)}{2p}\frac{1}{-2ips_0+s_0^2} + \frac{\theta(p-\mu)}{2p}\frac{1}{2ips_0+s_0^2}\\
    &= \frac{\theta(p-\mu)}{p(4p^2+s_0^2)}
    \end{split}
\end{equation}
On the other hand, introducing the Feynman parametrization to the first line of equation \eqref{eq:naiveresiduesection4}, we find
\begin{equation}
\begin{split}
     \int_0^1 \text{d} x\, i\text{Res} \left\{ \frac{1}{[(p_0+i\mu)^2+p^2+x(1-x)s_0^2]^2}\right\}_{p_0 = -i\mu+\sqrt{p^2+x(1-x)s_0^2}}
    &=\int_0^1 \text{d} x\frac{\theta \left[\sqrt{p^2+x(1-x)s_0^2}-\mu \right]}{4 [p^2+x(1-x)s_0^2]^\frac{3}{2}},
    \end{split}
\end{equation}
wherein we can isolate the contributions of the above residue prescription straightforwardly by adding and subtracting the corresponding Heaviside step function:
\begin{equation}
\begin{split}
    &\int_0^1 \text{d} x\frac{\theta (p-\mu) }{4 [p^2+x(1-x)s_0^2]^\frac{3}{2}}+\int_0^1 \text{d} x\frac{\theta \left[\sqrt{p^2+x(1-x)s_0^2}-\mu \right]-\theta (p-\mu)}{4 [p^2+x(1-x)s_0^2]^\frac{3}{2}}\\
    &=\frac{\theta(p-\mu)}{p(4p^2+s_0^2)} +\int_0^1 \text{d} x\frac{ \theta\left[\sqrt{p^2+x(1-x)s_0^2}-\mu \right]-\theta (p-\mu)}{4 [p^2+x(1-x)s_0^2]^\frac{3}{2}}.
    \end{split}
\end{equation}
This result implies that Feynman parametrization (in isolation) introduces corrections from non-vanishing temperature, but as seen from eq.~\eqref{eq:finTress0}, the Fermi-Dirac distribution function is required for the missing subset proportional to Dirac delta function at the limit of vanishing temperature
\begin{equation}
\begin{split}
   &\int_0^1 \text{d} x \left\{  \text{n-Res} \left[ \frac{1}{(P-xS)^4}  \right]_{s=0}-  i\text{Res} \left\{ \frac{1}{[(p_0+i\mu)^2+p^2+x(1-x)s_0^2]^2}\right\}_{p_0 = -i\mu+\sqrt{p^2+x(1-x)s_0^2}}\right\}\\
    %&\left(\overset{T\to0}{=}-\int_0^1 \text{d} x\frac{  \beta n_\text{F}'(-\beta \sqrt{p^2+x(1-x)s_0^2} + \beta \mu )}{4 [p^2+ x(1-x)s_0^2]}\right)
    &\underset{T \to 0}{\longrightarrow}-\int_0^1 \text{d} x\frac{ \delta \left[ \sqrt{p^2+x(1-x)s_0^2} -\mu\right]}{4 [p^2+ x(1-x)s_0^2]}
    \end{split}
\end{equation}
Let us lastly comment, on the difference between naive residue expression (of eq.~\eqref{eq:naiveresiduesection4}) and the full infinitesimal-temperature structure:
\begin{equation}
\begin{split}
&\int_0^1  \text{d} x\, \underset{T \to 0^+}{\text{lim}}  \text{n-Res}  \left[ \frac{1}{(P-xS)^4}  \right]_{s=0}- \frac{\theta(p-\mu)}{p(4p^2+s_0^2)}  \\
&= \int_0^1 \text{d} x \left\{\frac{ \theta\left[\sqrt{p^2+x(1-x)s_0^2}-\mu \right]-\theta (p-\mu)}{4 [p^2+x(1-x)s_0^2]^\frac{3}{2}}- \frac{ \delta \left[ \sqrt{p^2+x(1-x)s_0^2} -\mu\right]}{4 [p^2+ x(1-x)s_0^2]} \right\}.
\end{split}
\end{equation}
Each term herein -- as seen through the $\frac{s_0^2}{\mu^2}$ power series -- is directly proportional to $\delta(p-\mu)$ or its derivatives (computed utilizing partial integration). This is an indication of how the strict residue expressions neglect all contributions from the temperature regulation, as discussed first in \cite{Gorda:2022yex}.

 While the second scenario ($s \neq s_0 = 0$) can be found to be formally in agreement with the leading temperature limit, the strict residue computation was performed in section \ref{sec:spatialexpansionresidue} by moving the external momentum fully from Heaviside step functions to the propagator, leading to the need to introduce additional regularization to evaluate the corresponding spatial integrals. Such differences (lack and presence of Dirac delta functions) makes comparing the two types of expansions (and identifying equality through the integrand thereof) exceedingly unpleasant. For this reason, we choose to instead work with an implicit structure of $q \equiv |\vec{p}-\vec{s}|$ generating a new spatial expansion parameter $q-p = \mathcal{O}(s)$. Utilizing this convention, we follow the steps shown directly above.

Studying the strict $T=0$ residue structure yields us a simple subset of eq.~\eqref{eq:residueT0fulllist}:
\begin{equation}
\begin{split}
\label{eq:naiveresiduespatial}
    & i \sum_{n=0}^1\text{Res} \left\{ \frac{1}{[(p_0+i\mu)^2+p^2][(p_0+i\mu)^2+q^2]} \right\}_{p_0 = -i\mu+inp+i(1-n)q}\\
    &=\frac{1}{q^2-p^2} \left[\frac{\theta(p-\mu)}{2p }-\frac{\theta(q-\mu)}{2q} \right]\\
    &= \frac{\theta(p-\mu)}{2 p q(q+p)}-\frac{\theta(q-\mu)-\theta(p-\mu)}{2q(q-p)(p+q)},
    \end{split}
\end{equation}
where we isolated a structure combining the sum of the individual residues by subtracting and adding the Heaviside step function with explicit dependence on the loop momentum $p$. This is further motivated by noting that $\forall p,q \neq 0$ 
\begin{equation}
\label{eq:basicFPpurelyspatial}
    \int_0^1 \text{d} x \frac{1}{4 \left[xp^2+(1-x)q^2 \right]^\frac{3}{2}} = \frac{1}{2 p q (p+q)}.
\end{equation}
Utilizing this also for the Feynman parametrized $T=0$ residue computation, we find
\begin{equation}
\begin{split}
    &\int_0^1 \text{d} x \int_{-\infty}^\infty \frac{\text{d} p_0}{2\pi} \frac{1}{[(p_0+i\mu)^2 +xp^2+(1-x)q^2]^2}\\
    &=\int_0^1 \text{d} x \frac{\theta \left[\sqrt{xp^2+(1-x)q^2}-\mu\right]}{4 \left[xp^2+(1-x)q^2 \right]^\frac{3}{2}}\\
    &= \frac{\theta(p-\mu)}{2 p q(q+p)} + \int_0^1 \text{d} x \frac{\theta(\sqrt{p^2+(1-x)(q^2-p^2)}-\mu)-\theta(p-\mu)}{4 \left[p^2+(1-x)(q^2-p^2) \right]^\frac{3}{2}},
    \end{split}
\end{equation}
where we not only isolated the piece combining the two poles, but also re-wrote the parametrized expressions to contain structures proportional to the small expansion parameter $q-p$, specifically $(1-x)(q^2-p^2) = (1-x)(p+q)(q-p) = \mathcal{O}(ps)$. While $q^2-p^2$ is significantly larger (and of different energy dimension) than $q-p = \mathcal{O}(s)$, we recognize that it manifests alongside $p^2$ in each functional structure. This facilitates the small parameter expansion given by $\frac{|q^2-p^2|}{p^2} = \mathcal{O}\left( \frac{s}{p}\right)$, with $p \gg s$. By directly computing the Feynman parametrized residue equipped with the Fermi-Dirac distribution function, we find utilizing the above notation
\begin{equation}
    \begin{split}
    \label{eq:fullresidueFPspatialpqexp}
       &i \text{Res} \left[\frac{\tilde{n}_\text{F} (p_0+i\mu)}{[(p_0+i\mu)^2+xp^2+(1-x)q^2]^2} \right]_{p_0=-i \mu+i\sqrt{xp^2+(1-x)q^2}} \\
       &=-\frac{  \beta n_\text{F}'\left[-\beta \sqrt{p^2+(1-x)(q^2-p^2)} + \beta \mu \right]}{4 [p^2+(1-x)(q^2-p^2)]} +\frac{   n_\text{F}\left[-\beta \sqrt{p^2+(1-x)(q^2-p^2)} + \beta \mu \right]}{4 [p^2+(1-x)(q^2-p^2)]^\frac{3}{2}}\\
       &\underset{T \to 0}{\longrightarrow} -\frac{  \delta\left[\sqrt{p^2+(1-x)(q^2-p^2)} - \mu \right]}{4 [p^2+(1-x)(q^2-p^2)]} +\frac{   \theta\left[\sqrt{p^2+(1-x)(q^2-p^2)} - \mu \right]}{4 [p^2+(1-x)(q^2-p^2)]^\frac{3}{2}}
    \end{split}
\end{equation}
 We observe that when subtracting the terms not proportional to $\delta (p-\mu)$ from both the naive expression of \eqref{eq:naiveresiduespatial} and \eqref{eq:fullresidueFPspatialpqexp}, the remainders match in each order of $\mathcal{O} (p-q)^n$:
\begin{equation}
\begin{split}
\label{eq:differencetermequalitynos0}
    &\int_0^1 \text{d} x \left\{\frac{\theta\left[\sqrt{p^2+(1-x)(q^2-p^2)}-\mu\right]-\theta(p-\mu)}{4 \left[p^2 +(1-x) (q^2-p^2)\right]^\frac{3}{2}} -\frac{\delta\left[\sqrt{p^2+(1-x)(q^2-p^2)}-\mu\right]}{4 \left[p^2 + (1-x)(q^2-p^2)\right]} \right\}\\
    &= -\frac{\theta[p-\mu+(q-p)]-\theta(p-\mu)}{2 [p+(q-p)](q-p)[2p+(q-p)]}. \\
    %&=-\int_0^1 \text{d} x \sum_{n=0}^\infty  \frac{p \delta^{(n)} (p-\mu) (p-q)^n  }{4 (n+1)! \left[p^2 +(1-x) (q^2-p^2)\right]^\frac{3}{2}}. \\
    \end{split}
\end{equation}
This result comes with a rather amusing implication: inclusion of Feynman parametrization (alone) to the strict $T=0$ residue theorem solution breaks the one-to-one correspondence of the expansion to the one arising from the non-vanishing temperature expansion. This emphasizes the exceedingly interesting and nontrivial correspondence with the Heaviside theta functions from the strict residue theorem compensating in all orders of expansion for both the Feynman parametrization and the low-temperature limit of Fermi-Dirac distribution even on the level of integrands, rather than just the leading part observed in \cite{Gorda:2022yex}. For a more complete study on the power series structures above and equivalence thereof, we refer an interested reader to appendix \ref{app:powerseries}.

\section{Fermionic pole structures from general external momentum}
\label{sec:generalfermionicproblem}
\noindent
In this section, we perform three computations in parallel\footnote{The computation involving integration order starting with spatial integration is essentially identical with the similarly ordered computation utilizing the Fermi-Dirac distribution function and Feynman parametrization, so we need not bother writing it separately.} involving arbitrary real-valued (apart from being bound by chemical potential $s,|s_0| \ll \mu$) components of external momenta following the concepts of the previous section. Specifically, we compare the strict $T = 0$ residue result against one that is derived utilizing contributions from non-vanishing temperature via both Feynman parametrization and the Fermi-Dirac distribution function. In a manner similar to the extrema presented in section \ref{sec:momentumasymptotes}, we extract expansions corresponding to each computation method, and specific to the strict residue result at $T = 0$, we additionally need to study hierarchies between $|s_0|$ and $s$ to find the corresponding closed form result. This is in stark contrast with how any order of loop integration involving Feynman parametrized propagator structure at infinitesimal temperature yields a consistent result, structurally independent of any mutual hierarchy between components of the external momentum\footnote{The hierarchy-free regularity of the re-parametrized expression largely owes to the translation properties of the temporal integration, leading all shifts to loop momenta being absorbed (cf. eq.~\eqref{eq:complexmumasterbubble}).} . The distinct results between the hierarchical naive residue strategy and the full leading order structure from infinitesimal temperature are qualitatively compared, and as a key observation following the structure of section \ref{sec:differencetermasymptotesresidue}, the two expansions contain only partial overlap with terms involving both $s_0$ and radial delta functions $\delta (p-\mu)$ (or its derivatives)\footnote{This implies that thermal contributions (from distributions functions) involving $s_0$ are neglected in the strict residue prescription. The details of this are more carefully discussed in section \ref{sec:generalcomparison}.}. 
\subsection{Feynman parametrization at infinitesimal temperature}
\noindent
Let $(p_0+i\mu)^2+p^2 \neq 0$ and $(p_0+i\mu-s_0)^2+|\vec{p}-\vec{s}|^2 \neq 0$. Then we obtain
\begin{equation}
    \frac{1}{[(p_0+i\mu)^2+p^2][(p_0+i\mu-s_0)^2+|\vec{p}-\vec{s}|^2]} = \int_0^1 \frac{\text{d} x}{[|\vec{p}-x \vec{s}|^2+(p_0+i\mu-xs_0)^2+x(1-x)S^2]^2},
\end{equation}
where we again utilize the shorthand $S^2 = s_0^2+s^2$. Considering first the integration order beginning with the dimensionally regularized spatial integration, we need not include the Fermi-Dirac distribution function to extract the leading part of the expression at non-vanishing (infinitesimal) temperature. In this order of integration, the conditions from the Feynman parametrization and the spatial integration -- $p_0 \neq x s_0$ to each $x$ individually \footnote{This is similar to how in section \ref{sec:spatialexpansasympts0} the convergence of radial integrals onto the Euler beta function adjacent results could be facilitated using constrained exponent parameters.} -- are moved to the temporal principal value integration directly. With the trivial change of spatial integration variable, $\vec{p}- x \vec{s} \mapsto \vec{p}$, we find a result reminiscent of eq.~\eqref{eq:afterspatilprincipalvalue1}, which reads
\begin{equation}
\begin{split}
\label{eq:FPPVfullintinitial}
\mathcal{J}_{11} (\mu, \vec{s}, s_0)&= \int_0^1 \text{d} x \text{P.V.} \int_{-\infty}^\infty \frac{\text{d}p_0}{2\pi} 
 \int_p\frac{1}{[|\vec{p}-x \vec{s}|^2+(p_0+i\mu-xs_0)^2+x(1-x)S^2]^2}\\
 &= \left( \frac{e^\gamma \Lambda^2}{4 \pi} \right)^\frac{3-d}{2} \frac{\Gamma \left(2-\frac{d}{2} \right)}{(4 \pi)^\frac{d}{2}}\int_0^1 \text{d} x \text{P.V.} \int_{-\infty}^\infty \frac{\text{d}p_0}{2\pi}   \left[ (p_0+i\mu-xs_0)^2+x(1-x) S^2 \right]^\frac{d-4}{2}.
\end{split}
\end{equation}
 We can observe that as a consequence of our assumptions $|p_0-xs_0+i\mu| = \sqrt{(p_0-xs_0)^2+\mu^2}>\mu \gg x(1-x)|S|$, which enables us to expand the power function straightforwardly before performing either of the integrations. Similar to the steps shown following eq.~\eqref{eq:afterspatilprincipalvalue1}, this leads -- with only $x(1-x)s_0^2$ replaced by $x(1-x)S^2$ in the expansion parameter -- to the result
 \begin{equation}
\begin{split}
\label{eq:FPtempexpansionfullS}
   \mathcal{J}_{11} (\mu, \vec{s}, s_0)&=    \left( \frac{e^\gamma \Lambda^2}{4 \pi} \right)^\frac{3-d}{2} \frac{\Gamma \left(2-\frac{d}{2} \right)}{(4 \pi)^\frac{d}{2}} \int_0^1 \text{d} x \text{P.V.} \int_{-\infty}^\infty \frac{\text{d}p_0}{2\pi} \left[ (p_0+i\mu-xs_0)^2\right]^\frac{d-4}{2}  \left[ 1 +\frac{x(1-x) S^2}{(p_0+i\mu-xs_0)^2} \right]^\frac{d-4}{2}\\
      &= \sum_{n=0}^\infty \left[ \binom{2n+1}{n} \right]^{-1} (-S^2)^n \mathcal{I}_{2+n} (\mu),
      \end{split}
\end{equation}
which we note to be symmetric with respect to $s_0$ and $s$.

As a direct follow-up, we continue by confirming that the strict residue theorem applied to the Feynman parametrized expression at non-vanishing temperature leads to identical result. Simplifying the parametric dependence of the loop integrand akin to eq.~\eqref{eq:remedyconstrunction}, we find ourselves with
 \begin{equation}
 \begin{split}
 \label{eq:finiteTfullSresidue}
     &\int_0^1 \text{d} x \int_p \underset{T \to 0}{\text{lim}} i \text{Res} \left[ \frac{\tilde{n}_\text{F} (p_0+i\mu)}{[(p_0+i\mu)^2+|\vec{p}-x \vec{s}|^2+x(1-x)S^2]}\right]_{p_0 = -i\mu + i \sqrt{|\vec{p}-x \vec{s}|^2+x(1-x)S^2}}\\
     &=\int_0^1 \text{d} x \int_p \underset{T \to 0}{\text{lim}} \text{n-Res}  \left[ \frac{1}{(P-xS)^4}  \right],
     \end{split}
 \end{equation}
where we changed the loop integration variable as previously  between the two lines. The residue operation  sets an explicit condition onto the radial integral of the form
\begin{equation}
     \sqrt{p^2+x(1-x)S^2}-\mu >0,
 \end{equation}
which is also enforced by the low-temperature limit of the Fermi-Dirac distribution function (acting as a Heaviside step function). The expression of eq.~\eqref{eq:finiteTfullSresidue} can be observed to be structurally identical with that of eq.~\eqref{eq:finiteTresiduespatials} or in alignment with eq.~\eqref{eq:finTress0}, with the sole difference arising from trivial substitutions in expansion parameters. Thus, either following the steps below eq.~\eqref{eq:finTress0} for expansion or seeking hypergeometric representation via steps onward from eq. \eqref{eq:hypergeomderivationtos0expansion}, we find 
 \begin{equation}
 \begin{split}
 \label{eq:FPresiduefullexpansioncaptialS}
 \int_0^1 \text{d} x \int_p \underset{T \to 0}{\text{lim}} \text{n-Res}  \left[ \frac{1}{(P-xS)^4}  \right]
     %&\int_0^1 \text{d} x \int_p \underset{T \to 0}{\text{lim}} i \text{Res} \left[ \frac{\tilde{n}_\text{F} (p_0+i\mu)}{[(p_0+i\mu)^2+|\vec{p}-x \vec{s}|^2+x(1-x)S^2]^2}\right]_{p_0 = -i\mu + i \sqrt{|\vec{p}-x \vec{s}|^2+x(1-x)S^2}}\\
     &=\sum_{n=0}^\infty \left[ \binom{2n+1}{n} \right]^{-1} (-S^2)^n \mathcal{I}_{2+n} (\mu),
     \end{split}
 \end{equation}
 in agreement -- as expected -- with eq.~\eqref{eq:FPtempexpansionfullS}.

\subsection{Strict residue theorem at vanishing temperature}
\label{sec:generalcasestrictresidue}
\noindent
To study the strict residue expressions, we start with eq.~\eqref{eq:residuecollection} and fix $\mu > 0$ and $\text{Im} s_0 = 0$. Then the sought-after integral reads
\begin{equation}
    \begin{split}
    \label{eq:fullexpansionimplicitresidueS}
        &\int_p \frac{\theta(p-\mu)}{2p}  \left[  \frac{1}{2p(i s_0+sz)+S^2}-\frac{1}{2p(i s_0+sz)-S^2} \right]\\
        &=\int_p \frac{\theta(p-\mu)}{4p^2 (is_0 + sz)} \sum_{n=0}^\infty \left[ \frac{S^2}{2p(is_0+sz)}\right]^n \left[(-1)^n - 1\right]\\
        &=-\int_p \frac{\theta(p-\mu)}{p} \sum_{n=0}^\infty  \frac{(s_0^2+s^2)^{2n+1}}{[4p^2(-s_0^2+2i s_0 sz+s^2 z^2)]^{n+1}}\\
        &\equiv\sum_{n=0}^\infty \mathfrak{J}_n(\mu,s,s_0),
    \end{split}
\end{equation}
where we utilize the hierarchy $p \gg |S|$ to consider $\frac{S^2}{p(is_0+sz)} = \mathcal{O} \left( \frac{|S|}{p} \right)$ as a proper expansion parameter. We can easily observe that the clearly distinct coefficients for $s_0$ and $s$ break the symmetry present in eq.~\eqref{eq:FPresiduefullexpansioncaptialS}. Below we will illustrate the details of how this takes place.

Permitting here the additional angular regulator $b$ of section \ref{sec:spatialexpansionresidue}, we can extract either of the two asymptotes studied in sections \ref{sec:spatialexpansionresidue} and \ref{sec:temporalexpansionresidueasymps0} by explicitly setting either of the two parameters $s$ or $s_0$ to vanish. As such, for a consistent result, we expect to be --- and indeed are --- able to observe these result from the corresponding limits also after the results with general external momentum have been constructed. 

Also given that the results characterized by subsequent powers of $\mu$ become increasingly convoluted, let us compute here explicitly the two leading orders, facilitating explicit comparison against the results of previous subsection. Then let us start with the $n= 0$ indexed term of eq.~\eqref{eq:fullexpansionimplicitresidueS}, given by
\begin{equation}
   \mathfrak{J}_0 (\mu,s,s_0)= -\int_p \frac{\theta(p-\mu)}{4 p^3} \frac{s_0^2+s^2}{-s_0^2+2i s_0 sz+s^2 z^2}
\end{equation}

To study properly study each part of expansions, we continue to utilize eq.~\eqref{eq:angularbetafunctionhelp}, which most notably enables us to study (by construction) negative powers of $z$. We also remind the reader that all odd values of $z$ powers are systematically identified to vanish due to the symmetry of the angular integration, which here works to remove all imaginary components of the expansions listed below. Some of the integrals diverge near $z=0$ and therefore the $b$-regulator introduced in equation \eqref{eq:b_regulated_integral} is used.

Let us begin with the mathematically improper expansion (given the possible values of the angular coordinate), characterized by the hierarchy $|s z| > |s_0|$ and write the non-vanishing part of the expansion using indexation:
\begin{equation}
\begin{split}
\label{eq:reject1initial}
    \mathfrak{J}_0 (\mu,s,s_0) &=-\int_p^\mathcal{R}\frac{\theta(p-\mu)}{4 p^3 z^2} \left(1+\frac{s_0^2}{s^2}\right)\left(1-\frac{s_0^2}{s^2 z^2}+\frac{2 i s_0}{sz} \right)^{-1}\\
    &=-\int_p^\mathcal{R} \frac{\theta(p-\mu)}{4 p^3 z^2} \sum_{k=0}^\infty \left(-\frac{s_0^2}{s^2 z^2} \right)^k (2k+1) \left(1+\frac{s_0^2}{s^2}\right)\\
    %&=-\int_p \frac{\theta(p-\mu)}{4 p^3 z^2} +\int_p \frac{\theta(p-\mu)}{4 p^3}\sum_{k=1}^\infty \left(-\frac{s_0^2}{s^2 z^2} \right)^k\left(2k-1-\frac{2k+1}{z^2} \right).\\
    %&=-\int_p \frac{\theta(p-\mu)}{4 p^3 z^2} +\left( \frac{e^\gamma \Lambda^2}{4\pi} \right)^\frac{3-d}{2} \frac{\mu^{d-3}}{2 \pi (4\pi)^\frac{d-1}{2} (d-3) }  \sum_{k=1}^\infty \left(-\frac{s_0^2}{s^2} \right)^k\left[ \frac{k-\frac{1}{2}}{\frac{d}{2}-1-k}+1 \right] \frac{\Gamma \left(\frac{1}{2}-k \right)}{\Gamma  \left( \frac{d}{2}-k-1 \right)}\\
    \end{split}
    \end{equation}
 Let us also note that despite there being temptation to close the Taylor series to find an expression such that
 \begin{equation}
    \begin{split}
    \sum_{k=0}^\infty \left(-\frac{s_0^2}{s^2 z^2} \right)^k (2k+1) \left(1+\frac{s_0^2}{s^2} \right)  &= \frac{(s^2 z^2- s_0^2)(s^2 z^2)}{(s_0^2+s^2 z^2)^2} \left(1+\frac{s_0^2}{s^2} \right),
    \end{split}
\end{equation}
we find that simplifications after performing the angular $z$ integration give further insight to the structure of the terms. Accordingly, we utilize eq.~\eqref{eq:angularbetafunctionhelp} and the properties of Euler gamma function, leading straightforwardly to the solution
    \begin{equation}
    \begin{split}
    \label{eq:reject1}
     \mathfrak{J}_0 (\mu,s,s_0) 
    &=-\int_p^\mathcal{R} \frac{\theta(p-\mu)}{4 p^3 z^2} +\left( \frac{e^\gamma \Lambda^2}{4\pi} \right)^\frac{3-d}{2} \frac{\mu^{d-3}}{4 \pi (4\pi)^\frac{d-1}{2} }  \sum_{k=1}^\infty \left(-\frac{s_0^2}{s^2} \right)^k \frac{\Gamma \left(\frac{1}{2}-k \right)}{\Gamma  \left( \frac{d}{2}-k \right)}\\
    %&= \mathcal{I}_2 (\mu) \left[1+\frac{s_0^2}{s^2} (d-3) \dots \right]\\
    &= \mathcal{I}_2 (\mu) + (d-3) \int_p^\mathcal{R}\frac{\theta(p-\mu)}{4 p^3} \left[\left(1+\frac{s_0^2}{s^2 z^2} \right)^{-1}-1\right]\\
     &=\mathcal{I}_2 (\mu) -(d-3)  \int_p^\mathcal{R}\frac{\theta(p-\mu)s_0^2}{4 p^3 (s_0^2+s^2 z^2)}, 
     \end{split}
\end{equation}
where the expression on the last line has been re-cast to most easily correspond to the assumed hierarchy (and hence extract the leading solution $\mathcal{I}_2 (\mu)$). We again emphasize that the full expansion seen in the above steps is only admissible with the utilized angular regularization, but the opposite expansion assuming $|s_0| \gg s$ is accessible without it. The solution on the last line(s) can also be seen to generate equivalent $|s_0| \gg s$ expansion as one finds by directly utilizing eq.~\eqref{eq:fullexpansionimplicitresidueS}:
\begin{equation}
    \begin{split}
    \label{eq:reject3}
        \mathfrak{J}_0 (\mu,s,s_0) &=\int_p \frac{\theta(p-\mu)}{4 p^3} \sum_{n=0}^\infty\left(1+\frac{s^2}{s_0^2} \right) \left[1-\frac{2i  sz}{s_0}-\frac{s^2 z^2}{s_0^2} \right]^{-1}\\
         &=\int_p \frac{\theta(p-\mu)}{4 p^3}\sum_{k=0}^\infty \left(\frac{-s^2 z^2}{s_0^2} \right)^k (2k+1) \left(1+\frac{s^2}{s_0^2} \right)\\
        %&=\int_p \frac{\theta(p-\mu)}{4 p^3} -\int_p \frac{\theta(p-\mu)}{4 p^3}\sum_{k=1}^\infty \left(-\frac{s^2 z^2 }{s_0^2} \right)^k\left[\frac{2k-1}{z^2}-2k+1 \right]\\
        &=\text{Res}\left[\mathcal{I}_2 (\mu) \right]+ \left( \frac{e^\gamma \Lambda^2}{4\pi} \right)^\frac{3-d}{2} \frac{\mu^{d-3}}{4 \pi (4\pi)^\frac{d-1}{2} }  \sum_{k=1}^\infty \left(-\frac{s^2}{s_0^2} \right)^k  \frac{\Gamma \left(\frac{1}{2}+k \right)}{\Gamma  \left( \frac{d}{2}+k \right)} \\
         &=\text{Res}\left[\mathcal{I}_2 \right] (\mu) + (d-3)\int_p  \frac{\theta(p-\mu)s^2 z^2 }{4 p^3 (s_0^2+s^2 z^2)}.\\
        %&=\mathcal{I}_2 (\mu) - (d-3)\int_p  \frac{\theta(p-\mu)s_0^2}{4 p^3 (s_0^2+s^2 z^2)}
    \end{split}
\end{equation}
It is trivial to discern that the workload of simplifying the coefficients of the descending powers of $\mu$ to closed form expressions is essentially a binomially growing process. For this purpose, we settle to illustrating the next-to-leading order computation of eq.~\eqref{eq:fullexpansionimplicitresidueS}  in either hierarchy. Following the same kind of steps as above, the hierarchy $|sz| \gg |s_0|$ yields us 
\begin{equation}
\begin{split}
\label{eq:reject2}
    %&-\int_p \frac{\theta(p-\mu)}{16 p^5} \frac{(s_0^2+s^2)^3}{(-s_0^2+2i s_0 sz+s^2 z^2)^2}\\
   &\mathfrak{J}_1(\mu,s,s_0)= -\int_p^\mathcal{R}\frac{\theta(p-\mu)s^2}{16 p^5 z^4} \left(1+\frac{s_0^2}{s^2}\right)^3 \left(1-\frac{s_0^2}{s^2 z^2}+\frac{2i s_0}{sz} \right)^{-2}\\
    &=-\int_p^\mathcal{R} \frac{\theta(p-\mu)}{16 p^5 z^4}  \sum_{n=0}^\infty \binom{2n+3}{3}  \left(-\frac{s_0^2}{s^2 z^2}\right)^n \left[  \left(s^2+\frac{s_0^6}{s^4 } \right) +3\left(s_0^2+\frac{s_0^4}{s^2} \right)\right]\\
   &=-\frac{S^2}{3} \mathcal{I}_3 (\mu) -\int_p^\mathcal{R} \frac{\theta(p-\mu)}{16 p^5} \left\{\frac{2(d-5)s_0^2}{z^4}+\frac{(d-3)(d-5)s_0^2}{3}\sum_{k=0}^\infty\left(-\frac{s_0^2}{s^2 z^2}\right)^{k+1} \left[k-\frac{3+k}{z^{2}} \right] \right\}\\
    &= -\frac{S^2}{3}\mathcal{I}_3 (\mu)-\frac{2(d-5)s_0^2}{3} \mathcal{I}_3 (\mu) -\frac{(d-3)(d-5)s_0^2}{3}\int_p^\mathcal{R}\frac{\theta(p-\mu)}{16 p^5} \left[\frac{s_0^4}{(s_0^2+s^2 z^2)^2} \left( 1-\frac{1}{z^2}\right)+\frac{3s_0^2}{(s_0^2+s^2 z^2)z^2}\right],
    %&= -\frac{S^2}{3}\mathcal{I}_3 (\mu)-\frac{2(d-5)s_0^2}{3} \mathcal{I}_3 (\mu) -\int_p \frac{\theta(p-\mu)}{16 p^5} \left\{\frac{(d-3)(d-5)s_0^2}{3}\left[\frac{x^2}{(1+x)^2} \left( 1-\frac{1}{z^2}\right)+\frac{3x}{(1+x)z^2}\right]_{x= \frac{s_0^2}{s^2 z^2}} \right\}
    \end{split}
\end{equation}
where the expression on the third row is found by simplifying the Euler functions arising from the angular $z$ integrals alongside with the indexed coefficients (cf. the first and second row of eq.~\eqref{eq:reject1}). The opposite hierarchy of $|s_0| \gg s$, can be studied in identical manner, and then be utilized to generate an equivalent closed form result from the distinct expansion of 
\begin{equation}
\begin{split}
\label{eq:reject4}
    %&-\int_p \frac{\theta(p-\mu)}{16 p^5} \frac{(s_0^2+s^2)^3}{(-s_0^2+2i s_0 sz+s^2 z^2)^2}\\
    &\mathfrak{J}_1(\mu,s,s_0)= -\int_p \frac{\theta(p-\mu)s_0^2}{16 p^5 } \left(1+\frac{s^2}{s_0^2}\right)^3 \left(1-\frac{s^2 z^2}{s_0^2 }-\frac{2i sz}{s_0} \right)^{-2}\\
    &=-\int_p \frac{\theta(p-\mu)}{16 p^5}  \sum_{n=0}^\infty \binom{2n+3}{3}  \left(-\frac{s^2 z^2}{s_0^2}\right)^n \left[  \left(s_0^2+\frac{s^6}{s_0^4 } \right) +3\left(s^2+\frac{s^4}{s_0^2} \right)\right]\\
  &=-\frac{S^2}{3} \text{Res} \left[\mathcal{I}_3 \right] (\mu) -\int_p \frac{\theta(p-\mu)}{16 p^5} \left\{2(d-5)s^2 z^2+\frac{(d-3)(d-5)s^2}{3}\sum_{k=0}^\infty \left(-\frac{s^2 z^2}{s_0^2 }\right)^{k+1} \left[k-(3+k)z^2 \right] \right\}\\
  &\overset{}{=}-\frac{S^2}{3} \text{Res} \left[\mathcal{I}_3 \right] (\mu) -\int_p \frac{\theta(p-\mu)}{16 p^5} \left\{2(d-5)s^2 z^2+\frac{(d-3)(d-5)s^2}{3}\left[\frac{s^4z^4}{(s_0^2+s^2 z^2)^2} \left(1-z^2 \right)+\frac{3 s^2 z^4}{s_0^2+s^2 z^2} \right]\right\}
    \end{split}
\end{equation}

However, the most significant structural information can be noted by studying the leading order solution seen in either eq.~\eqref{eq:reject1} or eq.~\eqref{eq:reject3}. While we do indeed see the consistency of the residue results tending towards $\mathcal{I}_2 (\mu)$, when $s_0 \rightarrow 0$ and $\text{Res} \left[\mathcal{I}_2 \right] (\mu)$, when $s \rightarrow 0$, we again see that this consistency goes against the behavior demanded in eq.~\eqref{eq:leadinglimit}. Furthermore, the external parts proportional to $d-3$ in either expansion introduce even more problems in this regard. Specifically, for eq.~\eqref{eq:leadinglimit} to hold, we should as well be able to fix for example $s_0 = 2s$, which would explicitly lead to the residue result of
\begin{equation}
    \begin{split}
      \mathfrak{J}_0(\mu,s,2s) =   \mathcal{I}_2 (\mu) - (d-3)\int_p  \frac{\theta(p-\mu)}{ p^3 (4+ z^2)}.
    \end{split}
\end{equation}
This implies an explicit non-vanishing hypergeometric excess term (of the same order in $\mathcal{O}(\mu^{d-3})$) in addition to the sought-after low-temperature limit. Let us also emphasize that the differences from the inifinitesimal-temperature expressions are consistently characterized by overall $d-3$ (see eq.~\eqref{eq:reject1}) or $(d-3)(d-5)$ (see eq.~\eqref{eq:reject2}) coefficients. This essentially ties the corresponding integrands to being proportional to the (derivatives of) Dirac delta functions $\delta (p-\mu)$. This signifies that the parts of the expansions not proportional to the delta functions are systematically identical in the integrand level between the two approaches. Instead, the parts associated with the delta function signify at least partially distinct contributions, which explicitly match for the $s_0 = 0$ case. For $s_0 \neq 0$, the tower of differences between integration orders can be seen as $s_0$ working as a regulator in place of infinitesimal temperature or dimension $d$.

\subsection{Comparison}
\label{sec:generalcomparison}
\noindent
Let us finally study the two computation strategies in a manner similar to the discussion in section \ref{sec:differencetermasymptotesresidue}. Given that we identified all the relevant integrand level differences of the expansions, eqs.~\eqref{eq:FPresiduefullexpansioncaptialS} and \eqref{eq:fullexpansionimplicitresidueS}, to take place in such parts that are characterized by spatial delta functions, we work utilizing the shorthand $q \equiv |\vec{p}-\vec{s}|$ and $q-p$ as the small spatial expansion parameters. Then the spatial integrand from the prior temporal residue integration reads
\begin{equation}
\begin{split}
\label{eq:naiveresiduegeneral}
    &\sum_{n=0}^1 i \text{Res} \left[\frac{1}{[(p_0+i\mu)^2+p^2][(p_0+i\mu-s_0)^2+q^2]}\right]_{p_0 = -i \mu+i n p + (1-n)(s_0+i q)}\\
    &=\frac{\theta(p-\mu)}{2p}\frac{1}{(ip-s_0)^2+q^2} + \frac{\theta(q-\mu)}{2q}\frac{1}{(iq +s_0)^2+p^2} \\
    &=\theta(p-\mu)\left[\frac{1}{2p}\frac{1}{(ip-s_0)^2+q^2} + \frac{1}{2q}\frac{1}{(iq +s_0)^2+p^2}\right] +\frac{\theta(q-\mu)-\theta(p-\mu)}{2q}\frac{1}{(iq +s_0)^2+p^2} \\
     &= \frac{(p+q)\theta(p-\mu)}{2pq [(p+q)^2+s_0^2]}+\frac{\theta(q-\mu)-\theta(p-\mu)}{2q}\frac{1}{(iq +s_0)^2+p^2}\\
    %&= \frac{(p+q)\theta(p-\mu)}{2pq [(p+q)^2+s_0^2]}+\frac{\theta[p-\mu+(q-p)]-\theta(p-\mu)}{2[p+(q-p)]} \frac{1}{[i(q-p)+s_0][2ip+i(q-p)+s_0]},
    \end{split}
\end{equation}
where we again note that by extension of eq.~\eqref{eq:basicFPpurelyspatial}, 
\begin{equation}
\label{eq:thetalevelFP}
    \int_0^1  \frac{\text{d} x}{4[x p^2+ (1-x)q^2 +x (1-x)s_0^2]^\frac{3}{2}} = \frac{(p+q)}{2 pq [(p+q)^2+s_0^2]},
\end{equation}
we can observe parts proportional to $\theta(p-\mu)$ to be (fully) in common parts between the two expansions. The remaining (partially distinct) terms are notably proportional to differentiated Dirac delta functions $\delta^{(k)}(p-\mu)$, and utilizing the expansion parameters $\{q-p, s_0\}$ we can write them such that
\begin{equation}
\label{eq:generalfalseresidueexpansion1}
   \frac{\theta(q-\mu)-\theta(p-\mu)}{2q}\frac{1}{(iq +s_0)^2+p^2}= \frac{\theta[p-\mu+(q-p)]-\theta(p-\mu)}{2[p+(q-p)]} \frac{1}{[i(q-p)+s_0][2ip+i(q-p)+s_0]}
\end{equation}
When expanding this structure, we encounter two specific properties: terms that are purely imaginary and terms that manifest via a chosen hierarchy between expansion parameters, either $|s_0| > |p-q|$ or $|s_0| < |p-q|$.

Neither of these properties occurs in the full infinitesimal-temperature Feynman-parametrized integration scheme. To illustrate, we explicitly find the corresponding spatial integrand via
\begin{equation}
  \frac{1}{[(p_0+i\mu)^2+p^2][(p_0+i\mu-s_0)^2+q^2]}  \longmapsto \int_0^1 \text{d} x  \frac{\tilde{n}_\text{F} (p_0+i\mu)}{[(p_0+i\mu)^2+xp^2+(1-x)q^2+x(1-x)s_0^2]^2}
\end{equation}
and calculate the corresponding $p_0$-integral with the residue theorem
\begin{equation}
    \begin{split}
    \label{eq:residueofFPfullS}
       &i \text{Res} \left[\frac{\tilde{n}_\text{F} (p_0+i\mu)}{[(p_0+i\mu)^2+xp^2+(1-x)q^2+x(1-x)s_0^2]^2} \right]_{p_0=-i \mu+i\sqrt{xp^2+(1-x)q^2+x(1-x)s_0^2}} \\
       &=-\frac{  \beta n_\text{F}'\left[-\beta \sqrt{x p^2+(1-x)q^2 + x(1-x)s_0^2} + \beta \mu \right]}{4 [x p^2+(1-x)q^2 +x(1-x)s_0^2]} +\frac{   n_\text{F}\left[-\beta \sqrt{x p^2+(1-x) q^2+x(1-x)s_0^2} + \beta \mu \right]}{4 [x p^2+(1-x) q^2 + x(1-x)s_0^2]^\frac{3}{2}}\\
       &\underset{T \to 0}{\longrightarrow}\frac{\theta(p-\mu)}{4 [x p^2+(1-x) q^2 + x(1-x)s_0^2]^\frac{3}{2}} \\
       &-\frac{  \delta\left[\sqrt{p^2+(1-x)(q^2-p^2)+x(1-x)s_0^2} - \mu \right]}{4 [p^2+(1-x)(q^2-p^2)+x(1-x)s_0^2]} +\frac{   \theta\left[\sqrt{p^2+(1-x)(q^2-p^2)+x(1-x)s_0^2} - \mu \right]-\theta(p-\mu)}{4 [p^2+(1-x)(q^2-p^2)+x(1-x)s_0^2]^\frac{3}{2}},
    \end{split}
\end{equation}
where in the last step we have not only taken the explicit vanishing-temperature limit, and also isolated the parts explicitly proportional to $\theta(p-\mu)$ that fully overlap between eqs.~\eqref{eq:naiveresiduegeneral} and \eqref{eq:residueofFPfullS}. Furthermore, we again emphasize that the two expansion parameters can be expanded individually, leading only to real-valued terms, albeit the expansion is enacted using more convoluted expressions. Specifically, the expansion parameters are equipped with additional prefactors such that $x(1-x)s_0^2$ and $(1-x)(q^2-p^2) = (1-x) [2p+(q-p)](q-p)$. While we will mostly avoid explicit discussions of the expansions, let us note that in alignment with previous sections and appendix \ref{app:powerseries}, we can straightforwardly write the expanded (and parametrically integrated) spatial integrand implicitly -- utilizing eq.~\eqref{eq:smallparameterexpansions02} -- as
\begin{equation}
\begin{split}
\label{eq:implicitdifferentiatedthetaseries}
&\int_0^1 \text{d} x \left\{-\frac{  \delta\left[\sqrt{p^2+(1-x)(q^2-p^2)+x(1-x)s_0^2} - \mu \right]}{4 [p^2+(1-x)(q^2-p^2)+x(1-x)s_0^2]} +\frac{   \theta\left[\sqrt{p^2+(1-x)(q^2-p^2)+x(1-x)s_0^2} - \mu \right]}{4 [p^2+(1-x)(q^2-p^2)+x(1-x)s_0^2]^\frac{3}{2}}\right\}\\
    &=\int_0^1 \text{d} x\sum_{n=0}^\infty \frac{(-1)^n}{n!} \left[\left(-\frac{1}{2p}\frac{\text{d}}{\text{d} p} \right)^{n+1}\frac{ \theta ( p- \mu)}{2 p}\right]  [(1-x)(q^2-p^2)+x(1-x)s_0^2]^n\\
    &=\sum_{n=0}^\infty \sum_{k=0}^n \frac{(-1)^n n!}{(2n-k+1)!k!}  \left[\left(-\frac{1}{2p}\frac{\text{d}}{\text{d} p} \right)^{n+1}\frac{ \theta ( p- \mu)}{2 p}\right] (q^2-p^2)^k (s_0^2)^{n-k},
    \end{split}
\end{equation}
for which the differentiated distribution term has been explicitly written in eq.~\eqref{eq:closedformsquarebrackets}.

Given our earlier observations concerning the differences of the two expansion schemes, to isolate terms in common between eq.~\eqref{eq:generalfalseresidueexpansion1} the bottom row of eq.~\eqref{eq:residueofFPfullS}, we begin by isolating the minimal set of imaginary and scale hierarchy dependent parts of eq.~\eqref{eq:generalfalseresidueexpansion1}. To illustrate this, we can write
\begin{equation}
\begin{split}
    &-\frac{\theta(q-\mu)-\theta(p-\mu)}{2q} \frac{1}{[(q-p)-is_0][p+q-is_0]}\\
    %&=- \frac{\theta(q-\mu)-\theta(p-\mu)}{2pq} \frac{p(p+q+is_0)}{[(q-p)-is_0][(p+q)^2+s_0^2]}\\
    %&=- \frac{\theta(q-\mu)-\theta(p-\mu)}{2pq} \frac{2pq}{[(q-p)-is_0][(p+q)^2+s_0^2]}+\frac{\theta(q-\mu)-\theta(p-\mu)}{2pq} \frac{p}{(p+q)^2+s_0^2}\\
   % &=-  \frac{\theta(q-\mu)-\theta(p-\mu)}{[(q-p)-is_0][(p+q)^2+s_0^2]} +\frac{p[\theta(q-\mu)-\theta(p-\mu)]}{p+q} \frac{p+q}{2pq}\frac{1}{(p+q)^2+s_0^2}\\
    &=- \frac{\theta(q-\mu)-\theta(p-\mu)}{q-p}  \frac{1}{(p+q)^2+s_0^2} + \frac{\theta(q-\mu)-\theta(p-\mu)}{2q}\frac{1}{(p+q)^2+s_0^2}\\
    &-  \frac{is_0 [\theta(q-\mu)-\theta(p-\mu)]}{(q-p)[(q-p)-is_0][(p+q)^2+s_0^2]},
    \end{split}
\end{equation}
where the bottom row contains exclusively terms that are either generated by scale hierarchy from the structure $[(q-p)-i s_0]^{-1}$ or are imaginary. Thus, we find all interesting parts on the second to last row, and to alleviate the effort in comparison, we should re-insert parts proportional to $\theta(p-\mu)$ and express terms utilizing any proportionality to Feynman parametrization formula in eq.~\eqref{eq:thetalevelFP}:
\begin{equation}
\begin{split}
\label{eq:naiveresiduegeneratorhierarchyfree}
    &\frac{p+q}{2pq [(p+q)^2+s_0^2]} \left\{\theta(p-\mu)- \frac{2pq [\theta(q-\mu)-\theta(p-\mu)]}{(p+q)(q-p)}  +\frac{p[\theta(q-\mu)-\theta(p-\mu)]}{p+q} \right\}
    \end{split}
\end{equation}
All individual terms in the expansion(s) can be expressed as
\begin{equation}
    \frac{\theta^{(n)} \textcolor{black}{(p-\mu)} (s_0^2)^l(q-p)^j}{p^{k}}
\end{equation}
for $n \in \mathbb{Z}_+$. Then eq.~\eqref{eq:naiveresiduegeneratorhierarchyfree} constrains parameters $j,l \in \mathbb{N}_0$ such that $j \geq n-1$ and $k \geq 2l+j-n+3$. Eq.~\eqref{eq:residueofFPfullS} (or eq.~\eqref{eq:implicitdifferentiatedthetaseries}) sets the constrains more loosely, instead demanding $0\leq j$ and $2+l \leq k$ --- or alternatively introducing additional parametric region characterized by $0\leq j < n-1$ and $2+l \leq k < 2l+j-n+3$.

Interestingly, all terms generated by eq.~\eqref{eq:naiveresiduegeneratorhierarchyfree} are contained in the same small parameter expansion of eq.~\eqref{eq:residueofFPfullS}, in other words it can be seen as a subset characterized by the constrains given above. We also refer the reader to appendix \ref{app:powerseries}, for an illustration of simpler subset problem as an indication of how the terms match one another.

\subsection{Summary}
\noindent
 First, the strict residue prescription at $T=0$ fails to take into consideration all contributions from non-vanishing temperature. This breakdown manifests most clearly in the subset case of $s_0 \neq s = 0$, where all terms of the expansion with (spatial) integrands corresponding to the Dirac delta function (and its derivatives) $\delta (p-\mu)$ are exclusive to the Feynman parametrized expression at non-vanishing temperature.  The terms without $s_0$ components align perfectly with either computation method, but only as a consequence of the Heaviside step function acting identically (in terms of contained scales) to the correct low-temperature limit of the Fermi-Dirac distribution function.

For the case with all non-vanishing external momentum scales $\{s_0, s\}$, the two hierarchies obey either $|s_0| > s$ or $|s_0| < |sz|$. The hierarchies lead in the strict $T=0$ limit to a tower of $s_0$ dependent terms that share partial similarities with thermal corrections from the Feynman parametrized expressions, essentially with all the hierarchy free real-valued parts of the naive residue expansion occurring also in the Feynman parametrized expression at infinitesimal temperature. We again emphasize that the expansions are not fully distinct, but any two fermionic poles that can overlap (as faciliated by a small temporal shift $s_0$) generate a numerably infinite set of distinct expansion terms between the two strategies. 

Most damningly, characteristic to the leading order contributions to any expansion containing $s_0$ for the strict residue theorem, eq.~\eqref{eq:leadinglimit} can not be achieved. Despite the above, we acknowledge that the strict residue theorem has self-consistent expansions when utilizing the angular generalization of dimensional regularization (this is to say that the small parameter expansions lead to equivalent analytic continuations in either hierarchy).

As for the Feynman parametrization scheme, it establishes itself through multiple noteworthy qualities emphasizing its role as the definitive approach to study fermionic overlapping poles at non-vanishing temperature. First, let us note that Feynman parametrization is essential to computing the spatial integrations prior to the temporal ones, and that the scheme facilitates the correct way of studying expansions and the leading temperature behavior particular to this order of integrations\footnote{As noted in \cite{Gorda:2022yex}, the order of integration with spatial prior to temporal is viewed as a legitimate means of extracting the dimensionally regulated leading temperature behavior, given that the spatial integration can be performed at non-vanishing temperature at no additional complexity and with only minimal constraints set on the outermost thermal Matsubara sums (as well as due to the general assumption that the correct analytic continuation facilitates any integration order).}. In addition to this, the scheme yields identical results in any ordering of the loop integrals (spatial to temporal or temporal to spatial). These results facilitate the exchange of the order of integration and (external momentum) limit(s) as described by eq.~\eqref{eq:leadinglimit}. 

To contrast the computation strategies, we also emphasize that the two residue methods  agree -- in the context of the fermionic example -- only when the strict $T=0$ residue prescription yields the correct leading term as eq.~\eqref{eq:leadinglimit}, i.e. with scales $\{s_0 = 0, s>0\}$. Specific to the Feynman parametrization scheme, we also observe a correct expansion  (much akin to the vacuum theories) independent of any mutual hierarchy of $\{s_0, s\}$, explicitly absent from the strict residue approach.

Lastly, let us note that while the  bosonic counterpart of the studied integral is not the focus of this work, the properties of related expansions do not carry over from the fully fermionic study. Specifically, for a case with one bosonic and one fermionic propagator, the poles can not overlap, and hence the residue theorem automatically predicts the leading order correctly. This reliability extends to general small parameter expansions in terms of external loop momenta. For illustrative computations, we direct the reader to  appendix \ref{sec:bosons}. 

\section{Conclusions}
\label{sec:conclusions}
\noindent

\noindent
In this work, we have highlighted a previously overlooked ambiguity related to integration orders in finite density perturbative computations. This ambiguity emerges even (or particularly) in one-loop Feynman integrals containing external momenta, $S^\nu = (s_0, \vec{s})$. It becomes particularly apparent when comparing the leading order contributions in the following set of integrals:
\begin{equation}
\begin{split}
   \underset{T \to 0^+}{\text{lim}}\SumInt_P^f\underset{|S| \to 0^+}{\text{lim}}  \frac{1}{P^2 (P-S)^2} = \mathcal{I}_2(\mu),
   \end{split}
\end{equation}
\begin{equation}
\label{eq:conclusionsspatialfirst}
\begin{split}
  \underset{|S| \to 0^+}{\text{lim}} \underset{T \to 0^+}{\text{lim}}\SumInt_P^f \frac{1}{P^2 (P-S)^2}\longmapsto    \text{P.V.}  \int_{-\infty}^\infty \frac{\text{d} p_0}{2\pi} \underset{|S| \to 0^+}{\text{lim}} \int_p \frac{1}{(p_0+i\mu)^2+p^2} \frac{1}{(p_0+i\mu-s_0)^2+|\vec{p}-\vec{s}|^2} = \mathcal{I}_2 (\mu)
  \end{split}
\end{equation}
and
\begin{equation}
\begin{split}
\label{eq:conclusionsresiduefirst}
  \underset{|S| \to 0^+}{\text{lim}} \underset{T \to 0^+}{\text{lim}}\SumInt_P^f \frac{1}{P^2 (P-S)^2}\longmapsto    \int_p \underset{|S| \to 0^+}{\text{lim}} \underset{i \text{Res} \left[ \frac{1}{(p_0+i\mu)^2+p^2} \frac{1}{(p_0+i\mu-s_0)^2+|\vec{p}-\vec{s}|^2}\right]}{\underbrace{\int_{-\infty}^\infty \frac{\text{d} p_0}{2\pi} \frac{1}{(p_0+i\mu)^2+p^2}\frac{1}{(p_0+i\mu-s_0)^2+|\vec{p}-\vec{s}|^2}}} \neq \mathcal{I}_2 (\mu),
  \end{split}
\end{equation}
specifically assuming that $\exists c \in \mathbb{R}_+$, a fixed parameter, for which $|s_0| \geq c s$. We also observe that --- in line with \cite{Gorda:2022yex} --- fixing $s_0 = 0$ does not exhibit any ambiguity such that
\begin{equation}
\begin{split}
  \underset{s \to 0^+}{\text{lim}} \underset{T \to 0^+}{\text{lim}}\left.\SumInt_P^f \frac{1}{P^2 (P-S)^2}\right|_{s_0 = 0} \longmapsto    \int_p \underset{s \to 0^+}{\text{lim}} \underset{i \text{Res} \left[\frac{1}{(p_0+i\mu)^2+p^2}  \frac{1}{(p_0+i\mu)^2+|\vec{p}-\vec{s}|^2}\right]}{\underbrace{\int_{-\infty}^\infty \frac{\text{d} p_0}{2\pi} \frac{1}{(p_0+i\mu)^2+p^2} \frac{1}{(p_0+i\mu)^2+|\vec{p}-\vec{s}|^2}}} = \mathcal{I}_2 (\mu).
  \end{split}
\end{equation}
Moreover, this absence of ambiguity persists to all orders in the small parameter expansion $\left( \frac{s}{\mu} \right)^{2n}$ as discussed in section \ref{sec:comparison}. This behavior is in sharp contrast with the case of $s_0 \neq 0$ and $s=0$. In that scenario, the full small parameter expansions with respect $\frac{s_0}{\mu}$ match between integration orders only in the sense of coefficients multiplying respective master integral structures in the expansions (see eqs.~\eqref{eq:naiveresidueexp} and \eqref{eq:FPtempexpansion}). We further emphasize that the types of hierachies discussed here are most notably prevalent in the context of HTL (or Hard Dense Loop, HDL) computations.

The observed discrepancy is conceptually in line with the problems presented in \cite{Gorda:2022yex}. Accordingly, a generalization of the bubble Feynman integral results presented therein (see section \ref{sec:extensiontocomplex}) contribute significantly to our computations. More precisely, the discrepancy is associated with the application of the residue theorem to studying temporal real-valued shifts separating fermionic poles. Such shifts can replace the infinitesimally small temperature or the integration dimension $d$ as an infrared regulator, leading to undesired results even in the leading order of momentum expansions. This manifests most prominently in the subset case discussed above, with $s=0$ and $s_0 \neq 0$, where the overlap between fermionic poles is not accessible using standard residue approach at vanishing temperature. While we have kept the examples at one-loop order, similar issues organically appear in further loop orders. Furthermore we note that no thermally generated discrepancies are seen in the results for cases involving poles that can not overlap, or should the overlap be bosonic (for more related discussion see appendices \ref{sec:bosons} and \ref{app:distinctchemicalpotentials}).

The fermionic ambiguity can be remedied by explicitly re-introducing the Fermi-Dirac distribution function to the computations alongside utilizing Feynman parametrization for any fermionic poles that can overlap. This computation scheme casts the simplest one-loop structure to a form such that 
\begin{equation}
\begin{split}
 \SumInt_{P}^f\frac{1}{P^2 (P-S)^2}\longmapsto   \int_0^1 \text{d} x \int_p \oint_{p_0}^f  \frac{\tilde{n}_\text{F} (p_0)}{[(p_0-x s_0)^2 + |\vec{p}-x \vec{s}|^2 +x(1-x) S^2]^2},
 \end{split}
\end{equation}
where $\tilde{n}_\text{F}(p_0) \equiv n_\text{F} [i \beta (p_0-i\mu)] $. The corresponding low-temperature limits are schematically given by
\begin{equation}
    \int_0^1 \text{d}x \text{P.V.} \int_{-\infty}^\infty \frac{\text{d} p_0}{2\pi} \underset{T\to 0^+}{\text{lim}}\int_p \frac{\tilde{n}_\text{F} (p_0+i\mu)}{[(p_0+i\mu)^2 + p^2 +x(1-x) S^2]^2}
\end{equation}
and
\begin{equation}
\label{eq:residueexpressionconclusion}
    \int_0^1 \text{d} x \int_p \underset{T\to 0^+}{\text{lim}} i \text{Res}\left\{  \frac{\tilde{n}_\text{F} (p_0+i\mu)}{[(p_0+i\mu)^2 + p^2 +x(1-x) S^2]^2}\right\},
\end{equation}
which have been shown to yield equivalent results in sections \ref{sec:momentumasymptotes} and \ref{sec:generalfermionicproblem}. This approach is structurally straightforward to generalize to arbitrary number of propagators (and further loop orders), yielding reliable access to the full physically interesting momentum expansions thereof. We also refer the reader to appendix \ref{app:powerseries} and section \ref{sec:generalcomparison} for a somewhat detailed overview on the spatial integrands generated by expressions of the form seen in eq.~\eqref{eq:residueexpressionconclusion} and its strict $T=0$ residue theorem counterparts.

Finally, let us briefly comment on two interpretations outside the presence of conventional external momenta. First, we note that should a temporal momentum in place of $s_0$ (acting as a external momentum at one-loop computation) be a loop momentum, any corrections of the form $\mathcal{O} (s_0^{2n}) = \mathcal{O} (T^{2n})$ are comparable to terms that are removed from the leading low-temperature limit of the Euler-Maclaurin formula in eq.~\eqref{eq:EUlerMacLaurinformulatextbook}. This -- alongside the assumed hierarchy $|s_0| \ll \mu$ -- can be used to justify not only to consider computations of section \ref{sec:generalcasestrictresidue} using hierarchy $|s_0| \leq s$ but to also reject all terms of the form $\left(\frac{s_0^2}{s^2} \right)^n$ as thermally subleading. Through this kind of reasoning, the corresponding thermally leading expansions -- now containing exclusively spatial components of external momentum -- would match as discussed in section \ref{sec:comparison}. However, extending this to actual multi-loop computations is counter-productive, given that fermionic temporal loop momenta play a major role in evaluating bubble Feynman integrals \cite{Osterman:2023tnt,Ostermanthesis}.

Thus, let us instead discuss how the observed discrepancies --- and in particular the structures they coincide with --- affect the results obtained using the standard residue theorem in evaluation of multi-loop bubble Feynman integrals. To illustrate this, we consider the simple example of the sunset integral with two fermionic loop momenta $\{P^\nu,Q^\kappa\}$. The corresponding integral is given by
\begin{equation}
    \mathcal{S}_{111} (\mu) = \underset{T \to 0^+}{\text{lim}}\SumInt_{P,Q}^f \frac{1}{P^2 Q^2(P-Q)^2}
\end{equation}
and has been explicitly shown to yield the physically motivated (correct) result using the standard residue theorem at vanishing temperature (see for example \cite{Osterman:2023tnt}). In our analysis, we observed that by recasting the expressions to be given in terms of potentially problematic temporal integrands (with pathologically overlapping fermionic poles) does not affect the resulting (post temporal integration) spatial integrand. Notably all structural differences canceled out without any application of symmetries between the spatial loop momenta (for more details see appendix \ref{app:sunset}). Similar to the sunset integral we note that significant number of other multi-loop Feynman integrals without external scales have been shown to agree with the results from strict $T=0$ residue theorem \cite{Vuorinen:2003fs}. These structures are systematically characterized by the corresponding fermionic poles being exclusively first order (and consequent symmetries between loop momenta) \cite{Gorda:2022yex}. As such we recognize that sufficiently simple and symmetric bubble integrals do not manifest even intermediate excess terms from the fermionic pole overlap. However, we propose further research into the full extent of the observed problematic effects in  bubble Feynman integral expressions involving asymmetry between loop momenta.

\section*{Acknowlegdements}
\noindent
The authors are grateful to Aapeli Kärkkäinen and Aleksi Vuorinen  for enlightening discussions and insightful comments on an early version of the manuscript. JÖ has been supported by the funding from Research Council of Finland project 354533. MN acknowledges financial support from Magnus Ehrnrooth foundation.
\appendix

\section{Spatial regulation and Euler-Maclaurin formulae}
\label{sec:EM-subsec}
\noindent
Let us fix $\mu \in \mathbb{R}_+$ -- but note that in the example computation one retains symmetry $\mu \leftrightarrow -\mu$ --  and begin by considering the integration order starting with the dimensionally regularized spatial integration and  keeping the thermal sum intact:
\begin{equation}
 \mathcal{I}_\alpha (\mu) =  \underset{T \to 0^+}{\text{lim}} T \sum_{n=-\infty}^\infty \int_p \frac{1}{\{[(2n+1)\pi T+i\mu]^2+p^2\}^\alpha}. 
\end{equation}
The spatial integral herein is an analytic continuation of the Euler beta function on the complex plane -- as argued in \cite{Gorda:2022yex, Ostermanthesis} -- and is finite when regulated by the dimension $d$\footnote{If the exponent $\alpha$ is utilized as an additional regulator such that $\alpha < 1$, even a vanishing temperature permit analytic continuation. This, however, formally excludes all integer values for powers of denominators and makes all relevant algebra convoluted (including any comparison to results from the residue theorem) \cite{Gorda:2022yex,Ostermanthesis}.} if $(2n+1)\pi T \neq 0$ or equivalently $T \neq 0$. Thus, for any non-vanishing $T$, we find 
\begin{equation}
\begin{split}
  \mathcal{I}_\alpha (\mu) =  & \left( \frac{e^\gamma \Lambda^2}{4 \pi} \right)^\frac{3-d}{2} \frac{\Gamma \left(\alpha-\frac{d}{2} \right)}{(4\pi)^\frac{d}{2}} \underset{T \to 0^+}{\text{lim}} T \sum_{n=-\infty}^\infty \left\{[(2n+1)\pi T+i\mu]^2\right\}^{\frac{d}{2}-\alpha} \\
   \end{split}
\end{equation}
Following the schematic of \cite{Ostermanthesis,lainebook}, we can write
\begin{equation}
    \begin{split}
    \label{eq:simplestMatsubarcase}
       &T \sum_{n=-\infty}^\infty \left\{\left[(2 n +1)\pi T + i \mu \right]^2 \right\}^\frac{d-2\alpha}{2}\\
        &=2T \left( \pi T \right)^{d-2\alpha}\Re \left[ \sum_{n=0}^\infty \left\{\left[n  + \frac{i \mu}{ \pi T} \right]^2 \right\}^\frac{d-2\alpha}{2}-  2 ^{d-2\alpha}\sum_{n=0}^\infty \left\{\left[n  + \frac{i \mu}{2 \pi T} \right]^2 \right\}^\frac{d-2\alpha}{2}\right],\\
    \end{split}
\end{equation}
where we have utilized the real part operator to signify a sum between complex conjugates. A closed form expression for these sums can be found in terms of Hurwitz zeta function \cite{Vuorinen:2003fs}, but here we are more enticed to extract the zero-temperature limit. The Euler-Maclaurin formula allows us to write an asymptotic expansion of a summation with a corresponding integral as the leading term such that\cite{Stegun}
\begin{equation}
\label{eq:EUlerMacLaurinformulatextbook}
    \sum_{n=0}^\infty f(n) = \int_0^\infty \text{d} n\: f(n) + \frac{f(0)+f(\infty)}{2}+\sum_{k=1}^\infty \frac{B_{2k}}{(2k)!} \left(f^{(2k-1)}(\infty)-f^{(2k-1)}(0) \right)
\end{equation}
where $B_k$ corresponds to Bernoulli numbers. With the average being already thermally suppressed in the case of interest, considering only the foremost integral part on the right-hand side is sufficient, leaving us with
\begin{equation}
    \begin{split}
    \label{eq:quadraticbubblematsubara}
   & T \sum_{n=-\infty}^\infty \left\{\left[(2 n +1)\pi T + i \mu \right]^2 \right\}^\frac{d-2\alpha}{2}\\
   &= 2T \left( \pi T \right)^{d-2\alpha} \re \left[\int_0^\infty \text{d} n  \left\{\left[n  + \frac{i \mu}{ \pi T} \right]^2 \right\}^\frac{d-2\alpha}{2}-  2^{d-2\alpha} \int_0^\infty \text{d} n \left\{\left[n  + \frac{i \mu}{2 \pi T} \right]^2 \right\}^\frac{d-2\alpha}{2}\right]+\mathcal{O}(T)\\
       &=\frac{1}{\pi}  \int_0^\infty \text{d}n \left[ \left(1-\frac{1}{2} \right) \left\{\left[n + i \mu \right]^2 \right\}^\frac{d-2\alpha}{2} + \left(1-\frac{1}{2} \right) \left\{\left[n - i \mu \right]^2 \right\}^\frac{d-2\alpha}{2} \right] + \mathcal{O}(T).
    \end{split}
\end{equation}
On the last row we have chosen to express the real part explicitly in terms of complex conjugates to better convey further steps.

To compute such quadratic integrals, we utilize partial integration -- justified and facilitated by dimensional regularization of the parameter $d$ -- and find
\begin{equation}
\label{eq:quadratichelpintegral}
    \int_a^b \text{d} x [(x+A)^2]^{k} = \frac{1}{2k+1} \left\{(x+A)[(x+A)^2]^{k} \right\}_{x=a}^{x=b}.
\end{equation}
Applying this closed-form expression carefully, the low-temperature limit of the Matsubara summation in eq.~\eqref{eq:simplestMatsubarcase} becomes
\begin{equation}
\begin{split}
\label{eq:protoIalphasum}
     &\underset{T\to 0^+}{\text{lim}} T \sum_{n=-\infty}^\infty \left\{\left[(2 n +1)\pi T + i \mu \right]^2 \right\}^\frac{d-2\alpha}{2} = - \frac{i\mu}{d+1-2\alpha}\frac{1}{2\pi}\left[   \left( i\mu  \right)^{d-2\alpha}-(-i\mu)^{d-2\alpha} \right]\\
     &= -\frac{\mu^{d+1-2\alpha}}{d+1-2\alpha} \frac{1}{2\pi}\left[ \exp \left( i \pi \frac{d+1-2\alpha}{2} \right)+\exp \left( -i \pi \frac{d+1-2\alpha}{2} \right) \right]\\
     &\left(\overset{\mu \in \mathbb{R}}{=}  -\frac{|\mu|^{d+1-2\alpha}}{d+1-2\alpha} \frac{1}{2\pi}\left[ \exp \left( i \pi \frac{d+1-2\alpha}{2} \right)+\exp \left( -i \pi \frac{d+1-2\alpha}{2} \right) \right] \right),
     \end{split}
\end{equation}
where in the last row we have written the generalization for the formula to also include negative values of $\mu$. Utilizing Euler's reflection formula \cite{Stegun}, we can write the full result as
\begin{equation}
\begin{split}
\label{eq:masterformualIa}
    \mathcal{I}_\alpha (\mu) &=-\left( \frac{e^\gamma \Lambda^2}{4 \pi} \right)^\frac{3-d}{2} \frac{\Gamma \left(\alpha-\frac{d}{2}\right)}{(4\pi)^\frac{d}{2}\Gamma (\alpha) }  \frac{\mu^{d+1-2\alpha}}{d+1-2\alpha} \frac{1}{2\pi}
    \left[\exp \left( i \pi \frac{d+1-2\alpha}{2} \right)+\exp \left( -i \pi \frac{d+1-2\alpha}{2} \right) \right]\\
    &=-\left( \frac{e^\gamma \Lambda^2}{4 \pi} \right)^\frac{3-d}{2} \frac{1}{(4\pi)^\frac{d}{2}\Gamma (\alpha) \Gamma\left(\frac{d}{2}+1-\alpha\right)}  \frac{\mu^{d+1-2\alpha}}{d+1-2\alpha}.
    \end{split}
\end{equation}

Owing to dimensional regularization, this result can be observed to align with the strict limit of the distribution function after the spatial integral has been performed:
\begin{equation}
\begin{split}
    \text{P.V.} \int_{-\infty}^\infty \frac{\text{d} p_0}{2\pi} \int_p \frac{1}{[(p_0+i\mu)^2+p^2]^\alpha}
    &= \left( \frac{e^\gamma \Lambda^2}{4 \pi} \right)^\frac{3-d}{2} \frac{\Gamma \left(\alpha-\frac{d}{2}\right)}{(4\pi)^\frac{d}{2}\Gamma (\alpha) } \frac{1}{\pi} \text{Re} \left\{\int_0^\infty \text{d} p_0 [(p_0+i\mu)^2]^\frac{d-2\alpha}{2} \right\}\\
    &= \mathcal{I}_\alpha (\mu),
    \end{split}
\end{equation}
as seen easily upon comparing the integral structure against that of eq.~\eqref{eq:quadraticbubblematsubara}. In alignment with the last row of eq.~\eqref{eq:protoIalphasum}, we emphasize that for a more general value $\mu \in \mathbb{R}$, this result would be straightforwardly extend to $\mathcal{I}_\alpha (|\mu|)$.

\section{Temporal integration and the residue theorem}
\label{sec:temporalintegrationandresidue}
\noindent
\label{sec:residuenotation}
The conventional residue methods in finite density computations are characterized by
\begin{itemize}
    \item the strict $T\to 0$ limit applied onto the distribution function before either of the integrals has been performed,
    \item the temporal integration performed prior to spatial integration and
    \item the solid line extension to the temporal integration interval \cite{ghisou,Ostermanthesis,Gorda:2022yex}.
\end{itemize}
Working initially using the leading limit of the Euler-Maclaurin formula, the relevant mapping to the conventional residue expression can be written such that
\begin{equation}
     \text{P.V.} \int_{-\infty}^\infty  \frac{\text{d} p_0}{2 \pi} f(p_0,p) \longmapsto  \int_{-\infty}^\infty  \frac{\text{d} p_0}{2 \pi} f(p_0,p),
\end{equation} 
wherein the previously removed $p_0 = 0$ value is included and the target function of interest reads 
\begin{equation}
    f(p_0,p) = \frac{1}{[(p_0+i\mu)^2+p^2]^\alpha}.
\end{equation}

As discussed in \cite{Gorda:2022yex, Ostermanthesis}, the residue approach necessitates the exclusion of particular values of radial loop momenta. Continuing with the above $f(p_0, p)$ admitting now values $p_0 \in \mathbb{R}$, the corresponding spatial integral reads 
\begin{equation}
\label{eq:master1}
    \int_p \frac{1}{[(p_0+i\mu)^2+p^2]^\alpha} \overset{p_0 = 0}{\longmapsto} \int_p \frac{1}{(p^2-\mu^2)^\alpha},
\end{equation}
where explicit convergence requires the assumption $p \neq \mu$ if $\alpha \geq 1$. Superficially this is resolved by the selection criteria of the residues, as the actual computation is performed utilizing a semi-circular residue contour on the upper complex half-plane of Fig. \ref{fig:semicirclecontour} (with only upper line integrals parallel to real axis contributing in eq.~\eqref{eq:gammacontour}, cf. eq.~\eqref{eq:expdecayoflineintegrals}).
\begin{figure}[h!]
\centering
\begin{tikzpicture}[scale=0.8]
\def\blockh{1.2}
\def\blockl{3.0}
  \draw[->,thick] (-\blockl-1,-2*\blockh) -- (\blockl+1,-2*\blockh) node[above] {$\re(p_0)$};
  \draw[->,thick] (0,-2*\blockh-0.5) -- (0,2.5*\blockh-0.5) node[right] {$\im(p_0)$};
   \draw[-latex,thick,blue!75] (-\blockl,-1) -- (\blockl,-1) node[black, above left] {$i\mu+i\eta$};
    \draw[-, loosely dotted, line width = 0.4mm,black!75] (-\blockl+0.1,-1.7) -- (-0.1,-1.7) node[black, above left] {};
   \draw[-, loosely dotted, line width = 0.4mm,black!75] (0.1,-1.7) -- (\blockl,-1.7) node[black, above left] {};
  \clip (-\blockl, -1) rectangle (\blockl, 3 ); 
 \draw[-, thick, red!75] (0,-1) circle (\blockl);
  \draw[->,thick, red!75] (0.05,\blockl-1) -- (-0.2,\blockl-1) ;

\end{tikzpicture}

\caption[]{%
In the strict residue approach at $T=0$ the thermally suppressed part of image \ref{fig:contouroverlap} is removed, and thus only the upper line integral -- above the black dotted poles of the Fermi-Dirac distribution function -- remains, indicated by the blue solid line above. We emphasize that the blue solid line includes every point along its length, in particular the previously excluded point $\text{Re}(p_0)=0$. The line integral is studied in conjunction with an added semicircular arc -- the red line in image above -- covering the upper half of complex plane, specifically above the line parallel to real axis $\text{Im}(p_0) = i\mu + i\eta$ and therefore avoiding the black dotted line. Here $\eta > T$ is an arbitrarily small regulator needed to treat the low-temperature limit properly, and can be removed completely after the vanishing-temperature limit has been taken. The residue theorem is then studied within the corresponding closed contour.
  }
  \label{fig:semicirclecontour}
\end{figure}
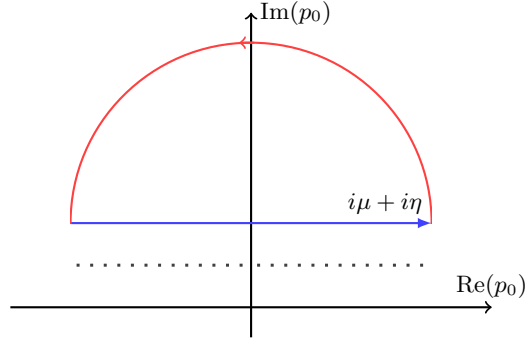

Now performing the corresponding residue computation (prior to spatial integration), one finds for $\alpha \in \mathbb{Z}_+$ (with the parameter constrained by the residue prescription)\cite{Gorda:2022yex, Ostermanthesis} 
\begin{equation}
\begin{split}
\label{eq:residuebubblemaster}
  \text{Res}[\mathcal{I}_\alpha] (\mu) &\equiv  \int_p i  \text{Res} \left[ \frac{1}{[(p_0+i\mu-ip)(p_0+i\mu+ip)]^\alpha}\right]_{p_0=-i\mu+ip}\\
  &=\frac{ (-1)^{\alpha-1} (2 \alpha-2)!}{ [(\alpha-1)!]^2}  \int_p    \frac{i \theta(p-\mu)}{(2ip)^{2\alpha-1}}\\
  &= -\left(\frac{e^\gamma \Lambda^2}{4\pi} \right)^\frac{3-d}{2} \frac{\Gamma \left(\alpha-\frac{1}{2}\right)}{\sqrt{\pi} (4\pi)^\frac{d}{2} \Gamma (\alpha) \Gamma \left( \frac{d}{2} \right)} \frac{\mu^{d+1-2\alpha}}{d+1-2\alpha},
    \end{split}
\end{equation}
where the factor of $\theta(p-\mu)$ is introduced by the requirement that the pole must lie in inside the integration contour as required by the Residue theorem. We can observe this result to agree with eq.~\eqref{eq:masterformualIa} only when $\alpha = 1$\footnote{The result $\mathcal{I}_\alpha (\mu)$ of eq~\eqref{eq:masterformualIa} is the physical structure of interest given that it corresponds to the true leading behavior with the vanishing temperature limit applied after the loop integration. (Because the vanishing temperature limit is taken as late as possible, the true limit of finite T result is obtained.)}. Note also that this result must by definition (and the steps leading to the close form solution) obey the symmetry $\mu \leftrightarrow -\mu$, similar to the result of $\mathcal{I}_\alpha (\mu)$. Thus, the extensions to $\mu \in \mathbb{R}$ reads $\text{Res} \left[ \mathcal{I}_\alpha \right] (|\mu|)$ . We can also relate the two different result to one another by writing for $n\in \mathbb{Z}_+$
\begin{equation}
    \mathcal{I}_n\left(\mu\right) = \frac{(-1)^{n}}{2n-3}\frac{\left(1-\frac{d}{2}\right)_{n-1}}{\left(-\frac{1}{2}\right)_{n-1}}\text{Res}\left[\mathcal{I}_n\left(\mu\right)\right], 
\end{equation}

where we use the Pochhammer symbol convention (for $n \in \mathbb{N}_0$) of 
\begin{equation}
    (a)_n = \frac{1}{a-1} \prod_{k=0}^n (a+k-1)
\end{equation}
to provide the result compactly.

Furthermore, it was shown in \cite{Gorda:2022yex}  that the two results differ by a set of terms originating from the neighborhood of $p = \mu$. These surface terms arise through partial integration enacted on the highest (sole) order structure resulting in agreement with either integration order, as exemplified by 
\begin{equation}
    \text{P.V.} \int_{-\infty}^\infty  \frac{\text{d} p_0}{2 \pi} \int_p \frac{1}{[(p_0+i\mu)^2+p^2]^2}= -\int_p \frac{\text{d}}{\text{d} p^2} \left[\int_{-\infty}^\infty  \frac{\text{d} p_0}{2 \pi} \frac{1}{(p_0+i\mu)^2+p^2} \right]_\text{Residue}.
\end{equation}
This result was also observed and argued to align with common residue theorem enacted on an integrand still equipped with the Fermi-Dirac distribution function --- associated with nonvanishing temperature --- such that 
\begin{equation}
    \begin{split}
    \label{eq:expdecayoflineintegrals}
        \oint_P^f \frac{\tilde{n}_\text{F} (p_0)}{P^4} &= \int_p \int_{-\infty+i\mu+ i\eta}^{\infty+i\mu+i\eta} \frac{\text{d} p_0}{2\pi} \frac{\tilde{n}_\text{F} (p_0)}{P^4} + \mathcal{O} (e^{-\beta \eta})\\
        &\overset{T \to 0}{\longrightarrow}  \int_p \underset{T \to 0}{\text{lim}} i \text{Res}\left[\frac{\tilde{n}_{\text{F}} (p_0+i\mu)}{[(p_0+i\mu-ip)(p_0+i\mu+ip)]^2}\right]_{p_0=-i\mu+ip}.
    \end{split}
\end{equation}
Herein the difference term between the non-vanishing and strict vanishing temperature expressions is generated by the derivative of the distribution function \cite{Gorda:2022yex}, explicitly a nascent delta function given by
 \begin{equation}
     \frac{1}{4T \cosh^2 \left[ \frac{p-\mu}{2T} \right]} \equiv \delta_{2T} (p-\mu).
 \end{equation}
Higher order poles can be similarly associated to multiple derivatives of the distribution function at nonvanishing temperature, and easiest interpreted through partial integration (with the weak derivative interpretation facilitated again by the regulated dimension $d$) of expressions such that 
    \begin{equation}
    \label{eq:partialintegrationbubble}
    \text{P.V.} \int_{-\infty}^\infty  \frac{\text{d} p_0}{2 \pi} \int_p \frac{1}{[(p_0+i\mu)^2+p^2]^{1+n}}= \frac{(-1)^n
}{n!}\int_p \left(\frac{\text{d}}{\text{d} p^2} \right)^n \left[\int_{-\infty}^\infty  \frac{\text{d} p_0}{2 \pi} \frac{1}{(p_0+i\mu)^2+p^2} \right]_\text{Residue}.
\end{equation}
Notably, closing the contour on the complex half-plane introduces a factor of $\theta(p-\mu)$ before the differentiation, which in turn produces terms proportional to $\delta(p-\mu)$ and its derivatives compared to the naive residue result.
\section{Avoidable sources of mistakes in expansions}
\label{sec:FalseExpansions}
\noindent
An interesting tangent of eqs.~\eqref{eq:PVcomplexsplit} and \eqref{eq:complexmumasterbubble} is that careless expansions can generate false subleading behaviour of the order $\mathcal{O} (\text{Im} \mu)$ for the hierarchy $\text{Im} \mu \ll \text{Re} \mu$. To illustrate, note that $|p_0+i\text{Re} \mu| = \sqrt{p_0^2+(\text{Re} \mu)^2} \gg |\text{Im} \mu|$, and expand the temporal integrand (post spatial integration) directly naively to find
\begin{equation}
\begin{split}
   \left[ \left(p_0+i \text{Re} \mu-\text{Im} \mu  \right)^2\right]^\frac{d-2\alpha}{2} &\overset{}{=}   \left[ \left(p_0+i \text{Re} \mu  \right)^2\right]^\frac{d-2\alpha}{2} \sum_{n=0}^\infty \frac{(2\alpha-d)_n}{n!} \left(\frac{\text{Im} \mu}{p_0 + i \text{Re} \mu} \right)^n \\
   &= \sum_{n=0}^\infty \frac{(2\alpha-d)_n}{n!} \left[ \left(p_0+i \text{Re} \mu  \right)^2\right]^{\frac{d-2\alpha}{2}-\lceil \frac{n}{2} \rceil} (p_0+i \text{Re}\mu)^{2\lceil \frac{n}{2} \rceil-n} \left( \text{Im} \mu \right)^n,
   \end{split}
\end{equation}
where we utilize ceiling operators to compactly write quadratic elements in terms of both the dimensionally regulated sector and terms from the expansion. To move forward, we again utilize  eq.~\eqref{eq:quadratichelpintegral2} and its differentiated version:
\begin{equation}
\label{eq:diffquadrataticintegralhelper}
   \int_a^b \text{d} x (x+A) [(x+A)^2]^{k-1} = \frac{1}{2k} \left\{[(x+A)^2]^k \right\}_{x=a}^{x=b}.
\end{equation}
Thus, integrating each individual part (for additional discussion see also appendices of \cite{Osterman:2023tnt}) of this expansion (again utilizing Euler-Maclaurin formula to extract the leading behaviour) leads to 
\begin{equation}
\begin{split}
    &\text{P.V.} \int_{-\infty}^\infty \frac{\text{d} p_0}{2\pi}\sum_{n=0}^\infty \frac{(2\alpha-d)_n}{n!} \left[ \left(p_0+i \text{Re} \mu  \right)^2\right]^{\frac{d-2\alpha}{2}-\lceil \frac{n}{2} \rceil} (p_0+i \text{Re}\mu)^{2\lceil \frac{n}{2} \rceil-n} \left( \text{Im} \mu \right)^n\\
    &=-\sum_{n=0}^\infty  \frac{(-i)^n(2\alpha-d)_{n}}{(n)! (d+1-2\alpha-n)}\frac{|\text{Re} \mu|^{d+1-2\alpha-n} (\text{Im} \mu)^{n}}{\Gamma \left(\alpha-  \frac{d}{2}\right)\Gamma \left( \frac{d}{2}+1-\alpha \right)}\\
    \label{eq:Ialpha_wrong}
    %&=\textcolor{red}{-}\sum_{n=0}^\infty  \frac{(\textcolor{red}{-}i)^n(2\alpha-d)_{n}}{(n)! (d+1-2\alpha-n)}\frac{|\text{Re} \mu|^{d+1-2\alpha-n} (\text{Im} \mu)^{n}}{\Gamma \left(\alpha-  \frac{d}{2}\right)\Gamma \left( \frac{d}{2}+1-\alpha \right)}\\
    %&\textcolor{blue}{\overset{\text{even}}{=}-\sum_{n=0}^\infty (-1)^n \frac{(2\alpha-d)_{2n}}{(2n)! (d+1-2\alpha-2n)}\frac{|\text{Re} \mu|^{d+1-2\alpha-2n} (\text{Im} \mu)^{2n}}{\Gamma \left(\alpha-  \frac{d}{2}\right)\Gamma \left( \frac{d}{2}+1-\alpha \right)}}\\
    %&\textcolor{blue}{\overset{\text{odd}}{+}i \sum_{n=0}^\infty (-1)^n \frac{(2\alpha-d)_{2n+1}}{(2n+1)! (d-2\alpha-2n)}\frac{|\text{Re} \mu|^{d-2\alpha-2n} (\text{Im} \mu)^{2n+1}}{\Gamma \left(\alpha-  \frac{d}{2}\right)\Gamma \left( \frac{d}{2}+1-\alpha \right)}}
    \end{split}
\end{equation}
This (structurally false) result is equivalent with a generalization to $\mu \in \mathbb{C}$ utilizing the following mapping\footnote{Note that, as can be seen from equation \eqref{eq:Ialpha_correct}, a pre-factor of $\Gamma(\alpha-d/2)/\left[(4\pi)^{d/2}\Gamma(\alpha)\right]$ needs to be inserted to equation \eqref{eq:Ialpha_wrong} for $\mathcal{I}_\alpha(\mu)$.}
\begin{equation}
\label{eq:falsegeneralizationoneloop}
\begin{split}
    %&\text{P.V.}\int_{-\infty}^\infty \frac{\text{d} p_0 }{2 \pi}  \int_p \frac{1}{[(p_0-\text{Im} \mu+i \text{Re}\mu)^2+p^2]^\alpha} =\\
    &\mathcal{I}_\alpha (|\text{Re}\mu|) \longmapsto  \mathcal{I}_\alpha (|\text{Re}\mu|+i\text{Im} \mu),
    \end{split}
\end{equation}
corresponding to a substitution of a complex-valued scale directly to the real-valued result naively\footnote{Specifically, this implies for $\text{Re} \mu > 0$ a substitution of the form $\text{Re} \mu \mapsto \mu$, while for $\text{Re} \mu < 0$ the substitution reads $\text{Re} \mu \mapsto \mu^*$.}. Between the two computations the agreement is only reached in the leading order of the expansion with respect to $\frac{\text{Im} \mu}{\text{Re} \mu}$. The mechanism of how this direct expansion fails to produce the (numerically confirmable) correct result relates to how eq.~\eqref{eq:minusp0oneloopintegral} can be seen to hold, facilitated by a change of variables $p_0 \mapsto - p_0$ (a property also hereditary from the Matsubara summation). Instead, in the naive integration (associated with the explicit expansion) the change of integration variable would be interpreted (again falsely) to add also a complex-valued phase factor to the second integral. Let us also further note that the solution for real-valued $\mu$, as seen in eq.~\eqref{eq:protoIalphasum}, explicitly manifests the symmetry between $\mu \leftrightarrow - \mu$, which is lost in the naive (and false) generalization scheme of eq.~\eqref{eq:falsegeneralizationoneloop}.

The issues taking place in the direct generalization of the $\mu \in \mathbb{R}_+$ result can also be observed by studying the complex-valued temporal integration of
\begin{equation}
    \int_0^\infty \frac{\text{d} p_0}{2\pi} \left\{[(p_0+i\mu)^2]^\frac{d-2\alpha}{2} +[(p_0-i\mu)^2]^\frac{d-2\alpha}{2} \right\},
\end{equation}
where the integration range has been achieved using a change of the temporal integration variable such that $p_0 \mapsto - p_0$. Computing this integral directly for a positive $\mu$ yields a closed form result such that 
\begin{equation}
    -\frac{i\mu}{2\pi(d+1-2\alpha)} \left[[(i\mu)^2]^\frac{d-2\alpha}{2}-[(-i\mu)^2]^\frac{d-2\alpha}{2} \right],
\end{equation}
which was further simplified in \cite{Gorda:2022yex,Osterman:2023tnt} by regulating both of the complex-valued power function parameters such that both  $i^2$ and $(-i)^2$ were assumed to belong to the same branch of complex numbers. Formally, one could write for example $i \mapsto \exp \left[ \frac{i \pi}{2} (1-\kappa) \right]$ to signify this. However, given any $\text{Im} \mu \neq 0$, this kind of simultaneous regulation is no longer possible for both $i \mu$ and $-i\mu$, with the squared value of one of them leading to a complex value in another branch. Thus, the naive generalization following eq.~\eqref{eq:falsegeneralizationoneloop} relies on the complex-conjugate symmetry between $i^2$ and $(-i)^2$ to be applied in the one branch facilitating the (false) result in the form 
\begin{equation}
    \int_0^\infty \frac{\text{d} p_0}{2\pi} \left\{[(p_0+i\mu)^2]^\frac{d-2\alpha}{2} +[(p_0-i\mu)^2]^\frac{d-2\alpha}{2} \right\} \longmapsto -\frac{\mu^{d+1-2\alpha}}{2\pi(d+1-2\alpha)} \cdot 2 \cos \left[ \frac{\pi (d+1-2\alpha)}{2} \right]. 
\end{equation}
for $\text{Re} \mu > 0$.

This type of issue describes a general challenge of dealing with real-valued scales associated with the temporal loop momentum, and expansions thereof. Notably this problem extends also to the integral of primary interest, with too careless expansions leading to not only false results but even introducing explicitly imaginary terms to fully real-valued dimensionally regularized integral(s).

In this context, let us continue by looking again at $\mathcal{J}_{11} (\mu, 0, s_0)$ as given in eq.~\eqref{eq:s0FPinitint}.  By utilizing partial fraction decomposition, we can compute the following spatial integral
\begin{equation}
\begin{split}
    \mathcal{J}_{11}(\mu, 0, s_0)&=\int_p \frac{1}{(p_0+i\mu)^2+p^2}\frac{1}{(p_0+i\mu-s_0)^2+ p^2}\\
    &= \left( \frac{e^\gamma \Lambda^2}{4 \pi} \right)^\frac{3-d}{2} \frac{\Gamma \left(1-\frac{d}{2} \right)}{(4\pi)^\frac{d}{2}} \frac{[(p_0+i\mu)^2]^{\frac{d}{2}-1}-[(p_0+i\mu-s_0)^2]^{\frac{d}{2}-1}}{(p_0+i\mu-s_0)^2-(p_0+i\mu)^2} 
    \end{split}
\end{equation}
to find a temporal integrand. Then expanding it directly through real-valued scales $s_0$ (alongside loop momenta), we find a false expansion characterized by imaginary terms and wrong coefficients after performing the temporal principal value integral\footnote{None of the subleading terms are in line with the thermal corrections from the Feynman parametrized expression, which facilitates the changes of integration variables to be carried out without added complex phases (as discussed at the end of section \ref{sec:preface}).}. While the leading, non-vanishing $\mathcal{O} (s_0^0)$ term does agree with the Feynman parametrized result, a notable difference is found by considering the now non-vanishing $\mathcal{O}(s_0)$ term of the misleading expansion. Utilizing eq.~\eqref{eq:diffquadrataticintegralhelper} we find
\begin{equation}
\begin{split}
    &\left( \frac{e^\gamma \Lambda^2}{4 \pi} \right)^\frac{3-d}{2} \frac{\Gamma \left(3-\frac{d}{2} \right)}{(4\pi)^\frac{d}{2}} \text{P.V.} \int_{-\infty}^\infty \frac{\text{d} p_0}{2\pi} (p_0+i\mu) \left[ (p_0+i\mu)^2 \right]^{\frac{d}{2}-3} s_0\\
    &= -\frac{i}{2}\left( \frac{e^\gamma \Lambda^2}{4 \pi} \right)^\frac{3-d}{2} \frac{\mu^{d-4}s_0}{(4\pi)^\frac{d}{2} \Gamma \left(\frac{d}{2}-1 \right)}\\
    &= \frac{\mathcal{I}_2 \left(\mu+is_0 \right)- \mathcal{I}_2 \left(\mu\right)}{2} + \mathcal{O}(s_0^2)\\
    \end{split}
\end{equation}
On the last line we identified the expression through the leading order structure 
 $\mathcal{I}_2 (\mu)$, where we again -- for convenience -- work with the assumption $\mu > 0$ (most of the time the extension is achieved with the mapping  $\mu \mapsto |\mu|$). Specifically, we can again associate the difference from desired result to a complex-valued scale insertion (enacted on one of the propagators), albeit due to its more difficult structure the full expansion can not be given as easily as earlier. In contrast with this issue, Feynman parametrization enables us to treat complex phases carefully (see section \ref{sec:momentumasymptotes}). This is most notably seen in cancellations of overlapping pieces, via changes of variables $p_0 \mapsto - p_0 $, not introducing excess imaginary structures. We emphasize that all imaginary contributions are absent even from the results arising from the results of the strict $T=0$ residue theorem (applied prior to spatial integrations).

\section{Cauchy theorem and vacuum results}
\label{app:Slimit}
\noindent
Let us briefly note also that agreement with the results from $T = \mu = 0$, theories or the leading \textcolor{black}{$|s_0|,|\Vec{s}| \gg \mu$} term is straightforward to extract utilizing the Cauchy theorem prescription introduced in \cite{Gorda:2022yex, Osterman:2023tnt}. Specifically, denoting the vacuum $D=d+1$ dimensional integration such that 
\begin{equation}
    \int_P \equiv \left( \frac{e^\gamma \Lambda^2}{4\pi}\right)^\frac{4-D}{2} \int_{\mathbb{R}^D} \frac{\text{d}^D P}{(2\pi)^d}
\end{equation}
and the standard vacuum result can be computed via straightforward dimensional regularization such that
\begin{equation}
\begin{split}
\mathcal{J}_{11}(0, \vec{s},s_0) =     \int_P \frac{1}{P^2(P-S)^2} &= \int_0^1 \text{dx} \int_P \frac{1}{[P^2+x(1-x)S^2]^2}\\
    &= \left( \frac{e^\gamma \Lambda^2}{4\pi}\right)^\frac{4-D}{2} \frac{\Gamma\left(2-\frac{D}{2} \right) (S^2)^{\frac{D}{2}-2}}{(4\pi)^D } \int_0^1 \text{d} x  [x(1-x)]^{\frac{D}{2}-2}\\
    &=  \left( \frac{e^\gamma \Lambda^2}{4\pi}\right)^\frac{4-D}{2} \frac{\Gamma\left(2-\frac{D}{2} \right)\Gamma^2\left(\frac{D}{2}-1 \right) }{(4\pi)^D \Gamma \left(D-2 \right) }(S^2)^{\frac{D}{2}-2}
    \end{split}
\end{equation}
which we can straightforwardly identify to agree with the partition to distinct temporal and spatial integration(s):
\begin{equation}
\begin{split}
\label{eq:splitvacuumnomu}
  \mathcal{J}_{11}(0, \vec{s},s_0) =  &\int_0^1 \text{d}x\, \text{P.V.} \int_{-\infty}^\infty \frac{\text{d} p_0}{2\pi} \int_p \frac{1}{[p^2+(p_0-xs_0)^2+x(1-x) S^2]^2}\\&= \left(  \frac{e^\gamma \Lambda^2}{4\pi}\right)^\frac{3-d}{2} \frac{\Gamma \left(2-\frac{d}{2} \right)}{(4\pi)^\frac{d}{2}} \int_0^1 \text{d}x\, \text{P.V.} \int_{-\infty}^\infty \frac{\text{d} p_0}{2\pi} [(p_0)^2+x(1-x)S^2]^{\frac{d}{2}-2}\\
   &= \left(  \frac{e^\gamma \Lambda^2}{4\pi}\right)^\frac{3-d}{2} \frac{\Gamma \left(2-\frac{d}{2} \right)(S^2)^{\frac{d+1}{2}-2}}{(4\pi)^\frac{d}{2}} \left[\int_0^1 \text{d}x x^{\frac{d+1}{2}-2}(1-x)^{\frac{d+1}{2}-2} \right] \int_0^\infty \frac{\text{d}y}{2\pi} \frac{y^{-\frac{1}{2}}}{(1+y)^{2-\frac{d}{2}}}\\
   &= \left( \frac{e^\gamma \Lambda^2}{4\pi}\right)^\frac{4-D}{2} \frac{\Gamma\left(2-\frac{D}{2} \right)\Gamma^2\left(\frac{D}{2}-1 \right) }{(4\pi)^D \Gamma \left(D-2 \right) }(S^2)^{\frac{D}{2}-2}.
    \end{split}
\end{equation}
To extract this solution of our Feynman parametrized expression of eq.~\eqref{eq:FPPVfullintinitial}
\begin{equation}
\begin{split}
\mathcal{J}_{11}(0, \vec{s},s_0) =  & \left( \frac{e^\gamma \Lambda^2}{4 \pi} \right)^\frac{3-d}{2} \frac{\Gamma \left(2-\frac{d}{2} \right)}{(4 \pi)^\frac{d}{2}}\int_0^1 \text{d} x\, \text{P.V.} \int_{-\infty}^\infty \frac{\text{d}p_0}{2\pi}   \left[ (p_0+i\mu-xs_0)^2+x(1-x) S^2 \right]^\frac{d-4}{2}\\
&\equiv\left( \frac{e^\gamma \Lambda^2}{4 \pi} \right)^\frac{3-d}{2} \frac{\Gamma \left(2-\frac{d}{2} \right)}{(4 \pi)^\frac{d}{2}}\int_0^1 \text{d} x\, \text{P.V.} \int_{-\infty}^\infty \frac{\text{d}p_0}{2\pi}   \left[ (p_0+i\mu)^2+x(1-x) S^2 \right]^\frac{d-4}{2}
\end{split}
\end{equation}
where we have noted that the only meaningful split of the principal value integral occurs at $p_0 \neq xs_0$ and perform a change of integration variable to move the point of interest to $p_0 = 0$. Utilizing Cauchy theorem to write the split temporal integral in terms of line integrals parallel and perpendicular to the real axis, we can write
\begin{equation}
    \int_0^\infty \frac{\text{d} p_0}{2\pi} \left[ (p_0\pm i \mu)^2 + x(1-x)S^2 \right]^{\frac{d}{2}-2} =  \int_0^\infty \frac{\text{d} p_0}{2\pi} \left[ p_0^2 + x(1-x)S^2 \right]^{\frac{d}{2}-2}-\int_0^{\pm i \mu} \frac{\text{d} p_0}{2\pi} \left[ p_0^2 + x(1-x)S^2 \right]^{\frac{d}{2}-2}.
\end{equation}
Thus, extracting the pieces (via the Cauchy theorem) parallel to the real axis, we find explicitly the steps presented in eq.~\eqref{eq:splitvacuumnomu}. By expanding the terms parallel to the imaginary axis naively by setting $\frac{S^2}{p_0^2}$ as the expansion parameter, and integrating carefully, we find the full expansion presented in eq.~\eqref{eq:FPresiduefullexpansioncaptialS}. Specifically this representation (utilizing conventions of \cite{Osterman:2023tnt}) reads
\begin{equation}
\begin{split}
    -\left( \frac{e^\gamma \Lambda^2}{4 \pi} \right)^\frac{3-d}{2} \frac{\Gamma \left(2-\frac{d}{2} \right)}{(4 \pi)^\frac{d}{2}} \sum_{n=0}^\infty \frac{(-1)^n\left(2-\frac{d}{2} \right)_n}{n!}\cdot 2 \text{Re} \left[\int_{0}^{i \mu}\frac{\text{d}p_0}{2\pi}   \left[ (p_0)^2\right]^\frac{d-4-2n}{2} \right] \int_0^1 \text{d} x [x(1-x)]^{n} (S^2)^n.
    \end{split}
\end{equation}
\section{Power series comparison}
\label{app:powerseries}
\noindent
In this section we study the equivalence of the naive residue result, as well as the method of introducing Feynman parametrization and the finite temperature effect.
In order to compare the series expressions on either side of the equality presented in eq.~\eqref{eq:differencetermequalitynos0} we find it convenient to actually take a step back and re-introduce the well-behaved naive expansion without delta function contributions to both sides. 
\subsection{Representations}
Accordingly, we find it more compelling to rather look into
\begin{equation}
\begin{split}
\label{eq:fullpowerseriescomparionformula}
    &\int_0^1 \text{d} x \left\{\frac{\theta\left[\sqrt{p^2+(1-x)(q^2-p^2)}-\mu\right]}{4 \left[p^2 +(1-x) (q^2-p^2)\right]^\frac{3}{2}} -\frac{\delta\left[\sqrt{p^2+(1-x)(q^2-p^2)}-\mu\right]}{4 \left[p^2 + (1-x)(q^2-p^2)\right]} \right\}\\
    &=\frac{\theta(p-\mu)}{2 p [p+(q-p)][2p+(q-p)]} -\frac{\theta[p-\mu+(q-p)]-\theta(p-\mu)}{2 [p+(q-p)](q-p)[2p+(q-p)]},
    %&=-\int_0^1 \text{d} x \sum_{n=0}^\infty  \frac{p \delta^{(n)} (p-\mu) (p-q)^n  }{4 (n+1)! \left[p^2 +(1-x) (q^2-p^2)\right]^\frac{3}{2}}. \\
    \end{split}
\end{equation}
where we immediately applied reverse Feynman parametrization on the representation of the term moved from the left-hand side to the right:
\begin{equation}
\label{eq:FeynmanparametleadingorderappB}
    \frac{\theta(p-\mu)}{2 p [p+(q-p)][2p+(q-p)]} = \int_0^1 \text{d} x \frac{\theta(p-\mu)}{4[p^2+(1-x)(q^2-p^2)]^\frac{3}{2}}
\end{equation}
The immediate utility here is the explicit expansion in terms of the expansion parameter $(1-x)(q^2-p^2)$, yielding us
\begin{equation}
\begin{split}
\label{eq:thetaexpansiontrivial}
     \frac{1}{2 p [p+(q-p)][2p+(q-p)]} &=  \sum_{k=0}^\infty \frac{(-1)^k \left( \frac{3}{2}\right)_k}{4p^3 (k+1)!} \frac{(q^2-p^2)^k}{p^{2k}} \\
     &= \sum_{k=0}^\infty \frac{(-1)^k (2k+1)!!}{2^{2+k}p^3 (k+1)!} \frac{(q^2-p^2)^k}{p^{2k}}, 
     \end{split}
\end{equation}
where we have performed the parametric integral after the expansion. This representation serves us exceedingly well given that we can express all the other terms arising from the expanded Heaviside function in terms of this expression:
\begin{equation}
\label{eq:pqexpansionrightside}
    -\frac{\theta[p-\mu+(q-p)]-\theta(p-\mu)}{2 [p+(q-p)](q-p)[2p+(q-p)]},
    =-\int_0^1 \text{d} x \sum_{n=0}^\infty  \frac{p \delta^{(n)} (p-\mu) (p-q)^n  }{4 (n+1)! \left[p^2 +(1-x) (q^2-p^2)\right]^\frac{3}{2}}.
\end{equation}
The most notable remaining hurdle is the left-hand side of eq.~\eqref{eq:fullpowerseriescomparionformula} and specifically the expansion of the distribution functions. However, having restored the full integrand structure, we can utilize the differentiation formulae of section \ref{sec:momentumasymptotes} (see eqs.~\eqref{eq:smallparameterexpansions02} and \eqref{eq:spatialvariantofnFtrepresentation}) and write the following representation:
\begin{equation}
\begin{split}
\label{eq:squarebracketformula}
&\int_0^1 \text{d} x \left\{\frac{\theta\left[\sqrt{p^2+(1-x)(q^2-p^2)}-\mu\right]}{4 \left[p^2 +(1-x) (q^2-p^2)\right]^\frac{3}{2}} -\frac{\delta\left[\sqrt{p^2+(1-x)(q^2-p^2)}-\mu\right]}{4 \left[p^2 + (1-x)(q^2-p^2)\right]} \right\}\\
    &=  \sum_{n=0}^\infty\frac{(-1)^n}{n!} \left[ \left(-\frac{1}{2p}\frac{\text{d}}{\text{d} p} \right)^{n+1}\frac{  \theta\left(p-\mu \right)}{2 p} \right]\left[\int_0^1 \text{d} x (1-x)^n\right] \left(q^2-p^2\right)^n \\
    &=\sum_{n=0}^\infty\frac{(-1)^n}{(n+1)!} \left[ \left(-\frac{1}{2p}\frac{\text{d}}{\text{d} p} \right)^{n+1}\frac{  \theta\left(p-\mu \right)}{2 p} \right]\left(q^2-p^2\right)^n.
    \end{split}
\end{equation}
While this expression was previously utilized exclusively through partial integration, here we aim to apply the derivatives directly to find a closed form summation of distributions to replace the implicit expression inside the square brackets. For this purpose let us study the following expression containing $n+1$ consecutive operations 
\begin{equation}
\begin{split}
(-1)^{n+1}\underset{n+1}{ \underbrace{\left(\frac{\text{d}}{\text{d}p} \frac{1}{2p} + \frac{1}{2p} \right)\left(\frac{\text{d}}{\text{d}p} \frac{1}{2p} + \frac{1}{2p} \right) \dots \left(\frac{\text{d}}{\text{d}p} \frac{1}{2p} + \frac{1}{2p} \right)  }}
\end{split}
\end{equation}
where each differential operator acts on the full previous integrand. To adjust it better to our problem, we may just as well add an additional $\frac{1}{2p}$ operator at the front, and observe specifically
\begin{equation}
\begin{split}
&\frac{(-1)^{n+1}}{2p}\underset{n+1}{ \underbrace{\left(\frac{\text{d}}{\text{d}p} \frac{1}{2p} + \frac{1}{2p} \right)\left(\frac{\text{d}}{\text{d}p} \frac{1}{2p} + \frac{1}{2p} \right) \dots \left(\frac{\text{d}}{\text{d}p} \frac{1}{2p} + \frac{1}{2p} \right)  }}\\
&= \frac{1}{2^{n+2}}\left[\frac{(2n+1)!!}{p^{2n+3}}-\frac{(2n+1)!!}{p^{2n+2}}+\frac{n(2n-1)!!}{p^{2n+1}}-\frac{(n-1)(2n-1)!!}{3p^{2n}}+\dots \frac{(-1)^{n} n(n+1)}{2p^{n+3}}+\frac{(-1)^{n+1}}{p^{n+2}}\right]\\
&=  \frac{1}{2^{n+2}} \sum_{k=0}^{n+1} \frac{(-1)^k(2n+2-k)!}{(2n+2-2k)!! k!p^{2n+3-k}}.
\end{split}
\end{equation}
Let us note that this (as well as many further summation formulae) are straightforward to prove via induction. For the purpose of illustrating this, we define a closely related expression without the prefactor such that 
\begin{equation}
\begin{split}
T_n \equiv(-1)^{n+1}\underset{n+1}{ \underbrace{\left(\frac{\text{d}}{\text{d}p} \frac{1}{2p} + \frac{1}{2p} \right)\left(\frac{\text{d}}{\text{d}p} \frac{1}{2p} + \frac{1}{2p} \right) \dots \left(\frac{\text{d}}{\text{d}p} \frac{1}{2p} + \frac{1}{2p} \right)  }} =  \frac{1}{2^{n+1}} \sum_{k=0}^{n+1} \frac{(-1)^k(2n+2-k)!}{(2n+2-2k)!! k!p^{2n+2-k}}.
\end{split}
\end{equation}
Then, we can observe the formula to work for $n=1$ trivially. Assuming it holds for an arbitrary integer $n$, we can easily show 
\begin{equation}
\begin{split}
    &-\left(\frac{\text{d}}{\text{d}p} \frac{1}{2p} + \frac{1}{2p}\right)T_{n}\\
    %&=-\left(\frac{\text{d}}{\text{d}p} \frac{1}{2p} + \frac{1}{2p}\right)\frac{1}{2^{n+1}} \sum_{k=0}^{n+1} \frac{(-1)^k (2n+2-k)!}{k!(2n+2-2k)!! }{p^{-2n-2+k} }\\
    &= -\frac{1}{2^{n+2}} \sum_{k=0}^{n+1} \frac{(-1)^k (2n+2-k)!}{k!(2n+2-2k)!! }{p^{-2n-3+k} } \left[1-\frac{2n+3-k}{p}\right]\\
   % &=  -\frac{1}{2^{n+2}} \sum_{k=0}^{n+1}\left[ \frac{(-1)^k (2n+2-k)!}{k!(2n+2-2k)!! }{p^{-2n-3+k} } -\frac{(-1)^k (2n+3-k)!}{k!(2n+2-2k)!! }{p^{-2n-4+k} }\right]\\
    %&= \frac{(-1)^{n+2}}{2^{n+2}p^{n+2}} + \frac{(2n+4)!}{2^{n+2}(2n+4)!!p^{2n+4}} \\
    %&-\frac{1}{2^{n+2}}\sum_{k=0}^{n} (-1)^k \left[ \frac{ (2n+2-k)!}{k!(2n+2-2k)!! } + \frac{ (2n+2-k)!}{(k+1)!(2n-2k)!! }\right]p^{-2n-3+k}\\
    &=\frac{(-1)^{n+2}}{2^{n+2}p^{n+2}} + \frac{(2n+4)!}{2^{n+2}(2n+4)!!p^{2n+4}} -\frac{1}{2^{n+2}}\sum_{k=0}^{n} (-1)^k \left[ \frac{ (2n+3-k)!}{(k+1)!(2n+2-2k)!! }\right]p^{-2n-3+k} \\
    &=  \frac{1}{2^{n+2}} \sum_{k=0}^{n+2} \frac{(-1)^k(2n+4-k)!}{k!(2n+4-2k)!! } \frac{1}{p^{2n+4-k}}=T_{n+1}.
    \end{split}
\end{equation}
We note that this operator formula exactly represents the action inside the square brackets of eq.~\eqref{eq:squarebracketformula}, with each power of $p$ associated with specific derivative of Heaviside step function. This allows us to identify explicitly
\begin{equation}
\label{eq:closedformsquarebrackets}
     \left[ \left(-\frac{1}{2p}\frac{\text{d}}{\text{d} p} \right)^{n+1}\frac{  \theta(p-\mu)}{2 p} \right] = \frac{1}{2^{n+2}} \sum_{k=0}^{n+1} \frac{(-1)^k (2n+2-k)!}{k!(2n+2-2k)!! } \frac{\theta^{(k)}(p-\mu)}{p^{2n+3-k} },
\end{equation}
where we emphasize that the summation is carried over $k$ times differentiated Heaviside step functions.
\subsection{First comparisons}
\noindent
As the very first observation of eq.~\eqref{eq:closedformsquarebrackets}, we can identify $k=0$ to yield in conjunction with eq.~\eqref{eq:squarebracketformula} an exactly identical power series expansion as seen to arise from eq.~\eqref{eq:thetaexpansiontrivial}. This is to say the leading order between the sides of eq.~\eqref{eq:fullpowerseriescomparionformula} match trivially.

Furthermore, the coefficient corresponding to $k=1$ is identical with that of $k= 0$ in eq.~\eqref{eq:closedformsquarebrackets}, apart from being multiplied once by $-p$. This can be directly seen to occur also by comparing the expressions in the right-hand sides of eqs.~\eqref{eq:FeynmanparametleadingorderappB} and \eqref{eq:pqexpansionrightside}. Thus, the first two orders of expansion (with respect to the distributions) have been observed to explicitly match without any additional computations.

To be able to deal with the order proportional to $\delta^{(1)} (p-\mu)$, we need to introduce the scale $p-q$ explicitly to the expansion. Specifically this requires us to split the previous expansion parameter such that $q^2 - p^2 = [2p+(q-p)](q-p)$, where $p \gg |(q-p)|$. For the right-hand side of eq.~\eqref{eq:fullpowerseriescomparionformula}, we can immediately identify from eq.~\eqref{eq:thetaexpansiontrivial} that the corresponding expression reads in terms of the previous expansions:
\begin{equation}
    -\frac{p\delta^{(1)} (p-\mu) (q-p)}{4p [p+(q-p)][2p+(q-p)]} =- \frac{p\delta^{(1)}(p-\mu)(q-p) }{2} \sum_{n=0}^\infty\frac{(-1)^n (2n+1)!!}{2^{2+n}p^3 (n+1)!} \frac{\{[2p+(q-p)](q-p)\}^n}{p^{2n}}.
\end{equation}

The left-hand side we can identify using again eq.~\eqref{eq:squarebracketformula} as 
\begin{equation}
\begin{split}
    &\sum_{n=0}^\infty \frac{(-1)^n}{(n+1)!}\left[ \left(-\frac{1}{2p}\frac{\text{d}}{\text{d} p} \right)^{n+1}\frac{  \theta(p-\mu)}{2 p} \right](q^2-p^2)^n\\
    %&=\sum_{n=0}^\infty \frac{(-1)^n}{(n+1)!} \frac{\left[2p(q-p)+(q-p)^2 \right]^n}{2^{n+2} p^{2n+3}} \sum_{k=0}^{n+1} \frac{(-1)^k (2n+2-k)!}{k!(2n+2-2k)!! } \theta^{(k)}(p-\mu) p^k\\
    &\overset{k=2}{\longmapsto} -\sum_{n=1}^\infty \frac{(-1)^n}{(n+1)!} \frac{\left[2p(q-p)+(q-p)^2 \right]^n}{2^{n+2} p^{2n+1}}  \frac{ (2n)! \delta^{(1)}(p-\mu) }{2!(2n-2)!! }. 
    \end{split}
\end{equation}
To subtract these two expressions from one another efficiently, order-by-order in terms of the new expansion parameter $q-p$ to explicitly show their equivalence, we need to first change the indexation of the latter equation such that it starts with $n=0$ (to align with the former) and utilize the binomial formula on both expressions. We conveniently divide the subtraction to commonly indexed terms in the binomial summation as follows
\begin{equation}
\label{eq:subtractionbinomial}
    -\frac{(-1)^n (2n+1)!!}{2^{n+3} (n+2)!} \frac{1}{p^{2n+2}} \sum_{k=0}^n \left[2 (n+1) \binom{n+1}{k}-(n+2)\binom{n}{k} \right] (2p)^{n-k} (q-p)^{n+1+k}
\end{equation}
and the single excess part from $(q^2-p^2)^{n+1}$
\begin{equation}
\label{eq:extracomponentdelta1}
   - \frac{(-1)^n (n+1) (2n+1)!!}{2^{n+3} (n+2)!} \frac{(q-p)^{2n+2}}{p^{2n+3}}.
\end{equation}
Upon summing both these structures over $n \geq 0$, cancellations occur for any $\mathcal{O}\left[(p-q)^j\right]$, with the above term summed up to $n = j-1$ and the latter for $n= \frac{j}{2}-1$ for an even $j = 2l$ (with the odd powers not affected). To illustrate, note that the indices $n$ contributing to the odd powers $(q-p)^{2j+1}$ can be determined by fixing $n+1+k = 2j+1$, and  substituting the extrema of the index $k$:
\begin{eqnarray}
    0 &=& 2j-n \Rightarrow n = 2j,\\
    n &=& 2j-n \Rightarrow n = j.
    \end{eqnarray}
Thus, all the terms contributing to the odd powers of $(p-q)^{2j+1}$ can be written such that
\begin{equation}
\begin{split}
\label{eq:2j1contributionsdelta1}
    \mathcal{O}\left[(p-q)^{2j+1} \right]& \reflectbox{$\in$} - \frac{(q-p)^{2j+1}}{p^{2j+2}}  \sum_{n=j}^{2j} \frac{(-1)^n (2n+1)!!}{2^{2j-n+3} (n+2)!} \left[ 2 (n+1) \binom{n+1}{2j-n}-(n+2)\binom{n}{2j-n}\right]\\
    &=- \frac{(q-p)^{2j+1}}{p^{2j+2}}  \sum_{n=0}^{j} \frac{(-1)^{n+j} (2n+2j+1)!!}{2^{j-n+3} (j-n)! (2n)! } \left[ \frac{ 2 (n+j+1)}{ (2n+1) (n+j+2)}-\frac{1}{ (n+j+1)}\right]
    \end{split}
\end{equation}
where the summation identically vanishes for all $j \in \mathbb{N}_0$, which can be seen through the associated (observed) non-trivial sum identities
\begin{eqnarray}
\label{eq:nontrivialsum1}
-\sum_{n=0}^{j}  \binom{2n+2j+2}{j-n,2n+1,n+j+1}\frac{(-1)^{j-n} (-1)^{2n+1} }{n+j+2} &=& 1, \\
-\sum_{n=0}^{j} (-1)^{j-n} (-1)^{2n+1}  \binom{2n+2j+2}{j-n,2n+1,n+j+1} &=& 2^{2j+1},\\
\label{eq:nontrivialsum3}
\sum_{n=0}^j (-1)^{j-n}(-1)^{2n}   \binom{2n+2j+1}{j-n,2n,n+j+1} &=& 2^{2j+1}-1.
\end{eqnarray}
Here all the rows utilize the compact representation of coefficients in trinomials generated by $(1-1-1)^{2n+2j+1}$ (and corresponding Pascal pyramid) on the left-hand-side.

For even powers of $(q-p)^{2j}$ we find instead via a similar substitution $k = 2j-n-1$ the following index subsets for $n$ summation of the subtraction sum in eq.~\eqref{eq:subtractionbinomial}
\begin{eqnarray}
    0 &=& 2j-n-1 \Rightarrow n = 2j-1,\\
    n-1 &=& 2j-n-1 \Rightarrow n = j,
    \end{eqnarray}
    where we note that the index $k= n$ would only contribute to odd powers $(p-q)^{2j+1}$, and hence the limit is fixed to $k= n-1$. For the contributions from eq.~\eqref{eq:extracomponentdelta1}, we fix $2n+2 = 2j \Rightarrow n = j-1$. Combining all these terms, we find
\begin{equation}
    \begin{split}
    \label{eq:2jcontributionsdelta1}
        \mathcal{O}\left[(p-q)^{2j} \right]& \reflectbox{$\in$}- \frac{(q-p)^{2j}}{p^{2j+1}} \sum_{n=j}^{2j-1} \frac{(-1)^n (2n+1)!!}{2^{2j-n+2} (n+2)!}   \left[2 (n+1) \binom{n+1}{2j-n-1}-(n+2)\binom{n}{2j-n-1} \right]\\
        &-\frac{(q-p)^{2j}}{p^{2j+1}} \frac{(-1)^{j-1}j (2j-1)!!}{2^{j+2}(j+1)!}\\
        &=- \frac{(q-p)^{2j}}{p^{2j+1}}  \sum_{n=0}^{j-1} \frac{(-1)^{n+j} (2n+2j+1)!!}{2^{j-n+2} (j-n-1)! (2n+1)! } \left[ \frac{ 2 (n+j+1)}{ (2n+2) (n+j+2)}-\frac{1}{ (n+j+1)}\right]\\
        &-\frac{(q-p)^{2j}}{p^{2j+1}} \frac{(-1)^{j-1}j (2j-1)!!}{2^{j+2}(j+1)!},
    \end{split}
\end{equation}
where the coefficients corresponding to each $j\in \mathbb{N}_0$ vanish identically, as indicated by the observed sum identities parallel to eqs.~\eqref{eq:nontrivialsum1}--\eqref{eq:nontrivialsum3}:
\begin{eqnarray}
\sum_{n=0}^{j} (-1)^{j-n} (-1)^{2n} \binom{2n+2j}{j-n,2n,n+j} \frac{1}{n+j+1} &=& 1,\\
 \sum_{n=0}^{j} (-1)^{j-n} (-1)^{2n} \binom{2n+2j}{j-n,2n,n+j} &=& 2^{2j},\\
      \sum_{n=0}^{j-1} (-1)^{j-n-1} (-1)^{2n+1} \binom{2n+2j+1}{j-n-1,2n+1,n+j+1} &=& 1-2^{2j}.
\end{eqnarray}
Thus, we can move forward, with the expansions proportional to $\delta^{(1)} (p-\mu)$ in eq.~\eqref{eq:fullpowerseriescomparionformula} being equivalent.
\subsection{Arbitrary order}
\noindent
To address terms contributing to the part of the power series characterized by an arbitrary derivative of Heaviside step function $\theta^{(b)} (p-\mu)$, we approach similar to the example above. Here we note that the steps leading up to terms replacing those given in eqs.~\eqref{eq:subtractionbinomial} and \eqref{eq:extracomponentdelta1} remain identical. A major difference (albeit trivial) is seen with the structure in place of eq.~\eqref{eq:extracomponentdelta1}, though, with in total $b-1$ terms appearing instead of just the one earlier. Their impact is also more general, with at most $\left \lceil\frac{b-1}{2} \right\rceil$ terms contributing to the cancellations of terms proportional to powers of $(q-p)^j$, where $j \geq b-1$. The contributions arising from the terms akin to those in eq.~\eqref{eq:subtractionbinomial}, can -- in alignment -- be seen to contribute to each power of $(q-p)^{2j+1}$ via exactly $j+1-b+\lfloor \frac{b}{2} \rfloor$ terms, and for powers $(q-p)^{2j}$ via $j-b+\lceil \frac{b}{2} \rceil$ terms in total, as we will demonstrate below. 

To begin with, the all contributions proportional to the arbitrary derivative of Heaviside step function $\theta^{(b)} (p-\mu)$ on the right-hand side of eq.~\eqref{eq:fullpowerseriescomparionformula} for all $b \geq 1$ (cf. eq~\eqref{eq:extracomponentdelta1}) as follows
\begin{equation}
\label{eq:thetabrightside}
    -\frac{p\theta^{(b)} (p-\mu) (q-p)^{b-1}}{2(b!)p [p+(q-p)][2p+(q-p)]} =-\frac{p\theta^{(b)}(p-\mu)(q-p)^{b-1} }{b!}\sum_{n=0}^\infty\frac{(-1)^n (2n+1)!!}{2^{2+n}p^3 (n+1)!} \frac{\{[2p+(q-p)](q-p)\}^n}{p^{2n}}.
\end{equation}
The corresponding expression on the left-hand side of eq.~\eqref{eq:fullpowerseriescomparionformula} can be written such that
\begin{equation}
\begin{split}
\label{eq:thetableftside}
    &\sum_{n=0}^\infty \frac{(-1)^n}{(n+1)!}\left[ \left(-\frac{1}{2p}\frac{\text{d}}{\text{d} p} \right)^{n+1}\frac{  \theta(p-\mu)}{2 p} \right] \left[2p(q-p)+(q-p)^2 \right]^n\\
    &=\sum_{n=0}^\infty \frac{(-1)^n}{(n+1)!} \frac{\left[2p(q-p)+(q-p)^2 \right]^n}{2^{n+2} p^{2n+3}} \sum_{k=0}^{n+1} \frac{(-1)^k (2n+2-k)!}{k!(2n+2-2k)!! } \theta^{(k)}(p-\mu) p^k\\
    &\overset{k=b}{\mapsto} \sum_{n=b-1}^\infty \frac{(-1)^{n+b}}{(n+1)!} \frac{\left[2p(q-p)+(q-p)^2 \right]^n}{2^{n+2} p^{2n+3-b}}  \frac{ (2n+2-b)!}{b!(2n+2-2b)!! } \theta^{(b)}(p-\mu) \\
    &=\sum_{n=0}^\infty \frac{(-1)^{n-1} (q-p)^{n+b-1}}{(n+b)!} \frac{\left[2p+(q-p) \right]^{n+b-1}}{2^{n+b+1} p^{2n+b+1}}  \frac{ (2n+b)!}{b!(2n)!! } \theta^{(b)}(p-\mu),
    \end{split}
\end{equation}
where we adjusted the expression on the lowest row to align through indexation (through the starting value $n=0$) with eq.~\eqref{eq:thetabrightside}. Expanding the powers of $(q^2-p^2)^{n}$ and $(q^2-p^2)^{n+b-1}$, we subtract these expressions from one another and find first the expression containing the first $n+1$ terms of either binomial expansion via (cf. eq.~\eqref{eq:subtractionbinomial})
\begin{equation}
\label{eq:subtractiontermthetab}
    \frac{(-1)^{n-1} }{2^{n+2} (n+b)! b!} \frac{1}{p^{2n+2}} \sum_{k=0}^n \left[ \frac{(2n+b)!}{(2n)!!} \binom{n+b-1}{k} - \frac{(n+b)!}{(n+1)!} \frac{(2n+2)!}{(2n+2)!!} \binom{n}{k}\right] (2p)^{n-k} (q-p)^{k+n+b-1}
\end{equation}
and additionally the $b-1$ terms in excess arising from the binomial expansion of $(q^2-p^2)^{n+b-1}$ given by
\begin{equation}
\label{eq:excesstermsthetab}
    \frac{(-1)^{n-1} }{2^{n+3} (n+b)! b!} \frac{1}{p^{2n+3}} \sum_{k=0}^{b-2}  \frac{(2n+b)!}{(2n)!!} \binom{n+b-1}{k+n+1}\left(\frac{1}{2p} \right)^{k}(q-p)^{2n+b+k},
\end{equation}
which we have re-indexed such that summation begins from $k=0$. Summing each of these two expressions over $n \in \mathbb{N}_0$, we find the former expression of eq.~\eqref{eq:subtractiontermthetab} to contribute to each power of $(p-q)^{j}$ with terms up to $n=j-b+1$ (corresponding to $k=0$), while the latter expression of eq.~\eqref{eq:excesstermsthetab} has at most $\lceil\frac{b-1}{2} \rceil$ terms contributing to each order $(p-q)^j$, and  occurrences of each contribution -- associated with index $k$ -- are situated at $\{n| j-2n = b+k \}$, with the very first additions contributing to $(q-p)^b$ (the next-to-leading order of the expansion). We emphasize that the contributions (pertaining to the $2n$ explicitly) are split separately to two distinct lists, those for even powers $(q-p)^{2j}$ and those for odd powers $(q-p)^{2j+1}$ as seen also in the example of the previous subsection.

To determine which terms from eq.~\eqref{eq:subtractiontermthetab} contribute to the odd powers $2j+1$, we fix $k=2j-n-b+2$ and note that the summation limits -- in alignment with those of eqs.~\eqref{eq:2j1contributionsdelta1} and \eqref{eq:2jcontributionsdelta1} -- arise from the limiting values of $0 \leq k \leq n$, or the closest indices able to generate the odd value of $2j+1$. The upper limit to the relevant values of $n$ arises from $k=0$ such that
\begin{eqnarray}
    0 &=& 2j-n-b+2 \Rightarrow n = 2j-b+2
    \end{eqnarray}
    while the lower limit corresponds to either of the following two equations depending on the even/odd nature of the integer variable $b$:
    \begin{eqnarray}
         n &\overset{b=2l}{=}& 2j-n-b+2 \Rightarrow n = j -\frac{b-2}{2} = j+1-l,\\
         n-1 &\overset{b=2l+1}{=}& 2j-n-b+2 \Rightarrow n = j -\frac{b-3}{2} = j+1-l.
    \end{eqnarray}
These can be condensed by writing the lower limit (corresponding again to $(q-p)^{2j+1}$) as follows
\begin{equation}
     n = j+1-\left\lfloor \frac{b}{2} \right\rfloor.
\end{equation}
For even powers of $(q-p)^{2j}$, we instead fix $k = 2j-n-b+1$. The corresponding upper limit of index $n$ is again found by setting $k=0$, and yields
\begin{eqnarray}
    0 &=& 2j-n-b+1 \Rightarrow n = 2j-b+1
    \end{eqnarray}
   while the lower limit reads
    \begin{eqnarray}
         %n &\overset{b=2l+1}{=}& 2j-n-b+1 \Rightarrow n = j -\frac{b-1}{2} = j-l\\
        % n-1 &\overset{b=2l}{=}& 2j-n-b+1 \Rightarrow n = j -\frac{b-2}{2} = j+1-l\\
         n &=& j+1-\left \lceil \frac{b}{2} \right \rceil,
    \end{eqnarray}
where we have already taken into account the specifics involving both the even and odd values of $b$. 

We note that is also convenient to split the expression of eq.~\eqref{eq:excesstermsthetab} into two distinct structures, with the one containing solely even powers $2k$, and the other one odd powers $2k-1$ from the sole summation. These read
    \begin{equation}
    \begin{split}
    \label{eq:distinctexcesssums}
    &\frac{(-1)^{n-1} }{2^{n+3} (n+b)! b!} \frac{1}{p^{2n+3}} \sum_{k=0}^{b-2}  \frac{(2n+b)!}{(2n)!!} \binom{n+b-1}{k+n+1}\left(\frac{1}{2p} \right)^{k}(q-p)^{2n+b+k}\\
    &=\frac{(-1)^{n-1} }{2^{n+3} (n+b)! b!} \frac{1}{p^{2n+3}} \sum_{k=0}^{\lfloor \frac{b-2}{2} \rfloor}  \frac{(2n+b)!}{(2n)!!} \binom{n+b-1}{2k+n+1}\left(\frac{1}{2p} \right)^{2k}(q-p)^{2n+b+2k}\\
    &+\frac{(-1)^{n-1} }{2^{n+3} (n+b)! b!} \frac{1}{p^{2n+3}} \sum_{k=1}^{\lceil \frac{b-2}{2} \rceil}  \frac{(2n+b)!}{(2n)!!} \binom{n+b-1}{2k+n}\left(\frac{1}{2p} \right)^{2k-1}(q-p)^{2n+b+2k-1}
    \end{split}
\end{equation}
and their mutually distinct contributions to powers $(q-p)^j$ is determined by the even/odd nature of $b$. Specifically, $b$ determines which of the two summations contributes to terms $(q-p)^{2j+1}$, directing the other summation to $(q-p)^{2j}$. Furthermore, we need to determine which $k$ indices contribute to each computation. For this purpose we move the $k$ summation (notably independent of the $n$ summation) outward, and fix $n$ to produce the sought-after power of $(q-p)^j$.

Let us list the 4 different combinations here. First let $b$ be even. Then $(q-p)^{2j+1}$ is contributed to by the last row of eq.~\eqref{eq:distinctexcesssums} and we can demand  $2n+b+2k-1 = 2j+1$. This implies that terms attributed to $k$ can be present at the order $(q-p)^{2j+1}$ only if $\exists n \geq 0$ such that $k =  j-n+1 - \frac{b}{2}$. This equation implies specifically that any $k \leq j+1-\frac{b}{2}$ is acceptable which -- along with original limits -- combines to $1 \leq k \leq \text{min} \left( j+1-\frac{b}{2} ,\lceil   \frac{b-2}{2} \rceil \right)$ simultaneously. Then note that the second to last row of eq.~\eqref{eq:distinctexcesssums} contributes to $(q-p)^{2j}$. Now demanding $2n+b+2k = 2j$, we find attributable terms if $\exists n$ such that $k = j-n-\frac{b}{2}$. Now the full limitations (following previous reasoning) read $0 \leq k \leq \text{min} \left(j-\frac{b}{2}, \left \lfloor \frac{b-2}{2} \right\rfloor \right)$.

Let us next assume instead that $b$ is odd. Now the powers which the summations contribute to are reversed from above. Specifically, for $(q-p)^{2j+1}$, this fixes index $k$ such that $k = j-n-\frac{b-1}{2}$, which implies the condition $0 \leq k \leq \text{min} \left( j-\frac{b-1}{2}, \lfloor \frac{b-2}{2} \rfloor \right) $. The remaining summation aligned with $(q-p)^{2j}$ enforces the equation $k = j - n-\frac{b-1}{2}$. This implies the full condition such that $1 \leq k \leq \text{min} \left (j-\frac{b-1}{2}, \lceil   \frac{b-2}{2}  \rceil \right)$.

To summarize the summation limits onto eq.~\eqref{eq:distinctexcesssums}, for $(q-p)^{2j+1}$ we find $\lceil \frac{b+1}{2} \rceil - \lfloor \frac{b+1}{2} \rfloor \leq k \leq \text{min} \left(j-\left \lfloor \frac{b-1}{2} \right \rfloor, \lfloor\frac{b-2}{2} \rfloor \right)$. Similarly for $(q-p)^{2j}$ the limits for $k$ summation read $\left \lceil \frac{b}{2} \right \rceil-\left \lfloor \frac{b}{2} \right \rfloor \leq k \leq \text{min} \left(j-\lfloor \frac{b}{2} \rfloor, \lceil \frac{b-2}{2} \rceil\right)$.

Following the reasoning dissected above, let us list the terms contributing to each power of $(q-p)^j$. First, we can write all contributions to odd powers such that
\begin{equation}
\begin{split}
    &\mathcal{O}\left[(p-q)^{2j+1} \right]\\
    &\reflectbox{$\in$} -\frac{(q-p)^{2j+1}}{(2p)^{2j+4-b}}  \sum_{n=j+1-\lfloor \frac{b}{2} \rfloor}^{2j-b+2} \frac{(-2)^n}{(n+b)! b!} \left[ \frac{(2n+b)!}{(2n)!!} \binom{n+b-1}{2j+2-n-b} - \frac{(n+b)!}{(n+1)!} \frac{(2n+2)!}{(2n+2)!!} \binom{n}{2j+2-n-b}\right]\\
    &+\frac{(q-p)^{2j+1}}{p^{2j+4-b}}\sum_{k= \left \lceil \frac{b+1}{2} \right \rceil - \left \lfloor \frac{b+1}{2} \right \rfloor}^{\text{min}\left(j-\lfloor\frac{b-1}{2} \rfloor ,\lfloor\frac{b-2}{2} \rfloor\right)}\frac{(-1)^{j -k-\lfloor\frac{b+1}{2} \rfloor} \left(2j-2k+1 + \left \lceil \frac{b+1}{2} \right \rceil - \left \lfloor \frac{b+1}{2} \right \rfloor \right)! }{2^{j +k+3-\lfloor\frac{b}{2}\rfloor} (j -k+\lceil\frac{b+1}{2} \rceil)! b! (2j -2k-2 \lfloor \frac{b-1}{2} \rfloor)!!}   \binom{j -k+ \lceil\frac{b-1}{2} \rceil}{j +k+1-\lceil\frac{b-1}{2} \rceil},
    \end{split}
\end{equation}
 where we combined the two different scenarios corresponding to even/odd values of $b$ on the last line. Utilizing similar conventions, the even powers can be written as
\begin{equation}
\begin{split}
    &\mathcal{O}\left[(p-q)^{2j} \right]\\
    &\reflectbox{$\in$}-\frac{(q-p)^{2j}}{(2p)^{2j-b+3}} \sum_{n=j+1-\lceil\frac{b}{2} \rceil}^{2j-b+1}\frac{(-2)^{n} }{ (n+b)! b!}  \left[ \frac{(2n+b)!}{(2n)!!} \binom{n+b-1}{2j-n-b+1} - \frac{(n+b)!}{(n+1)!} \frac{(2n+2)!}{(2n+2)!!} \binom{n}{2j-n-b+1}\right]\\
    &+\frac{(q-p)^{2j}}{\textcolor{black}{p}^{2j-b+3}}\sum_{k= \left \lceil \frac{b}{2} \right \rceil - \left \lfloor \frac{b}{2} \right \rfloor}^{\text{min}\left(j-\lfloor\textcolor{black}{\frac{b}{2}} \rfloor ,\lceil\frac{b-2}{2} \rceil\right)}\frac{(-1)^{j -k-1-\lfloor\frac{b}{2} \rfloor} \left(2j-2k + \left \lceil \frac{b}{2} \right \rceil - \left \lfloor \frac{b}{2} \right \rfloor \right)! }{2^{j +k+3- \left\lceil\textcolor{black}{\frac{b}{2}} \right\rceil} \left(j -k+ \left\lfloor \frac{b+1}{2} \right\rfloor \right)! b!(2j -2k-2 \lfloor \frac{b}{2} \rfloor)!!}  \binom{j -k+ \lfloor\frac{b-1}{2} \rfloor}{j +k-\lfloor\frac{b-1}{2} \rfloor}.
    \end{split}
\end{equation}
As previously, every single coefficient of powers $(q-p)^j$ can be observed to vanish (a formal proof via induction or combinatorics is deferred to another research paper). The impliciation of the general formula is clear: the two computation schemes are equivalent on integrand level and not just integral level. This aligns with and extends the assumptions and statements made earlier in \cite{Gorda:2022yex}, and has effectively led to this rather neat list of non-trivial summation identities.

\section{Bosonic propagators}
\label{sec:bosons}
\noindent 
In this appendix, we briefly demonstrate that strict residue theorem is sufficient approach to describe the bosonic variant of the fermionic one-loop example of the main text. Specifically, we walk through multiple subset examples, and compute (at least) the leading behaviour utilizing two different integration orders (again to recognize equal results). Here, even more than in the fermionic examples, it can be explicitly seen how much the strict $T=0$ residue theorem simplifies the computations, even at the one loop order, and we aim to justify its general usage (as opposed to the fully fermionic counterpart).

To streamline compactly communicate the following computations we introduce the following notation
\begin{equation}
    \mathcal{B}_{\alpha_1 \alpha_2} (\mu, \vec{s},s_0) =   \text{P.V.}  \int_{-\infty}^\infty \frac{\text{d} p_0}{2\pi} \int_p \frac{1}{[(p_0+i\mu)^2+p^2]^{\alpha_1}} \frac{1}{[(p_0-s_0)^2+|\vec{p}-\vec{s}|^2]^{\alpha_2}},
\end{equation}
\begin{equation}
    \text{Res} \left[\mathcal{B}_{\alpha_1 \alpha_2} \right] (\mu, \vec{s},s_0) =    \int_p i \text{Res} \left\{ \frac{1}{[(p_0+i\mu)^2+p^2]^{\alpha_1}} \frac{1}{[(p_0-s_0)^2+|\vec{p}-\vec{s}|^2]^{\alpha_2}}\right\},
\end{equation}
where we are most interested in cases involving $\{\alpha_1 = 1, \alpha_2 \in \mathbb{Z}_+\}$\footnote{We note that from the preceding discussion, $\alpha_1 \geq 2$ would experience some discrepancy between the results, as can be seen by setting for example $\alpha_1=2,\alpha_2=0$. The choice used here is therefore to study the effects of the bosonic propagator.}. 

\subsection{Simplest scenario}
\label{sec:simplestboson}
\noindent
Let us begin by studying the leading structure of 
\begin{equation}
    \text{Res} \left[ \mathcal{B}_{11} \right] (\mu, 0,0) =     \int_p i\sum_{n=0}^1 \text{Res} \left[ \frac{1}{(p_0+i\mu)^2+p^2}\frac{1}{p_0^2+ p^2} \right]_{p_0=-in\mu+ip},
\end{equation}
lacking formal external momenta, through the residue approach at strict $T=0$ limit. By straightforward steps -- involving simple changes of integration variables -- continuing from eq.~\eqref{eq:residuecollection}, we find
 \begin{equation}
   \begin{split}
   \label{eq:bosonpropnoexternalmomentumresidue}
   \text{Res} \left[ \mathcal{B}_{11} \right] (\mu, 0,0)
       &=\frac{1}{\mu} \int_p \frac{1}{2p} \left[ \frac{\theta(p-\mu)}{2p-\mu}-\frac{1}{2p+\mu} \right]\\
       &= \left( \frac{e^\gamma \Lambda^2}{4 \pi} \right)^\frac{3-d}{2}\frac{ \mu^{d-3}}{2(4\pi)^\frac{d}{2} \Gamma \left( \frac{d}{2} \right)} \left[\int_1^\infty \frac{\text{d}p p^{d-3}}{1-\frac{1}{2p}}  -\frac{1}{2^{d-2}} \int_0^\infty \frac{\text{d}p p^{d-2}}{1+p}\right]\\
       &= \left( \frac{e^\gamma \Lambda^2}{4 \pi} \right)^\frac{3-d}{2}\frac{ \mu^{d-3}}{2(4\pi)^\frac{d}{2}\Gamma \left( \frac{d}{2} \right)} \left[\int_0^1 \frac{\text{d}y y^{1-d}}{1-\frac{y}{2}}  -\frac{1}{2^{d-2}} \int_0^\infty \frac{\text{d}p p^{d-2}}{1+p}\right]\\
       &= \left( \frac{e^\gamma \Lambda^2}{4 \pi} \right)^\frac{3-d}{2}\frac{\mu^{d-3}}{2^{d-1} (4 \pi)^\frac{d}{2}\Gamma \left( \frac{d}{2} \right)} \left[B \left(\frac{1}{2}, 2-d, 0 \right)-\Gamma \left(d-1 \right)\Gamma \left(2-d \right) \right],
       \end{split}
   \end{equation}
where we have expressed the hypergeometric integral \footnote{\label{fn:hypergeom}Integral representation of hypergeometric function reads $\int_0^1 \text{d} x \, x^{b-1} (1-x)^{c-b-1} (1-z x)^{-a} = \frac{\Gamma \left( c \right)}{\Gamma \left( c -b \right)\Gamma \left( b \right)}{}_2 F_1 \left[a, b, c, z \right]$\cite{Stegun}.} -- characterized by integration interval $x \in (0,1)$ -- from the second to last row in terms of an incomplete beta function \footnote{Integral representation of incomplete beta function reads $\int_0^z \text{d} x \, x^{a-1} (1-x)^{b-1}  = B(z,a,b)$\cite{Stegun}.} on the last row. The formal constraints for convergence of this expression can be easily discerned from the latter Euler beta function (multiplication of two Euler gamma functions), implying that $1 < d < 2$. However, the closed form expressions utilized here easily enable analytic continuation to  $d \in \mathbb{R}_+ \backslash \mathbb{N}$.

With the first closed form result in place, let us next illustrate how the opposite,
\begin{equation}
  \mathcal{B}_{11}  (\mu, 0,0) =  \text{P.V.} \int_{-\infty}^\infty \frac{\text{d} p_0}{2 \pi}  \int_p \frac{1}{(p_0+i\mu)^2+p^2} \frac{1}{p_0^2+p^2},
\end{equation}
integration order leads to the exact same result. Specifically, applying first the dimensionally regularized spatial integration, we need not write the distribution function, and instead consider directly the outermost temporal integral as the principal value integration discussed in the main text sections. We additionally note that given the lack of small expansion parameters and external spatial momenta, Feynman parametrization is essentially optional, and one can equivalently use partial fraction decomposition to find
\begin{equation}
     \mathcal{B}_{11}  (\mu, 0,0)  = \left( \frac{e^\gamma \Lambda^2}{4 \pi} \right)^\frac{3-d}{2} \frac{\Gamma \left(1-\frac{d}{2} \right)}{(4\pi)^\frac{d}{2}} \frac{[(p_0+i\mu)^2]^{\frac{d}{2}-1}-(p_0^2)^{\frac{d}{2}-1}}{p_0^2-(p_0+i\mu)^2} 
\end{equation}
which is valid for all $p_0 \neq 0$. Then the remaining temporal integration (note again the split at $p_0 = 0$) reads 
\begin{equation}
    \begin{split}
     & \mathcal{B}_{11}  (\mu, 0,0)  \\
        &= \left( \frac{e^\gamma \Lambda^2}{4 \pi} \right)^\frac{3-d}{2} \frac{\Gamma \left(1-\frac{d}{2} \right)}{(4\pi)^\frac{d}{2}} \int_0^\infty \frac{\text{d} p_0}{2\pi} \left\{  \frac{[(p_0+i\mu)^2]^{\frac{d}{2}-1}-(p_0^2)^{\frac{d}{2}-1}}{p_0^2-(p_0+i\mu)^2} +  \frac{[(p_0-i\mu)^2]^{\frac{d}{2}-1}-(p_0^2)^{\frac{d}{2}-1}}{p_0^2-(p_0-i\mu)^2} \right\}\\
        &= \left( \frac{e^\gamma \Lambda^2}{4 \pi} \right)^\frac{3-d}{2} \frac{\Gamma \left(1-\frac{d}{2} \right)}{(4\pi)^\frac{d}{2}} \frac{1}{2i \mu}\int_0^\infty \frac{\text{d} p_0}{2\pi} \left\{-  \frac{[(p_0+i\mu)^2]^{\frac{d}{2}-1}-(p_0^2)^{\frac{d}{2}-1}}{p+\frac{i\mu}{2}} +  \frac{[(p_0-i\mu)^2]^{\frac{d}{2}-1}-(p_0^2)^{\frac{d}{2}-1}}{p_0-\frac{i\mu}{2}} \right\}
    \end{split}
\end{equation}
The most efficient method to extract the hypergeometric presentation hereof is to perform Feynman parametrization now acting on linear power terms of $p_0$. Writing all expressions carefully in the temporal integration interval $p_0 \in (0,\infty)$, we find the temporal integral to be given such that
\begin{equation}
    \begin{split}
    \label{eq:bosoniclowestorderFPintermediate}
        &\frac{1}{2i \mu}\int_0^\infty \frac{\text{d} p_0}{2\pi} \left\{-  \frac{[(p_0+i\mu)^2]^{\frac{d}{2}-1}-(p_0^2)^{\frac{d}{2}-1}}{p_0+\frac{i\mu}{2}} +  \frac{[(p_0-i\mu)^2]^{\frac{d}{2}-1}-(p_0^2)^{\frac{d}{2}-1}}{p_0-\frac{i\mu}{2}} \right\}\\
        &= (2-d) \int_0^1 \text{d} x (1-x)^{1-d}  \int_0^\infty \frac{i \text{d} p_0}{4 \pi \mu} \left\{\left[p_0+\left(1-\frac{x}{2}\right)i\mu\right]^{d-3}-\left[p_0+\frac{i x \mu}{2}\right]^{d-3}\right\}\\
    &-(2-d)\int_0^1 \text{d} x (1-x)^{1-d}  \int_0^\infty \frac{i \text{d} p_0}{4 \pi \mu} \left\{\left[p_0-\left(1-\frac{x}{2}\right)i\mu\right]^{d-3}-\left[p_0-\frac{i x \mu}{2}\right]^{d-3}\right\}\\
    &=\frac{2 \mu^{d-3}}{4 \pi}  \sin \left[\left(1- \frac{d}{2}  \right) \pi  \right]
 \left[\int_0^1 \text{d}x (1-x)^{1-d} \left(1- \frac{x}{2}\right)^{d-2}-\frac{1}{2^{d-2}}\int_0^1 \text{d}x (1-x)^{1-d} x^{d-2} \right]
    \end{split}
\end{equation}
After utilizing Euler's reflection formula and hypergeometric formulae from \cite{Stegun}, we find 
\begin{equation}
    \begin{split} 
   \mathcal{B}_{11}  (\mu, 0,0)  &= \left( \frac{e^\gamma \Lambda^2}{4 \pi} \right)^\frac{3-d}{2} \frac{\mu^{d-3}}{2^{d-1}(4\pi)^\frac{d}{2} \Gamma\left( \frac{d}{2} \right)} \left[B\left( \frac{1}{2},2-d,0 \right) -\Gamma \left(2-d \right)\Gamma \left(d-1 \right)\right],
    \end{split}
\end{equation}
which explicitly agrees with the result presented in eq.~\eqref{eq:bosonpropnoexternalmomentumresidue}.
\subsection{Leading contributions of raised bosonic poles}
\label{sec:raisedbosonicpoles}
\noindent
Let us next consider another case without explicit external momenta, characterized by second order bosonic poles such that
\begin{equation}
 \mathcal{B}_{12} (\mu, 0,0)=  \text{P.V.}  \int_{-\infty}^\infty \frac{\text{d} p_0}{2\pi} \int_p \frac{1}{(p_0+i\mu)^2+p^2} \frac{1}{[p_0^2+p^2]^2}
\end{equation}
and
\begin{equation}
     \text{Res}\left[\mathcal{B}_{12} \right] (\mu, 0,0)=   \int_p i \text{Res} \left[ \frac{1}{(p_0+i\mu)^2+p^2} \frac{1}{[p_0^2+p^2]^2} \right].
\end{equation}
This direct comparison is motivated by the discussion presented in \cite{Osterman:2023tnt}, wherein it was stated that the residue approach works equally well for any power of bosonic propagator, when studying the sunset integral. In a further subsection \ref{sec:raisedbosonsdeltafunctions}, we address this in a more complete sense, but here we will preface such discussion by explicit results showing equivalence between the two integration orders. 

Again, we begin by performing the temporal integrations through the residue theorem at $T=0$, which leads us to
\begin{equation}
    \begin{split}
    \label{eq:raisedbosonicpoleinitialresidue}
       \text{Res}\left[\mathcal{B}_{12} \right] (\mu, 0,0)  &=\int_p \left[ \frac{\theta(p-\mu)}{2p} \frac{1}{[-(p-\mu)^2+p^2]^2}-\frac{1}{4 \mu^2p^3}+ \frac{1}{2\mu^2 p (\mu+2p)^2}\right]\\
        &=\int_p \frac{1}{8p\mu^2}\left[  \frac{\theta(p-\mu)}{\left(p-\frac{\mu}{2}
        \right)^2}+\frac{1}{ \left(p+\frac{\mu}{2}\right)^2}\right],
    \end{split}
\end{equation}
where we have removed the middle contribution on the first row, given that it vanishes in dimensional regularization. The two resulting spatial integrals share a great deal of structural resemblence, with those seen in eq.~\eqref{eq:bosonpropnoexternalmomentumresidue}, which extends also to the corresponding special function representations. Indeed, utilizing the same representation, we find
\begin{eqnarray}
\label{eq:raisedbosonicgammma}
    \int_p \frac{1}{8p\mu^2}\frac{1}{ \left(p+\frac{\mu}{2}\right)^2}&=&-\left(\frac{e^\gamma \Lambda^2}{4\pi}\right)^\frac{3-d}{2} \frac{ 2^{1-d}(d-2) \mu^{d-5}}{(4\pi)^\frac{d}{2}} \frac{\Gamma \left(d-1 \right) \Gamma \left(2-d \right)}{ \Gamma \left(\frac{d}{2}\right)},\\
    \label{eq:raisedbosonicincompletebeta}
\int_p \frac{1}{8p\mu^2}  \frac{\theta(p-\mu)}{\left(p-\frac{\mu}{2}
        \right)^2}&=&\left(\frac{e^\gamma \Lambda^2}{4\pi}\right)^\frac{3-d}{2}\frac{\mu^{d-5}}{4 (4\pi)^\frac{d}{2} \Gamma \left( \frac{d}{2} \right)}  \left[ 2+ 2^{3-d}(d-2) B \left(\frac{1}{2}, 3-d,0 \right) \right].
\end{eqnarray}

In a manner similar to the previous subsection, let us study the opposite integration order by first utilizing partial fraction decomposition on the integrand. Then integrating the spatially, we find 
\begin{equation}
\begin{split}
  \mathcal{B}_{12}(\mu,0,0) &= (-i) \text{P.V.}\int_{-\infty}^\infty \frac{\text{d} p_0}{2\pi} \int_p \frac{1}{2\mu} \frac{1}{p_0 + \frac{i\mu}{2}}\left[\frac{1}{p_0^2+p^2}-\frac{1}{(p_0+i\mu)^2+p^2} \right] \frac{1}{p_0^2+p^2}\\
   &= (-i)\text{P.V.}\int_{-\infty}^\infty \frac{\text{d} p_0}{2\pi} \frac{1}{2\mu} \frac{1}{p_0 + \frac{i\mu}{2}} \int_p \frac{1}{(p_0^2+p^2)^2}\\
   &+\text{P.V.}\int_{-\infty}^\infty \frac{\text{d} p_0}{2\pi} \frac{1}{4\mu^2} \left[\frac{1}{p_0 + \frac{i\mu}{2}} \right]^2 \int_p\left[\frac{1}{p_0^2+p^2}-\frac{1}{(p_0+i\mu)^2+p^2} \right] \\
   &= (-i)\text{P.V.}\int_{-\infty}^\infty \frac{\text{d} p_0}{2\pi} \frac{1}{2\mu} \frac{1}{p_0 + \frac{i\mu}{2}} \left( \frac{e^\gamma \Lambda^2}{4\pi} \right)^\frac{3-d}{2} \frac{\Gamma \left( 2-\frac{d}{2}\right) (p_0^2)^\frac{d-4}{2}}{(4\pi)^\frac{d}{2} } \\
   &+\text{P.V.}\int_{-\infty}^\infty \frac{\text{d} p_0}{2\pi} \frac{1}{4\mu^2} \left[\frac{1}{p_0 + \frac{i\mu}{2}} \right]^2 \left( \frac{e^\gamma \Lambda^2}{4\pi} \right)^\frac{3-d}{2} \frac{\Gamma \left( 1-\frac{d}{2}\right) }{(4\pi)^\frac{d}{2} } \left\{(p_0^2)^\frac{d-2}{2}-[(p_0+i\mu)^2]^\frac{d-2}{2} \right\}.
   \end{split}
\end{equation}
Similar to the previous subsection, we both perform Feynman parametrization utilizing linear scales of the temporal momentum and express the resulting temporal integrations side-by-side in the interval $p_0 \in (0,\infty)$. After the temporal integrations, we are left with 
\begin{equation}
    \begin{split}
       \mathcal{B}_{12}(\mu,0,0)    &=-\frac{i}{4\pi\mu} \left( \frac{e^\gamma \Lambda^2}{4\pi} \right)^\frac{3-d}{2} \frac{\Gamma \left( 2-\frac{d}{2}\right)}{(4\pi)^\frac{d}{2} }   \int_0^1 \text{d} x (1-x)^{3-d} \left[ \left(\frac{i \mu x}{2} \right)^{d-4}- \left(-\frac{i \mu x}{2} \right)^{d-4} \right]\\
   &+ \frac{(2-d)}{8\pi \mu^2}\left( \frac{e^\gamma \Lambda^2}{4\pi} \right)^\frac{3-d}{2} \frac{\Gamma \left( 1-\frac{d}{2}\right) }{(4\pi)^\frac{d}{2} } \int_0^1 \text{d} x (1-x)^{1-d} x  \left[ \left(\frac{i \mu x}{2} \right)^{d-3}+ \left(-\frac{i \mu x}{2} \right)^{d-3} \right]\\
   &- \frac{(2-d)}{8\pi \mu^2}\left( \frac{e^\gamma \Lambda^2}{4\pi} \right)^\frac{3-d}{2} \frac{\Gamma \left( 1-\frac{d}{2}\right) }{(4\pi)^\frac{d}{2} } \int_0^1 \text{d} x (1-x)^{1-d} x  \left\{ \left[\left(1-\frac{x}{2}\right)i\mu \right]^{d-3}+\left[-\left(1-\frac{x}{2}\right)i\mu \right]^{d-3} \right\}
    \end{split}
\end{equation}
Utilizing Euler's reflection formula and beta function integral representations, we can straightforwardly associate the components of the above to eqs. \eqref{eq:raisedbosonicgammma} and \eqref{eq:raisedbosonicincompletebeta} such that
\begin{equation}
\begin{split}
   &-\frac{i}{4\pi\mu} \left( \frac{e^\gamma \Lambda^2}{4\pi} \right)^\frac{3-d}{2} \frac{\Gamma \left( 2-\frac{d}{2}\right)}{(4\pi)^\frac{d}{2} }   \int_0^1 \text{d} x (1-x)^{3-d} \left[ \left(\frac{i \mu x}{2} \right)^{d-4}- \left(-\frac{i \mu x}{2} \right)^{d-4} \right]\\
   &= -2 \cdot \left(\frac{e^\gamma \Lambda^2}{4\pi}\right)^\frac{3-d}{2} \frac{ 2^{1-d}(d-2) \mu^{d-5}}{(4\pi)^\frac{d}{2}} \frac{\Gamma \left(d-1 \right) \Gamma \left(2-d \right)}{ \Gamma \left(\frac{d}{2}\right)},
   \end{split}
   \end{equation}
   \begin{equation}
   \begin{split}
   \label{eq:raisedbosonicpoleFP2}
  &\frac{(2-d)}{8\pi \mu^2}\left( \frac{e^\gamma \Lambda^2}{4\pi} \right)^\frac{3-d}{2} \frac{\Gamma \left( 1-\frac{d}{2}\right) }{(4\pi)^\frac{d}{2} } \int_0^1 \text{d} x (1-x)^{1-d} x  \left[ \left(\frac{i \mu x}{2} \right)^{d-3}+ \left(-\frac{i \mu x}{2} \right)^{d-3} \right]\\
  &=+\left(\frac{e^\gamma \Lambda^2}{4\pi}\right)^\frac{3-d}{2} \frac{ 2^{1-d}(d-2) \mu^{d-5}}{(4\pi)^\frac{d}{2}} \frac{\Gamma \left(d-1 \right) \Gamma \left(2-d \right)}{ \Gamma \left(\frac{d}{2}\right)},
  \end{split}
  \end{equation}
  \begin{equation}
  \begin{split}
  \label{eq:raisedbosonicpoleFP3}
  &- \frac{(2-d)}{8\pi \mu^2}\left( \frac{e^\gamma \Lambda^2}{4\pi} \right)^\frac{3-d}{2} \frac{\Gamma \left( 1-\frac{d}{2}\right) }{(4\pi)^\frac{d}{2} } \int_0^1 \text{d} x (1-x)^{1-d} x  \left\{ \left[\left(1-\frac{x}{2}\right)i\mu \right]^{d-3}+\left[-\left(1-\frac{x}{2}\right)i\mu \right]^{d-3} \right\}\\
  &=\left(\frac{e^\gamma \Lambda^2}{4\pi}\right)^\frac{3-d}{2}\frac{\mu^{d-5}}{4 (4\pi)^\frac{d}{2} \Gamma \left( \frac{d}{2} \right)}  \left[ 2+ 2^{3-d}(d-2) B \left(\frac{1}{2}, 3-d,0 \right) \right].
  \end{split}
\end{equation}
Here we emphasized the relevant multiplicative differences to the expressions from the residue prescription by extracting any such factors at the front of the right-hand side in all these equations. By summing these, it is trivial to see that the results are equivalent. This is again a non-trivial example of systematic behaviour that extends to any arbitrary power of the bosonic propagator.

\subsection{Extension to bosonic temporal external momenta}
\label{sec:extensiontobostemporal}
\noindent
Let us return to consider first order bosonic poles, with now an added temporal shift such that
\begin{equation}
 \mathcal{B}_{11} (\mu, 0,k_0)=  \text{P.V.}  \int_{-\infty}^\infty \frac{\text{d} p_0}{2\pi} \int_p \frac{1}{(p_0+i\mu)^2+p^2} \frac{1}{[(p_0-k_0)^2+p^2]}
\end{equation}
and
\begin{equation}
 \text{Res}\left[\mathcal{B}_{11} \right] (\mu, 0,k_0)=  \int_p i \text{Res} \left[\frac{1}{(p_0+i\mu)^2+p^2} \frac{1}{[(p_0-k_0)^2+p^2]}\right].
\end{equation}
Carrying now out the strict $T=0$ residue prescription, we find in alignment with eq.~\eqref{eq:residuecollection} the spatial integrals such that
\begin{equation}
\begin{split}
    \text{Res}\left[\mathcal{B}_{11} \right] (\mu, 0,k_0) &=\frac{1}{\mu -ik_0}\int_p \left[\frac{\theta(p-\mu)}{2p}\frac{1}{ 2p-\mu+i k_0}-\frac{1}{2p}\frac{1}{2p + \mu - ik_0} \right]\\
    &=\frac{1}{\mu } \left(1-\frac{ik_0}{\mu} \right)^{-1}\int_p \left[\frac{\theta(p-\mu)}{2p(2p-\mu)} \left(1+\frac{ik_0}{2p-\mu} \right)^{-1}-\frac{1}{2p (2p+\mu)}\left(1-\frac{ik_0}{2p+\mu} \right)^{-1} \right],
    \end{split}
\end{equation}
where we have in anticipation of further steps explicitly written the well-defined binomial power expressions for generation of subleading contributions proportional to $\mathcal{O}(k_0^n)$. Given that the computation -- specifically in the opposite integration order -- becomes somewhat convoluted, we are satisfied by confirming the equivalence of results in the first subleading order of $\mathcal{O}(k_0)$. These read 
\begin{equation}
    \begin{split}
    \label{eq:bosonicpoletemporaextresidue}
   & \text{Res}\left[\mathcal{B}_{11} \right] (\mu, 0,k_0)\\ &=\frac{1}{\mu}\left( 1+\frac{ik_0}{\mu} \right)\int_p \left[\frac{\theta(p-\mu)}{2p}\frac{1}{ 2p-\mu}-\frac{1}{2p}\frac{1}{2p + \mu} \right]-\frac{ik_0}{\mu } \int_p \left[\frac{\theta(p-\mu)}{2p(2p-\mu)^2} +\frac{1}{2p (2p+\mu)^2} \right]\\
    &= \left(1+\frac{ik_0}{\mu} \right) \int_p  i \text{Res} \left\{ \frac{1}{[(p_0+i\mu)^2+p^2] [p_0^2+p^2]} \right\}-ik_0 \mu \int_p  i \text{Res} \left\{ \frac{1}{[(p_0+i\mu)^2+p^2] [p_0^2+p^2]^2} \right\},
    \end{split}
\end{equation}
where on the lowest row we have implicitly associated the terms to those discussed in previous two subsections, specifically for the former those seen in eq.~\eqref{eq:bosonpropnoexternalmomentumresidue} and eqs.~\eqref{eq:raisedbosonicgammma}-\eqref{eq:raisedbosonicincompletebeta}, respectively. 

Let us then consider the reverse order of integration again beginning with the computation by partial fraction decomposition. Performing afterwards the spatial integrating we find
\begin{equation}
\begin{split}
\label{eq:bosonproptexttemporaldefFP}
  \mathcal{B}_{11} (\mu,0,k_0) &= \left( \frac{e^\gamma \Lambda^2}{4 \pi} \right)^\frac{3-d}{2}\text{P.V.} \int_{-\infty}^\infty \frac{\text{d} p_0}{2\pi} \frac{\Gamma \left(1-\frac{d}{2} \right)}{(4\pi)^\frac{d}{2}} \frac{[(p_0+i\mu)^2]^{\frac{d}{2}-1}-[(p_0-k_0]^2)^{\frac{d}{2}-1}}{(p_0-k_0)^2-(p_0+i\mu)^2} \\
   &=\frac{i}{2(\mu-ik_0)}\left( \frac{e^\gamma \Lambda^2}{4 \pi} \right)^\frac{3-d}{2}\text{P.V.} \int_{-\infty}^\infty \frac{\text{d} p_0}{2\pi} \frac{\Gamma \left(1-\frac{d}{2} \right)}{(4\pi)^\frac{d}{2}} \frac{[(p_0+i\mu)^2]^{\frac{d}{2}-1}-[(p_0-k_0)^2]^{\frac{d}{2}-1}}{(p_0 +  \frac{i\mu}{2} )-\frac{k_0}{2}}.
    \end{split}
\end{equation}
For the purpose of properly isolating the subleading contributions, let us evaluate separately the two terms in the numerator detached from one another by the minus sign, $[(p_0+i\mu)^2]^{\frac{d}{2}-1}-[(p_0-k_0)^2]^{\frac{d}{2}-1}$. First considering the latter, the most convenient way to approach the expansion is to move all $k_0$ dependence to the denominator by performing a change of integration variable $p_0 \mapsto p_0+k_0$, which additionally moves the original position of the split in temporal integration. Then, we can re-express the terms up to $\mathcal{O}(k_0)$ such that 
\begin{equation}
\begin{split}
&-\frac{i}{2(\mu-ik_0)}\left( \frac{e^\gamma \Lambda^2}{4 \pi} \right)^\frac{3-d}{2} \text{P.V.}\int_{-\infty}^\infty \frac{\text{d} p_0}{2\pi} \frac{\Gamma \left(1-\frac{d}{2} \right)}{(4\pi)^\frac{d}{2}} \frac{[(p_0-k_0)^2]^{\frac{d}{2}-1}}{p_0 +  \frac{i\mu-k_0}{2} } \\
&=-\left(1+\frac{ik_0}{\mu} \right)\frac{i}{2\mu}\left( \frac{e^\gamma \Lambda^2}{4 \pi} \right)^\frac{3-d}{2} \text{P.V.}\int_{-\infty}^\infty \frac{\text{d} p_0}{2\pi} \frac{\Gamma \left(1-\frac{d}{2} \right)}{(4\pi)^\frac{d}{2}} \frac{[(p_0)^2]^{\frac{d}{2}-1}}{p_0 +  \frac{i\mu}{2} }\\
&+ik_0 \mu \cdot\frac{1}{4\mu^2}\left( \frac{e^\gamma \Lambda^2}{4 \pi} \right)^\frac{3-d}{2} \left\{ \int_{0}^\infty \frac{\text{d} p_0}{2\pi} \frac{\Gamma \left(1-\frac{d}{2} \right)}{(4\pi)^\frac{d}{2}} \frac{[(p_0)^2]^{\frac{d}{2}-1}}{\left(p_0 +  \frac{i\mu}{2} \right)^2 } +\int_{0}^\infty \frac{\text{d} p_0}{2\pi} \frac{\Gamma \left(1-\frac{d}{2} \right)}{(4\pi)^\frac{d}{2}} \frac{[(p_0)^2]^{\frac{d}{2}-1}}{\left(p_0 - \frac{i\mu}{2} \right)^2 } \right\}.
    \end{split}
\end{equation}
The remaining contributions from eq.~\eqref{eq:bosonproptexttemporaldefFP} can be directly expanded to yield instead
\begin{equation}
\begin{split}
 &\frac{i}{2(\mu-ik_0)}\left( \frac{e^\gamma \Lambda^2}{4 \pi} \right)^\frac{3-d}{2} \text{P.V.} \int_{-\infty}^\infty \frac{\text{d} p_0}{2\pi} \frac{\Gamma \left(1-\frac{d}{2} \right)}{(4\pi)^\frac{d}{2}} \frac{[(p_0+i\mu)^2]^{\frac{d}{2}-1}}{p_0 +  \frac{i\mu-k_0}{2} }\\
 &= \left(1+\frac{ik_0}{\mu} \right)\frac{i}{2\mu}\left( \frac{e^\gamma \Lambda^2}{4 \pi} \right)^\frac{3-d}{2} \text{P.V.} \int_{-\infty}^\infty \frac{\text{d} p_0}{2\pi} \frac{\Gamma \left(1-\frac{d}{2} \right)}{(4\pi)^\frac{d}{2}} \frac{[(p_0+i\mu)^2]^{\frac{d}{2}-1}}{p_0 +  \frac{i\mu}{2} }\\
 &+ik_0 \mu \cdot \frac{1}{4\mu^2}\left( \frac{e^\gamma \Lambda^2}{4 \pi} \right)^\frac{3-d}{2} \left\{ \int_{0}^\infty \frac{\text{d} p_0}{2\pi} \frac{\Gamma \left(1-\frac{d}{2} \right)}{(4\pi)^\frac{d}{2}} \frac{[(p_0+i\mu)^2]^{\frac{d}{2}-1}}{\left(p_0 +  \frac{i\mu}{2} \right)^2 } +\int_{0}^\infty \frac{\text{d} p_0}{2\pi} \frac{\Gamma \left(1-\frac{d}{2} \right)}{(4\pi)^\frac{d}{2}} \frac{[(p_0+i\mu)^2]^{\frac{d}{2}-1}}{\left(p_0 - \frac{i\mu}{2} \right)^2 } \right\}.
    \end{split}
\end{equation}
Combining these two expressions and comparing them against eq.~\eqref{eq:bosoniclowestorderFPintermediate} and eqs.~\eqref{eq:raisedbosonicpoleFP2} -- \eqref{eq:raisedbosonicpoleFP3}, we can immediately recognize the equivalence to eq.~\eqref{eq:bosonicpoletemporaextresidue} without any further computations. We again emphasize that this equivalence can be confirmed to extend to any arbitrary power of the temporal external momentum in the corresponding expansion. 

\subsection{Extension to spatial external momenta}
\label{sec:extensiontospatialextmomentaboson}
\noindent
We recognize that it is somewhat obvious that the spatial external momenta only contribute in ways that are respected in both integration orders (specifically referring to the residue approach), given that the distribution functions in full absorb the added scales. However, for the sake of completeness, we choose to be explicit and write below the most first subleading contributions, and find them to be indeed equal through either computation scheme. Thus, let us begin by defining the integral of interest as
\begin{equation}
  \mathcal{B}_{11} (\mu, \vec{s},0)=\, \text{P.V.} \int_{-\infty}^\infty \frac{\text{d} p_0}{2\pi} \int_p \frac{1}{(p_0+i\mu)^2+p^2}\frac{1}{p_0^2+ |\vec{p}-\vec{s}|^2}
\end{equation}
and
\begin{equation}
  \text{Res} \left[\mathcal{B}_{11} \right] (\mu, \vec{s},0)=\int_p i \text{Res} \left\{ \frac{1}{(p_0+i\mu)^2+p^2}\frac{1}{p_0^2+ |\vec{p}-\vec{s}|^2} \right\}. 
\end{equation}
Then following eq.~\eqref{eq:residuecollection}, we can explicitly write the spatial integrand for all orders of $\mathcal{O}(s^n)$. To perform the expansion, we utilize the fact that $s \ll \mu$ and therefore $\left|\theta(p-\frac{\mu \mp \mu}{2})\frac{2psz \pm s^2}{2p\mu\pm\mu^2} \right| \ll 1$. By noting that any order proportional to odd power of the angular coordinate $z$ has to vanish, we find the first non-vanishing subleading terms at $\mathcal{O}(s^2)$ by writing 
\begin{equation}
\begin{split}
\label{eq:fullresiduespatialboson1}
   \text{Res} \left[\mathcal{B}_{11} \right] (\mu, \vec{s},0)  &=\int_p \left[ \frac{\theta(p-\mu)}{2p} \frac{1}{-(p-\mu)^2+|\vec{p}-\vec{s}|^2}+\frac{\theta(|\vec{p}-\vec{s}|)}{2|\vec{p}-\vec{s} |} \frac{1}{-(|\vec{p}-\vec{s}|+\mu)^2+p^2} \right]\\
    %&= \int_p \left[ \frac{\theta(p-\mu)}{2p} \frac{1}{\mu (2p-\mu)-2psz+s^2}-\frac{\theta(p)}{2p} \frac{1}{\mu (2p+\mu)+2psz+s^2} \right]\\
    &=\int_p \frac{\theta(p-\mu)}{2p}  \frac{1}{2p-\mu}\left( 1-\frac{2psz-s^2}{2p\mu-\mu^2}\right)^{-1}-\int_p \frac{1}{2p} \frac{1}{2p\mu+\mu^2} \left(1+\frac{2psz+s^2}{2p\mu+\mu^2}\right)^{-1} \\
    &= \frac{1}{\mu} \int_p \frac{1}{2p} \left[ \frac{\theta(p-\mu)}{2p-\mu}-\frac{1}{2p+\mu} \right] - s^2 \int_p\left[\frac{\theta(p-\mu)}{2p}  \frac{1}{(2p\mu-\mu^2)^2}-\frac{1}{2p} \frac{1}{(2p \mu + \mu^2)^2}\right]\\
    &+s^2 \int_p \left[\frac{\theta(p-\mu)}{2p}  \frac{4p^2 z^2}{(2p\mu-\mu^2)^3}-\frac{1}{2p} \frac{4 p^2 z^2}{(2p \mu + \mu^2)^3}\right] + \mathcal{O}(s^4)
    \end{split}
\end{equation} 
The terms on the second to last row can be found explicitly from eqs.~\eqref{eq:bosonpropnoexternalmomentumresidue} and \eqref{eq:raisedbosonicgammma}--\eqref{eq:raisedbosonicincompletebeta}. Thus the sole additional contributions (containing explict angular dependence), can be computed straightforwardly using eq.~\eqref{eq:angularbetafunctionhelp}, which leads to 
\begin{equation}
\begin{split}
\label{eq:fullresiduespatialboson2}
   \int_p \frac{\theta(p-\mu)}{2p}  \frac{4p^2 z^2}{(2p\mu-\mu^2)^3} &=  \left( \frac{e^\gamma \Lambda^2{}}{4\pi} \right)^\frac{3-d}{2} \frac{\mu^{d-5}}{\pi (4\pi)^\frac{d-1}{2} \Gamma \left( \frac{d-1}{2} \right)} \left[\int_{-1}^1 \text{d} z (1-z^2)^\frac{3-d}{2} z^2 \right] \int_1^\infty \text{d}p\: \frac{ p^{d}}{ 4\left(p- \frac{1}{2} \right)^3}\\
   &=\left( \frac{e^\gamma \Lambda^2{}}{4\pi} \right)^\frac{3-d}{2} \frac{\mu^{d-5}}{4(4\pi)^\frac{d}{2} \Gamma \left( \frac{d}{2}+1 \right)} \left[2+d+2^{1-d}d(d-1) B \left(\frac{1}{2},2-d,0 \right) \right]
   \end{split}
\end{equation}
and 
\begin{equation}
\label{eq:fullresiduespatialboson3}
    \begin{split}
        -\int_p\frac{1}{2p} \frac{4 p^2 z^2}{(2p \mu + \mu^2)^3} &= \left( \frac{e^\gamma \Lambda^2{}}{4\pi} \right)^\frac{3-d}{2} \frac{\mu^{d-5}}{\pi (4\pi)^\frac{d-1}{2} \Gamma \left( \frac{d-1}{2} \right)} \left[\int_{-1}^1 \text{d} z (1-z^2)^\frac{3-d}{2} z^2 \right] \int_0^\infty \text{d}p\: \frac{ p^{d}}{ 2^{d}\left(p+ 1\right)^3}\\
        &=-\left( \frac{e^\gamma \Lambda^2}{4\pi} \right)^\frac{3-d}{2} \frac{\mu^{d-5}\Gamma(d+1) \Gamma(2-d)}{2^{1+d}(4\pi)^\frac{d}{2} \Gamma \left( \frac{d}{2}+1 \right)}. 
    \end{split}
\end{equation}

To compute the spatial integration first, we must begin with Feynman parametrization enacted onto the two propagator elements. We follow this up naturally with the spatial integration, leading to
\begin{equation}
    \begin{split}
    \mathcal{B}_{11} (\mu, \vec{s},0)   &=\left( \frac{e^\gamma \Lambda^2}{4 \pi} \right)^\frac{3-d}{2} \frac{\Gamma \left(2-\frac{d}{2} \right)}{(4\pi)^\frac{d}{2}} \text{P.V.}  \int_{-\infty}^\infty \frac{\text{d} p_0}{2\pi}\int_0^1 \text{d} x \left[x(p_0+i\mu)^2 + (1-x) p_0^2 + x(1-x)s^2 \right]^{\frac{d}{2}-2}.
      \end{split}
      \end{equation}
Before carrying out the temporal integration (or parametric integration), one should expand the power function, specifically noting that $x(1-x)s^2 < |x (p_0+i\mu)^2+(1-x)p_0^2|$ if $p_0 \neq 0$. This allows us to extract the two leading structures such that
\begin{equation}
    \begin{split}
      &\left( \frac{e^\gamma \Lambda^2}{4 \pi} \right)^\frac{3-d}{2} \frac{\Gamma \left(2-\frac{d}{2} \right)}{(4\pi)^\frac{d}{2}} \text{P.V.}  \int_{-\infty}^\infty \frac{\text{d} p_0}{2\pi}\int_0^1 \text{d} x \left[x(p_0+i\mu)^2 + (1-x) p_0^2 + x(1-x)s^2 \right]^{\frac{d}{2}-2}\\
      &=\left( \frac{e^\gamma \Lambda^2}{4 \pi} \right)^\frac{3-d}{2} \frac{\Gamma \left(1-\frac{d}{2} \right)}{(4\pi)^\frac{d}{2}}  \text{P.V.}   \int_{-\infty}^\infty \frac{\text{d} p_0}{2\pi} \frac{1}{2 i \mu} \frac{1}{p_0 + \frac{i\mu}{2}} \left[(p_0^2)^\frac{d-2}{2}-[(p_0+i\mu)^2]^\frac{d-2}{2} \right]\\
      &-\left( \frac{e^\gamma \Lambda^2}{4 \pi} \right)^\frac{3-d}{2} \frac{\Gamma \left(3-\frac{d}{2} \right)}{(4\pi)^\frac{d}{2}}  \text{P.V.}  \int_{-\infty}^\infty \frac{\text{d} p_0}{2\pi}\int_0^1 \text{d} x \left[x(p_0+i\mu) + (1-x) p_0^2 \right]^{\frac{d}{2}-3}x(1-x)s^2.
    \end{split}
\end{equation}
Here the subleading part can be further simplified by performing the Feynman parametrization explicitly, which leads to the following temporal integrand:
\begin{equation}
\begin{split}
    &\int_0^1\text{d} x x(1-x) \left[x(p_0+i\mu)^2+(1-x) p_0^2 \right]^\frac{d-6}{2}\\
    &=-\frac{2[(p_0+i\mu)^2]^\frac{d-4}{2}}{[(p_0+i\mu)^2-p_0^2]^3} \left\{\frac{[(p_0+i\mu)p_0]^2}{d-4}-\frac{(p_0+i\mu)^2[(p_0+i\mu)^2+p_0^2]}{d-2} +\frac{(p_0+i\mu)^4}{d}\right\}\\
    &+\frac{2[(p_0)^2]^\frac{d-4}{2}}{[(p_0+i\mu)^2-p_0^2]^3} \left\{\frac{[(p_0+i\mu)p_0]^2}{d-4}-\frac{(p_0)^2[(p_0+i\mu)^2+p_0^2]}{d-2} +\frac{(p_0)^4}{d}\right\}.
    \end{split}
\end{equation}
The simplifications from this expression are easiest carried out by re-casting all numerator expressions to be given in terms of the scale under the effect of the $d$ dependent power. Explicitly this means that all numerator expressions in the second to last row are written in terms of $p_0+i\mu$, specifically $p_0^2 = (p_0+i\mu)^2-2i\mu (p_0+i\mu)-\mu^2$. In the last row, the numerator exclusively utilizes $p_0$ and thus we would write $(p_0+i\mu)^2 = p_0^2+ 2i \mu p_0 -\mu^2$. After these simplifications, the remaining temporal integration is carried out after the Feynman parametrization of the integrand -- leading to the divisors combining with the numerators as linear shifts. We note that all these steps are straightforward and follow the steps carried out in sections \ref{sec:raisedbosonicpoles} and \ref{sec:extensiontobostemporal}, but given how lengthy the computation is, we are satisfied to note that the result explicit agrees with those listed in eqs.~\eqref{eq:fullresiduespatialboson1}--\eqref{eq:fullresiduespatialboson3}. Similar to this first subleading contribution, one can straightforwardly find agreement up to any order of $\mathcal{O}(s^{2n})$.

\subsection{General raised poles and vanishing delta function contributions}
\label{sec:raisedbosonsdeltafunctions}
\noindent
We have demonstrated through studying the lowest subleading orders the behaviour such that $\mathcal{B}_{11} (\mu, \vec{s},s_0) = \text{Res} \left[ \mathcal{B}_{11} \right] (\mu, \vec{s},s_0) $ or 
\begin{equation}
\begin{split}
   &\text{P.V.} \int_{-\infty}^\infty \frac{\text{d} p_0}{2 \pi}  \int_p \frac{1}{(p_0+i \mu)^2 + p^2} \frac{1}{(p_0-s_0)^2+|\vec{p}-\vec{s}|^2}\\
    &=  \int_p  i \text{Res} \left[\frac{1}{(p_0+i \mu)^2 + p^2} \frac{1}{(p_0-s_0)^2+|\vec{p}-\vec{s}|^2} \right]_{p_0 = i n (p -\mu)+i (1-n) |\vec{p}-\vec{s}| },
    \end{split}
\end{equation}
which explicitly does extend to all orders of the small parameter expansion with respect to $\frac{s}{\mu}$ and $\frac{s_0}{\mu}$. Furthermore, we showed that  $\mathcal{B}_{12} (\mu, 0,0) = \text{Res} \left[\mathcal{B}_{12} \right](\mu, 0,0)$ or
\begin{equation}
\begin{split}
  \text{P.V.} \int_{-\infty}^\infty \frac{\text{d} p_0}{2 \pi}  \int_p \frac{1}{(p_0+i \mu)^2 + p^2} \frac{1}{[p_0^2+p^2]^2}
    &=  \int_p  i \text{Res} \left\{\frac{1}{(p_0+i \mu)^2 + p^2} \frac{1}{[p_0^2+p^2]^2} \right\}_{p_0 = i n (p -\mu)+i (1-n)p },
    \end{split}
\end{equation}
which exemplifies a more general behavior of equivalence taking place with any arbitrary integer power of the bosonic propagator. Here, we argue how and why this extends even to allow further scales such as small external momenta. To this end, let $m\in \mathbb{Z}_+$, and consider the steps of showing agreement between the lines of the following (so-far unproven) identity 
\begin{equation}
    \begin{split}
    \label{eq:mostgeneralbosonequivalenceoneloop}
          \mathcal{B}_{1m} (\mu, \vec{s},s_0) &=\text{P.V.} \int_{-\infty}^\infty \frac{\text{d} p_0}{2 \pi}  \int_p \frac{1}{(p_0+i \mu)^2 + p^2} \frac{1}{[(p_0-s_0)^2+|\vec{p}-\vec{s}|^2]^m}\\
    &=  \int_p  i \sum_{n=0}^1 \text{Res} \left\{\frac{1}{(p_0+i \mu)^2 + p^2} \frac{1}{[(p_0-s_0)^2+|\vec{p}-\vec{s}|^2]^m} \right\}_{p_0 = i n (p -\mu)+ (1-n)(s_0+i |\vec{p}-\vec{s}|) }.
    \end{split}
\end{equation}
First let us look back to eq.~\eqref{eq:raisedbosonicpoleinitialresidue}, where we note that on the integrand level, there is an excess term besides the structures that match between the results of integration orders. That difference term, however, vanishes as a scaleless integral in dimensional regularization such that 
\begin{equation}
    \int_p \frac{1}{4\mu^2p^3} = 0.
\end{equation}
Let us next illustrate that a similar property of belonging to the kernel of spatial integration characterizes also any thermal contributions that residue theorem at strict $T=0$ neglects. For this purpose, we study the lowest order case of raised bosonic poles. Noting that since the bosonic and fermionic poles are distinct, we can find the leading temperature related corrections by adding fermionic distribution function (without any need to implement any additional Feynman parametrizations). This leads to
\begin{equation}
    \begin{split}
       \mathcal{B}_{12} (\mu, \vec{s},s_0) &=\int_p\underset{T \rightarrow 0^+}{\text{lim}}  i  \sum_{n=0}^1\text{Res} \left\{\frac{n_\text{F} (i \beta p_0)}{(p_0+i \mu)^2 + p^2} \frac{1}{[(p_0-s_0)^2+|\vec{p}-\vec{s}|^2]^2} \right\}_{p_0 = i n (p -\mu)+(1-n)(s_0+i |\vec{p}-\vec{s}|) }\\
       &=\int_p i \sum_{n=0}^1 \text{Res} \left\{\frac{1}{(p_0+i \mu)^2 + p^2} \frac{1}{[(p_0-s_0)^2+|\vec{p}-\vec{s}|^2]^2} \right\}_{p_0 = i n (p -\mu)+ (1-n)(s_0+i |\vec{p}-\vec{s}|) }\\
       &-\int_p \frac{ \delta( |\vec{p}-\vec{s}|)}{4|\vec{p}-\vec{s}|^2} \frac{1}{(s_0+i|\vec{p}-\vec{s}|+i\mu)^2+p^2}
    \end{split}
\end{equation}
or equivalently following \cite{Gorda:2022yex} through differentiation of the lowest order residue we find
\begin{equation}
\begin{split}
   \mathcal{B}_{12} (\mu, \vec{s},s_0)  &=-\int_p\underset{q \rightarrow |\vec{p}-\vec{s}|}{\text{lim}} \frac{\text{d}}{\text{d} q^2} i  \sum_{n=0}^1\text{Res} \left\{\frac{1}{(p_0+i \mu)^2 + p^2} \frac{1}{(p_0-s_0)^2+q^2} \right\}_{p_0 = i n (p -\mu)+(1-n)(s_0+i q) }\\
     &=\int_p i \sum_{n=0}^1 \text{Res} \left\{\frac{1}{(p_0+i \mu)^2 + p^2} \frac{1}{[(p_0-s_0)^2+|\vec{p}-\vec{s}|^2]^2} \right\}_{p_0 = i n (p -\mu)+ (1-n)(s_0+i |\vec{p}-\vec{s}|) }\\
       &-\int_p \frac{\delta( |\vec{p}-\vec{s}|)}{4|\vec{p}-\vec{s}|^2} \frac{1}{(s_0+i|\vec{p}-\vec{s}|+i\mu)^2+p^2}.
    \end{split}
\end{equation}
The last row of either equation above corresponds to the difference term from the strict residue theorem expression at $T=0$. To study the relevant spatial integral, let external momentum $s_k$ act as a small shift and perform (again) a change of integration variables such that $\text{d} p_k = \text{d} q_k $, where $\vec{q} + \vec{s} = \vec{p}$. Now, we can write
\begin{equation}
    \begin{split}
        -\int_p\frac{\delta( |\vec{p}-\vec{s}|)}{4|\vec{p}-\vec{s}|^2} \frac{1}{(s_0+i|\vec{p}-\vec{s}|+i\mu)^2+p^2} &=-\int_q\frac{\delta( q)}{4q^2} \frac{1}{(s_0+iq+i\mu)^2+|\vec{q}+\vec{s}|^2}.
        \end{split}
        \end{equation}
Given that the right-hand side is directly proportional to scaleless radial integral, we can recognize that it indeed vanishes in dimensional regularization such that
    \begin{equation}
        \int_0^\infty \text{d} q \delta(q) q^{d-3}  f(\vec{q}, \vec{s},s_0, \mu) =  f(\vec{0},\vec{s},s_0, \mu)\underset{q \rightarrow 0}{\text{lim}} q^{d-3}\overset{\text{dim. reg.}}{=} 0,
\end{equation}
where it is assumed that $f(\vec{q},\vec{s},s_0,\mu)$ is independent of the dimension $d$ and $|f(\vec{0}, \vec{s},s_0, \mu)|$ is convergent. Then the vanishing result is achieved when the dimension is kept implicit until after integration (allowing dimensional regularization to take place).

For a $n \in \mathbb{Z}_+$ exponent to the bosonic propagator, the non-vanishing temperature prescription would introduce $n-1$ boundary terms in addition to the naive residue at $T=0$, leading up to a set of corrections such that
\begin{equation}
    \sum_{k=0}^{n-2} \frac{\delta^{(k)} (q)}{q^{n+k}} f_k ( \vec{q}, \vec{s}, s_0, \mu)
\end{equation}
with each of them vanishing in radial dimensional regularization whenever the spatial integral is carried out. Given that the inclusion of thermal distribution functions acts as a bridge between the two integration orders and that the first-order pole integral yields identical values in either integration order, we can indeed deduce that the statement of eq.~\eqref{eq:mostgeneralbosonequivalenceoneloop} is valid.

\section{Distinct chemical potentials}
\label{app:distinctchemicalpotentials}
\noindent
In this appendix, we discuss an extension to the formulae of the previous appendix, with two explicitly distinct chemical potentials in the two propagator structures of the desired one loop integral:
\begin{equation}
\label{eq:twochemicalpotentialsdef}
  \mathcal{C}_{\alpha_1 \alpha_2}(\mu_1,\mu_2) =  \text{P.V.}  \int_{-\infty}^\infty \frac{\text{d} p_0}{2\pi}  \int_p \frac{1}{[(p_0+i\mu_1)^2+p^2]^{\alpha_1}}\frac{1}{[(p_0+i\mu_2)^2+p^2]^{\alpha_2}},
\end{equation}
and
\begin{equation}
\label{eq:twochemicalpotentialsdefres}
  \text{Res}\left[\mathcal{C}_{\alpha_1 \alpha_2}\right](\mu_1,\mu_2) =   \int_p i \text{Res}\left[\frac{1}{[(p_0+i\mu_1)^2+p^2]^{\alpha_1}}\frac{1}{[(p_0+i\mu_2)^2+p^2]^{\alpha_2}} \right],
\end{equation}
where we specifically study parameters $\alpha_1 = \alpha_2 = 1$.

We emphasize that while we do not explicitly consider extensions to external spatial momenta $\vec{q} = \vec{p}-\vec{s}$ here, such cases would align with all the statements discussed here and previously about the special cases in sections \ref{sec:spatialexpansionresidue} and \ref{sec:Fpspatialdirect} as well as appendix \ref{sec:extensiontospatialextmomentaboson}. This specifically refers to how in all these cases the Heaviside step functions fully carry out the role of Fermi-Dirac distribution functions of infinitesimal temperature, and accordingly either integration order yields equivalent results. However, any raised propagator powers -- outside of those affecting purely bosonic poles -- break this property, and require either spatial differentiation or an explicit Fermi-Dirac distribution to be implemented as part of the calculations \cite{Gorda:2022yex}.

Our demonstrative calculations go through two cases involving two signatures of chemical potentials: one with both positive and the other with one of each sign. The remaining case with both chemical potentials assigned to negative values can be trivially associated with a case with only positive values due to the quadratic symmetry of $(p_0+i\mu)^2$ as well as the range of temporal momenta being (almost) full real line.
\subsection{Two positive chemical potentials}
\noindent
For the purpose of this computation we fix $\mu_1 \geq\mu_2 > 0$ without any true loss of generality. Then, let us utilize the residue theorem onto the integrand of eq.~\eqref{eq:twochemicalpotentialsdefres} along the upper half of the complex plane to find    
\begin{equation}
\begin{split}
\label{eq:twopositivechempotsres}
   \text{Res}\left[ \mathcal{C}_{11} \right](\mu_1,\mu_2)  &=\int_p \left[\frac{\theta(p-\mu_1)}{2p} \frac{1}{(ip - i \mu_1+i\mu_2)^2+p^2}+\frac{\theta(p-\mu_2)}{2p} \frac{1}{(ip - i \mu_2+i\mu_1)^2+p^2} \right]\\
    &=\frac{1}{\mu_1-\mu_2} \int_p \left[\frac{\theta(p-\mu_1)}{2p} \frac{1}{2p -(\mu_1-\mu_2)}-\frac{\theta(p-\mu_2)}{2p} \frac{1}{2p +(\mu_1-\mu_2)} \right]
    \end{split}
\end{equation}
The remaining spatial integrals can be computed straightforwardly by utilizing the following formula
\begin{equation}
\begin{split}
\label{eq:hypergeomplusminustransform}
    \int_{\mu_\pm}^\infty \frac{\text{d} p p^{d-2}}{p\mp\frac{\mu_+-\mu_{-}}{2}}&= \frac{\mu_\pm^{d-2}}{2-d} {}_2 F_1 \left[1,2-d,3-d, \pm \frac{\mu_+-\mu_{-}}{2 \mu_\pm}\right]\\
    &=\frac{1}{2-d} \left[ \frac{\mu_++\mu_{-}}{2 } \right]^{d-2} {}_2 F_1 \left[2-d,2-d,3-d, \frac{\mp (\mu_+-\mu_{-})}{\mu_++\mu_{-}}\right],
    \end{split}
\end{equation}
where the second equality is achieved using a Pfaff transformation \cite{Stegun}. Via substitutions $\mu_+ = \mu_1$ and $\mu_{-} = \mu_2$, we can write the full solution as
\begin{equation}
    \begin{split}
    \text{Res}\left[\mathcal{C}_{11} \right](\mu_1,\mu_2)  &=\left( \frac{e^\gamma \Lambda^2}{4 \pi} \right)^\frac{3-d}{2} \frac{(\mu_1+\mu_2)^{d-2}}{2^{d-1}  (4\pi)^\frac{d}{2} \Gamma\left( \frac{d}{2} \right)(\mu_1-\mu_2) (2-d)} {}_2 F_1 \left[2-d,2-d,3-d, \frac{\mu_2-\mu_1}{\mu_1+\mu_2}\right]\\
    &+(\mu_1 \leftrightarrow \mu_2)
    \end{split}
\end{equation}
Then let us continue with opposite integration order, in alignment of how eq.~\eqref{eq:twochemicalpotentialsdef} has been written. Following previously established strategy, we again begin with utilizing partial fraction decomposition (or equivalently Feynman parametrization) before spatial integration. After this, we can write each remaining temporal integral in the range of $p_0 \in (0, \infty)$ leading stepwise to
\begin{equation}
    \begin{split}
         \mathcal{C}_{11}(\mu_1,\mu_2)
         &= \left( \frac{e^\gamma \Lambda^2}{4 \pi} \right)^\frac{3-d}{2} \frac{\Gamma \left(1-\frac{d}{2} \right)}{(4\pi)^\frac{d}{2}} \text{Re} \left\{ \int_0^\infty \frac{\text{d} p_0}{\pi} \frac{[(p_0+i\mu_2)^2]^{\frac{d}{2}-1}-[(p_0+i\mu_1)^2]^{\frac{d}{2}-1}}{(p_0+i\mu_1)^2-(p_0+i\mu_2)^2}\right\}\\
        &=- \left( \frac{e^\gamma \Lambda^2}{4 \pi} \right)^\frac{3-d}{2} \frac{\Gamma \left(1-\frac{d}{2} \right)}{(4\pi)^\frac{d}{2}} \text{Re} \left\{ \frac{1}{2i (\mu_2-\mu_1)} \int_0^\infty \frac{\text{d} p_0}{\pi}   \frac{[(p_0+i\mu_2)^2]^{\frac{d}{2}-1}-[(p_0+i\mu_1)^2]^{\frac{d}{2}-1}}{p_0+\frac{i(\mu_1+\mu_2)}{2}}\right\}
    \end{split}
\end{equation}
where we again utilize the real part operator. We further continue by using Feynman parametrization enacted on the linear scales of the form $p_0 + c \cdot i\mu  $. Then leaving the out the preamble of $\left[ e^\gamma \Lambda^2/(4\pi) \right]^\frac{3-d}{2}(4\pi)^{-\frac{d}{2}} \Gamma \left(1-\frac{d}{2} \right)$, we find
\begin{equation}
\begin{split}
    &-\text{Re} \left\{\frac{i(d-2)}{2(\mu_2-\mu_1)}\int_0^1 \text{d} x x^{1-d} \int_0^\infty \frac{\text{d} p_0}{\pi}  \left[p_0 + \frac{i}{2} \left( \mu_1 + \mu_2-x\mu_1+x \mu_2 \right) \right]^{d-3} \right\} + \left(\mu_1 \leftrightarrow \mu_2 \right)\\
    &=\frac{\sin \left[ \left(1-\frac{d}{2} \right) \pi\right]}{2^{d-1}\pi (\mu_2-\mu_1)}\int_0^1 \text{d} x x^{1-d}  \left( \mu_1 + \mu_2-x\mu_1+x \mu_2 \right)^{d-2} + \left(\mu_1 \leftrightarrow \mu_2 \right).
    \end{split}
\end{equation}
Utilizing this structure, we can write the full solution as
\begin{equation}
\begin{split}
\mathcal{C}_{11}(\mu_1,\mu_2)
    %&\text{P.V} \int_{-\infty}^\infty \frac{\text{d} p_0}{2\pi}\int_p \frac{\text{d} p_0}{2\pi} \frac{1}{(p_0+i\mu_1)^2+p^2}\frac{1}{(p_0+i\mu_2)^2+p^2}\\
    %&= \left( \frac{e^\gamma \Lambda^2}{4 \pi} \right)^\frac{3-d}{2} \frac{(\mu_1+\mu_2)^{d-2}}{2^{d-1}  (4\pi)^\frac{d}{2} \Gamma\left( \frac{d}{2} \right)(\mu_2-\mu_1)}\left\{\int_0^1 \text{d} x x^{1-d}  \left(1-x \frac{\mu_1-\mu_2}{\mu_1+\mu_2}\ \right)^{d-2}- \int_0^1 \text{d} x x^{1-d}  \left(1-x\frac{\mu_2-\mu_1}{\mu_1+\mu_2} \right)^{d-2}   \right\}\\
     &= \left( \frac{e^\gamma \Lambda^2}{4 \pi} \right)^\frac{3-d}{2} \frac{(\mu_1+\mu_2)^{d-2}}{2^{d-1}  (4\pi)^\frac{d}{2} \Gamma\left( \frac{d}{2} \right)(\mu_2-\mu_1)}\int_0^1 \text{d} x x^{1-d}  \left(1-x \frac{\mu_1-\mu_2}{\mu_1+\mu_2}\ \right)^{d-2} + (\mu_1 \leftrightarrow\mu_2)\\
      &=\left( \frac{e^\gamma \Lambda^2}{4 \pi} \right)^\frac{3-d}{2} \frac{(\mu_1+\mu_2)^{d-2}}{2^{d-1}  (4\pi)^\frac{d}{2} \Gamma\left( \frac{d}{2} \right)(\mu_1-\mu_2) (2-d)} {}_2 F_1 \left[2-d,2-d,3-d, \frac{\mu_2-\mu_1}{\mu_1+\mu_2}\right] + (\mu_1 \leftrightarrow\mu_2)
    %&=\left( \frac{e^\gamma \Lambda^2}{4 \pi} \right)^\frac{3-d}{2} \frac{(\mu_1+\mu_2)^{d-2}}{2^{d-1}  (4\pi)^\frac{d}{2} \Gamma\left( \frac{d}{2} \right)(\mu_1-\mu_2) (2-d)} \left\{{}_2 F_1 \left[2-d,2-d,3-d, \frac{\mu_2-\mu_1}{\mu_1+\mu_2}\right] -{}_2 F_1 \left[2-d,2-d,3-d, \frac{\mu_1-\mu_2}{\mu_1+\mu_2}\right]\right\}.
    \end{split}
\end{equation}
This is -- as expected -- in alignment with the result from the strict residue theorem applied to the integrand in the opposite integration order.

\subsection{One positive and one negative chemical potential}
For the purpose of removing ambiguity from the following steps, we redifine the integrand of interest such that 
\begin{equation}
    \mathcal{C}_{11}(\mu_1,-\mu_2) =  \text{P.V.}  \int_{-\infty}^\infty \frac{\text{d} p_0}{2\pi}  \int_p  \frac{1}{(p_0+i\mu_1)^2+p^2} \frac{1}{(p_0-i\mu_2)^2+p^2},
\end{equation}
\begin{equation}
     \text{Res}\left[\mathcal{C}_{11}\right](\mu_1,-\mu_2) =   \int_p i \text{Res}\left[ \frac{1}{(p_0+i\mu_1)^2+p^2} \frac{1}{(p_0-i\mu_2)^2+p^2} \right],
\end{equation}
where we work in the hierarchy $\mu_1 \geq \mu_2 >0$ (with the opposite hierarchy trivial to show using similar steps). Following the same order of computations as above, let us first consider the integration order that starts with strict residue theorem. We begin with performing residues along the upper half-plane (of the complex plane) to find the spatial integrand as 
\begin{equation}
    \begin{split}
    \label{eq:posnegchempotres}
     \text{Res} \left[\mathcal{C}_{11} \right] (\mu_1, -\mu_2) =  \int_p \frac{1}{2p(\mu_1+\mu_2)} \left[ \frac{\theta(p-\mu_1)}{2p-(\mu_1+\mu_2)}-\frac{1}{2p+(\mu_1+\mu_2)}+\frac{\theta(\mu_2-p)}{2p-(\mu_1+\mu_2)} \right],
    \end{split}
\end{equation}
where we emphasize the structural difference from the result in eq.~\eqref{eq:twopositivechempotsres} specifically seen in the very last term proportional to $\theta(\mu_2-p)$. Each of these is somewhat straightforward to write in terms of widely-known special functions. They read from left to right 
\begin{equation}
    \begin{split}
    \label{eq:minuspluschempotresiduepart1}
        \int_p \frac{1}{2p(\mu_1+\mu_2)}  \frac{\theta(p-\mu_1)}{2p-(\mu_1+\mu_2)} = \left( \frac{e^\gamma \Lambda^2}{4 \pi} \right)^\frac{3-d}{2} \frac{\mu_1^{d-2}}{2 (4\pi)^\frac{d}{2} \Gamma\left( \frac{d}{2} \right)(\mu_1+\mu_2) (2-d)} {}_2 F_1 \left[1,2-d,3-d, \frac{\mu_1+\mu_2}{2\mu_1}\right],
    \end{split}
\end{equation}
\begin{equation}
    \begin{split}
    \label{eq:minuspluschempotresiduepart2}
        &\frac{1}{4(\mu_1+\mu_2)}\int_p \frac{1}{p} \frac{1}{p+\frac{\mu_1+\mu_2}{2}} = \left( \frac{e^\gamma \Lambda^2}{4\pi} \right)^\frac{3-d}{2}\frac{ \Gamma \left(2-d \right) \Gamma \left(d-1 \right)}{2^{d-1}  (4 \pi)^\frac{d}{2} \Gamma \left( \frac{d}{2}\right)}(\mu_1+\mu_2)^{d-3}
    \end{split}
\end{equation}
and 
\begin{equation}
    \begin{split}
    \label{eq:minuspluschempotresiduepart3}
       - \int_p \frac{1}{2p(\mu_1+\mu_2)} \frac{\theta(\mu_2-p)}{2p-(\mu_1+\mu_2)} = \left( \frac{e^\gamma \Lambda^2}{4 \pi} \right)^\frac{3-d}{2} \frac{\mu_2^{d-1}}{  (4\pi)^\frac{d}{2} \Gamma\left( \frac{d}{2} \right)(\mu_1+\mu_2)^2 (d-1)} {}_2 F_1 \left[1,d-1,d, \frac{2 \mu_2}{\mu_1+\mu_2}\right].
    \end{split}
\end{equation}

Then let us compute the corresponding integral in reverse order of integration and confirm the equivalence of the results. Again beginning with partial fraction decomposition of the integrand, we continue by writing the temporal integration to real positive range and find
\begin{equation}
    \begin{split}
          \mathcal{C}_{11} (\mu_1, -\mu_2)  &= \left( \frac{e^\gamma \Lambda^2}{4 \pi} \right)^\frac{3-d}{2} \frac{\Gamma \left(1-\frac{d}{2} \right)}{(4\pi)^\frac{d}{2}} \text{Re} \left\{\int_0^\infty \frac{\text{d} p_0}{\pi}  \frac{[(p_0-i\mu_2)^2]^{\frac{d}{2}-1}-[(p_0+i\mu_1)^2]^{\frac{d}{2}-1}}{(p_0+i\mu_1)^2-(p_0-i\mu_2)^2}\right\}\\
        &= \left( \frac{e^\gamma \Lambda^2}{4 \pi} \right)^\frac{3-d}{2} \frac{\Gamma \left(1-\frac{d}{2} \right)}{(4\pi)^\frac{d}{2}} \text{Re} \left\{ \frac{1}{2i (\mu_1+\mu_2)} \int_0^\infty \frac{\text{d} p_0}{\pi} \frac{[(p_0-i\mu_2)^2]^{\frac{d}{2}-1}-[(p_0+i\mu_1)^2]^{\frac{d}{2}-1}}{p_0+\frac{i(\mu_1-\mu_2)}{2}} \right\}
    \end{split}
\end{equation}
By considering the parts of the temporal integrand after the preamble $\left[ e^\gamma \Lambda^2/(4\pi) \right]^\frac{3-d}{2}(4\pi)^{-\frac{d}{2}} \Gamma \left(1-\frac{d}{2} \right)$, we find via Feynman parametrization (essentially following the same strategy as in previous subsection)
\begin{equation}
\begin{split}
\label{eq:minuspluschempotFPintermediate}
    &-\frac{(d-2)}{2(\mu_1+\mu_2)} \text{Im} \left\{\int_0^1 \text{d} x x^{1-d} \int_0^\infty \frac{\text{d} p_0}{ \pi}  \left[p_0 + \frac{i}{2} \left( \mu_1 - \mu_2-x\mu_1-x \mu_2 \right) \right]^{d-3}\right\}\\
    &+\frac{(d-2)}{2(\mu_1+\mu_2)} \text{Im} \left\{\int_0^1 \text{d} x x^{1-d} \int_0^\infty \frac{\text{d} p_0}{ \pi} \left[p_0 + \frac{i}{2} \left( \mu_1 - \mu_2+x\mu_2+x \mu_1 \right) \right]^{d-3}  \right\}\\
    &= \frac{ (\mu_1+\mu_2)^{d-3}}{2 \pi } \text{Im}\left\{\int_0^1 \text{d} x x^{1-d}  \left[ \frac{i}{2} \left( \frac{\mu_1 - \mu_2}{\mu_1+\mu_2}-x \right) \right]^{d-2}   \right\}\\
    &-\frac{ (\mu_1+\mu_2)^{d-3}}{4 \pi i} \text{Im} \left\{\int_0^1 \text{d} x x^{1-d}  \left[ \frac{i}{2} \left( \frac{\mu_1 - \mu_2}{\mu_1+\mu_2}+x \right) \right]^{d-2} \right\}
    \end{split}
\end{equation}
Here the hypergeometric integral of the lower term only contains positive scales in the chosen hierarchy and accordingly can be studied in a liberal manner. Specifically, we can write the part of the full integral it contributes to as
\begin{equation}
    \begin{split}
        &\left( \frac{e^\gamma \Lambda^2}{4\pi} \right)^\frac{3-d}{2}\frac{ (\mu_1+\mu_2)^{d-3}\Gamma\left(1-\frac{d}{2} \right) \sin \left[\left(1-\frac{d}{2} \right)\pi \right]}{2^{d-1} \pi (4 \pi)^\frac{d}{2} }\int_0^1 \text{d} x x^{1-d}  \left( \frac{\mu_1 - \mu_2}{\mu_1+\mu_2}+x \right) ^{d-2}\\
        &=\left( \frac{e^\gamma \Lambda^2}{4\pi} \right)^\frac{3-d}{2}\frac{ (\mu_1-\mu_2)^{d-2}}{2^{d-1} (\mu_1+\mu_2)  (4 \pi)^\frac{d}{2} \Gamma \left( \frac{d}{2}\right)}\int_0^1 \text{d} x x^{1-d}  \left( 1+\frac{\mu_1+\mu_2}{\mu_1-\mu_2}x \right) ^{d-2}\\
        &=\left( \frac{e^\gamma \Lambda^2}{4 \pi} \right)^\frac{3-d}{2} \frac{\mu_1^{d-2}}{2 (4\pi)^\frac{d}{2} \Gamma\left( \frac{d}{2} \right)(\mu_1+\mu_2) (2-d)} {}_2 F_1 \left[1,2-d,3-d, \frac{\mu_1+\mu_2}{2\mu_1}\right]
    \end{split}
\end{equation}
where the last equality follows from a Pfaff transform of hyperegeometric functions. The last form is chosen to show how this result matches exactly with that of eq.~\eqref{eq:minuspluschempotresiduepart1}. 

The second to last row of eq.~\eqref{eq:minuspluschempotFPintermediate} should be split in two to efficiently extract its special function representation. Specifically, its parametric integral is split here such that $x \in \left(0, \frac{\mu_1-\mu_2}{\mu_1+\mu_2} \right)$ characterizes (a) part and $x \in \left( \frac{\mu_1-\mu_2}{\mu_1+\mu_2}, 1 \right)$ characaterizes (b) part. Thus, part (a) reads
\begin{equation}
    \begin{split}
    \label{eq:betafromhypergeom}
       &-\left( \frac{e^\gamma \Lambda^2}{4\pi} \right)^\frac{3-d}{2}\frac{ (\mu_1+\mu_2)^{d-3}}{2^{d-1}  (4 \pi)^\frac{d}{2} \Gamma \left( \frac{d}{2}\right)}\int_0^ \frac{\mu_1-\mu_2}{\mu_1+\mu_2} \text{d} x x^{1-d}   \left( \frac{\mu_1 - \mu_2}{\mu_1+\mu_2}-x \right)^{d-2} \\
       &=-\left( \frac{e^\gamma \Lambda^2}{4\pi} \right)^\frac{3-d}{2}\frac{ \Gamma \left(2-d \right) \Gamma \left(d-1 \right)}{2^{d-1}  (4 \pi)^\frac{d}{2} \Gamma \left( \frac{d}{2}\right)}(\mu_1+\mu_2)^{d-3},
    \end{split}
\end{equation}
which perfectly aligns with eq.~\eqref{eq:minuspluschempotresiduepart2}. The remaining hypergeometric structure reads
\begin{equation}
    \begin{split}
       &-\left( \frac{e^\gamma \Lambda^2}{4\pi} \right)^\frac{3-d}{2}\frac{ (\mu_1+\mu_2)^{d-3}}{2^{d-1}  (4 \pi)^\frac{d}{2} \Gamma \left( \frac{d}{2}\right)}\int_\frac{\mu_1-\mu_2}{\mu_1+\mu_2}^1 \text{d} x x^{1-d}   \left(x- \frac{\mu_1 - \mu_2}{\mu_1+\mu_2}\right)^{d-2} \\
       &=-\left( \frac{e^\gamma \Lambda^2}{4\pi} \right)^\frac{3-d}{2}\frac{ (\mu_1+\mu_2)^{d-3}}{2^{d-1}  (4 \pi)^\frac{d}{2} \Gamma \left( \frac{d}{2}\right)}\int_0^\frac{2\mu_2}{\mu_1+\mu_2} \text{d} y \left(y +\frac{\mu_1-\mu_2}{\mu_1+\mu_2} \right)^{1-d}   y^{d-2} \\
       &=\left( \frac{e^\gamma \Lambda^2}{4 \pi} \right)^\frac{3-d}{2} \frac{\mu_2^{d-1}}{  (4\pi)^\frac{d}{2} \Gamma\left( \frac{d}{2} \right)(\mu_1+\mu_2)^2 (d-1)} {}_2 F_1 \left[1,d-1,d, \frac{2 \mu_2}{\mu_1+\mu_2}\right],
    \end{split}
\end{equation}
where we have identified the result as equivalent to eq.~\eqref{eq:minuspluschempotresiduepart3} after performing a Pfaff transform. Thus, the two integration orders do indeed yield identical results.  

\section{Sunset integral}
\label{app:sunset}
\noindent
In this appendix we discuss the mechanism of a specific example where the residue theorem yields the sought-after result in loop computations, despite the previously established issues associated with particular temporal integrand structures. Before studying explicitly integrands in terms of the strict residue theorem approach at vanishing temperature, let us briefly emphasize that in bubble Feynman integrals, all temporal momenta can be considered to be proportional to temperature. Pertaining specifically to hierarchy where the absolute value of temporal scale is smaller than the chemical potential, all contributions proportional to a positive power of such a scale could be seen as thermally subleading. With the observed differences arising from the temporal scales (see section \ref{sec:generalcomparison}), and the thermally subleading interpretation as justification, removing $s_0$  dependence completely can be seen as a means to match results from different integration orders in multi-loop computations. However, this would somewhat misleadingly imply factorization in the temporal coordinates, for which reason we will instead look more closely to explicit multi-loop residue computations.

Specifically, we study spatial integrands of so-called sunset integral at vanishing temperature 
\begin{equation}
\label{eq:sunsetdefinition}
\mathcal{S}_{111} (\mu) =     \int_{p,q} \underset{i \text{Res} \left[ \frac{1}{(p_0+i\mu)^2+p^2} \frac{1}{(q_0+i\mu)^2+q^2} \frac{1}{(p_0-q_0)^2+|\vec{p}-\vec{q}|^2}\right]}{\underbrace{\int_{-\infty}^\infty \frac{\text{d}p_0\text{d}q_0}{(2\pi)^2} \frac{1}{(p_0+i\mu)^2+p^2} \frac{1}{(q_0+i\mu)^2+q^2} \frac{1}{(p_0-q_0)^2+|\vec{p}-\vec{q}|^2}}}
\end{equation}
in terms of different residue prescriptions (here we emphasize the convention of the temporal integrals indicating the range as solid real line). First we utilize literature \cite{Osterman:2023tnt} to identify an integrand that leads to the sought-after value for the full integral, and compare that against the established one-loop properties from this work. Second we forcefully recast the integrand and temporal integration order to introduce a possibly problematic one-loop structure, and follow up by showing that the present symmetries remove any possible excess terms in the integrand.
\subsection{Conventional computation}
\noindent
Let us then consider the residue prescription of the sunset integral presented in eq.~\eqref{eq:sunsetdefinition} explicitly. Specifically, we consider such an integration order, in which the temporal integrations are performed prior to any of the spatial ones. To keep the discussion simpler, we omit the spatial integrals and focus solely on the temporal ones. There is no informational difference in choosing either of the temporal loop momenta first. Thus, beginning with the $p_0$ momentum, and working on the upper half of the complex plane, we find 
\begin{equation}
\begin{split}
\label{eq:sunsetstandardresidue1}
    &i\sum_{n=0}^1\text{Res} \left[ \frac{1}{(p_0+i\mu)^2+p^2} \frac{1}{(p_0-q_0)^2+|\vec{p}-\vec{q}|^2}\right]_{p_0=in(p-\mu)+(1-n)(q_0+i|\vec{p}-\vec{q}|)}\\
    &= \underset{S_A}{\underbrace{\frac{\theta(p-\mu)}{2p} \frac{1}{(q_0+i\mu-ip)^2+|\vec{p}-\vec{q}|^2}}} + \underset{S_B}{\underbrace{\frac{1}{2|\vec{p}-\vec{q}|} \frac{1}{(q_0+i\mu+i|\vec{p}-\vec{q}|)^2+p^2}}}.
    \end{split}
    \end{equation}
  Here we note that the relevant integrand contains exclusively first order poles and has both a fermionic and a bosonic propagator. This specific structure was noted in appendix \ref{sec:bosons} to be fully encompassed by the residue prescription, as far as the expansions with respect to scales akin to external momenta are concerned. Accordingly, this can be seen as a non-problematic temporal integral to consider as an intermediate step. Also from this point onward, we compactify the steps by utilizing the following shorthand $k \equiv |\vec{p}-\vec{q}|$

  Let us move forwards by considering the first part of the result in eq. \eqref{eq:sunsetstandardresidue1}, again working in the upper complex half-plane, we find (now using Kronecker delta function to compress the substitutions)
    \begin{equation}
    \begin{split}
    \label{eq:secondwellbehavingresiduesunset}
    & i \text{Res}\left[\frac{S_A}{(q_0+i\mu)^2+q^2}\right]\\
    &=i\sum_{n=0}^2\text{Res} \left[\frac{1}{2p}\frac{\theta(p-\mu)}{(q_0+i\mu)^2+q^2} \frac{1}{(q_0+i\mu-ip)^2+k^2}\right]_{p_0=i \delta_n (q-\mu)+i \delta_{n-1} (k+p-\mu)+i \delta_{n-1}(p-k-\mu)}\\
    &= \frac{\theta(q-\mu)}{4pq}\frac{\theta(p-\mu)}{(iq-ip)^2+k^2}+ \frac{\theta(p+k-\mu)}{4pk} \frac{\theta(p-\mu)}{(ip+ik)^2+q^2}-\frac{\theta(p-\mu-k)}{4pk} \frac{\theta(p-\mu)}{(ip-ik)^2+q^2}\\
    &= \frac{\theta(q-\mu)}{4pq}\frac{\theta(p-\mu)}{(iq-ip)^2+k^2}+ \frac{\theta(p-\mu)}{4p k} \frac{1}{(ip+ik)^2+q^2}-\frac{\theta(p-\mu-k)}{4pk} \frac{1}{(ip-ik)^2+q^2},
    \end{split}
    \end{equation}
    where we have simplified the expression using only the most constraining Heaviside step functions on the last line. We additionally note that the integrand is a slight generalization to those discussed in appendix \ref{app:distinctchemicalpotentials}, albeit of the type which can be still seen to lead to equivalent results given that all the scales are fully contained in the Heaviside step functions. This is to say, the integrand contains specifically no real-valued temporal shifts, but two fermionic propagators with two distinct chemical potentials, which are again to be considered well-behaving integrals for finding physically interesting results using the residue theorem.

    Looking lastly at the remaining contributions from eq.~\eqref{eq:sunsetstandardresidue1}, we find with explicitly similar steps and reasoning the following result
    \begin{equation}
    \begin{split}
    & i \text{Res}\left[\frac{S_B}{(q_0+i\mu)^2+q^2}\right]\\
    &=i\sum_{n=0}^1\text{Res}\left[\frac{1}{2k} \frac{1}{(q_0+i\mu)^2+q^2}  \frac{1}{(q_0+i\mu+ik)^2+p^2}\right]_{p_0 = i (1-n)  (q-\mu) + i n  (p-k-\mu)} \\
    &=\frac{\theta(q-\mu)}{4qk}\frac{1}{(iq+ik)^2+p^2}+ \frac{1}{4pk} \frac{\theta(p-\mu-k)}{(ip-ik)^2+q^2},
    \end{split}
    \end{equation}
    where the integrand is again found to be well-behaving type of fermionic expression with two fermionic propagators, in alignment with eq.~\eqref{eq:secondwellbehavingresiduesunset}.
Adding all the pieces together, we find in total 
\begin{equation}
\label{eq:soughtaftersunsetresidue}
    \mathcal{H}_1\equiv \  \frac{\theta(p-\mu)\theta(q-\mu)}{4pq}\frac{1}{(iq-ip)^2+k^2} + \frac{\theta(q-\mu)}{4qk}\frac{1}{(iq+ik)^2+p^2}+ \frac{\theta(p-\mu)}{4pk} \frac{1}{(ip+ik)^2+q^2},
\end{equation}
or after substituting back $k = |\vec{p}-\vec{q}|$
\begin{equation}
\label{eq:soughtaftersunsetresidue2}
    \mathcal{H}_1\equiv\  \frac{\theta(p-\mu)\theta(q-\mu)}{4pq}\frac{1}{(iq-ip)^2+|\vec{p}-\vec{q}|^2} + \frac{\theta(q-\mu)}{4q|\vec{p}-\vec{q}|}\frac{1}{(iq+i|\vec{p}-\vec{q}|)^2+p^2}+ \frac{\theta(p-\mu)}{4p|\vec{p}-\vec{q}|} \frac{1}{(ip+i|\vec{p}-\vec{q}|)^2+q^2},
\end{equation}
which notably aligns with eq. 2.5 of \cite{ghisou}. While we do not discuss in detail further simplifications, we note that by recasting the Heaviside step functions such that $\theta(p-\mu) = 1- \theta(\mu-p)$, two pieces that vanish in dimensional regularization can be found, with the only non-vanishing part characterized by proportionality to $\theta(\mu-p)\theta(\mu-q)$ (seen in eq.~\eqref{eq:soughtaftersunsetresidue} as the part proportional to $\theta(p-\mu)\theta(q-\mu)$). Additionally, we again emphasize that this residue result has been shown to agree with independent methods of derivation, for more details about this result and the IBP derivations, see for example \cite{Osterman:2023tnt,Ostermanthesis}.

\subsection{Pathological representation and integration order}
\noindent
Let us next study the stability of the residue operation in this type of bubble Feynman integral, by specifically introducing structurally distinct orders of temporal integration. First, let us re-write the sunset integral such that
\begin{equation}
    \frac{1}{(p_0+i\mu)^2+p^2} \frac{1}{(p_0+i\mu-k_0)^2+q^2} \frac{1}{k_0^2+|\vec{p}-\vec{q}|^2},
\end{equation}
where we have formally written the bosonic temporal momentum as the second temporal loop momentum. Also we continue to use the shorthand $k = |\vec{p}-\vec{q}|$ to abbreviate the steps to follow and introduced the change of variable $q_0=p_0-k_0$. First let us begin by performing the fermionic temporal loop integral as residue such that
\begin{equation}
\label{eq:breakdown}
\begin{split}
    &i\sum_{n=0}^1\text{Res} \left[ \frac{1}{(p_0+i\mu)^2+p^2} \frac{1}{(p_0+i\mu-k_0)^2+q^2}\right]_{p_0 = (1-n)i (p-\mu)+n(k_0+iq-i\mu)}\\
    &= \underset{S_C}{\underbrace{\frac{\theta(p-\mu)}{2p} \frac{1}{(k_0-ip)^2+q^2}}} + \underset{S_D}{\underbrace{\frac{\theta(q-\mu)}{2q} \frac{1}{(k_0+iq)^2+p^2}}},
    \end{split}
    \end{equation}
    where we specifically note that the temporal integrand is a type that we have observed to fail to describe the sought-after physical behavior through the residue theorem for expansions of external momenta (see sections \ref{sec:momentumasymptotes} and \ref{sec:generalfermionicproblem}). Following this notion, we expect to see something different manifesting in the spatial integrand. Continuing with the bosonic residue, we find 
    \begin{equation}
        \begin{split}
        &i \text{Res} \left[\frac{S_C}{k_0^2+k^2} \right]\\
            =&\:i\sum_{n=0}^2\text{Res} \left[\frac{1}{k_0^2+k^2}\frac{\theta(p-\mu)}{2p} \frac{1}{(k_0-ip)^2+q^2}\right]_{k_0 = i \delta_{n}k + i \delta_{n-1} (p+q) + i \delta_{n-2} (p-q) }\\
            =&\:\frac{\theta(p-\mu)}{4pk} \frac{1}{(ip-ik)^2+q^2}-\frac{\theta(p-\mu)\theta(p-q)}{4pq} \frac{1}{(ip-iq)^2 + k^2} +\frac{\theta(p-\mu)}{4pq}\frac{1}{(iq+ip)^2+k^2}
        \end{split}
    \end{equation}
    and 
    \begin{equation}
        \begin{split}
        &i \text{Res} \left[\frac{S_C}{k_0^2+k^2} \right]\\
            =\:&i\sum_{n=0}^1\text{Res} \left[\frac{1}{k_0^2+k^2}\frac{\theta(q-\mu)}{2q} \frac{1}{(k_0+iq)^2+p^2}\right]_{k_0 = i(1-n)k +n (ip-iq) }\\
            =\: &\frac{\theta(q-\mu)}{4qk} \frac{1}{(ik+iq)^2+p^2}+\frac{\theta(q-\mu) \theta(p-q)}{4pq} \frac{1}{(ip-iq)^2+k^2},
        \end{split}
    \end{equation}
    where we note that both the temporal integrands were non-problematic structures discussed in subsection \ref{sec:simplestboson}. These combine to yield
    \begin{equation}
        \begin{split}
            \mathcal{H}_2 \equiv \ &\frac{\theta(p-\mu)}{4pk} \frac{1}{(ip-ik)^2+q^2}+\frac{\theta(p-\mu)}{4pq}\left[\frac{1}{(ip+iq)^2+k^2}-\frac{1}{(ip-iq)^2+k^2} \right] \\
          &+\frac{\theta(p-\mu)\theta(q-\mu)}{4pq} \frac{1}{(ip-iq)^2 + k^2}+\frac{\theta(q-\mu)}{4qk} \frac{1}{(ik+iq)^2+p^2},
        \end{split}
    \end{equation}
from which we can additionally subtract the sought-after value of eq.~\eqref{eq:soughtaftersunsetresidue} to find the difference term as 
    \begin{equation}
        \begin{split}
            \mathcal{H}_2-\mathcal{H}_1 = \ &\frac{\theta(p-\mu)}{4pk} \left[\frac{1}{(ip-ik)^2+q^2}-\frac{1}{(ip+ik)^2+q^2} \right]+\frac{\theta(p-\mu)}{4pq}\left[\frac{1}{(ip+iq)^2+k^2}-\frac{1}{(ip-iq)^2+k^2} \right] \\
        \end{split}
    \end{equation}    
We can simplify this by studying the subtractions inside square brackets. Specifically, we observe that 
    \begin{equation}
    \begin{split}
        \frac{\theta(p-\mu)}{4pk} \left[ \frac{1}{(ip-ik)^2+q^2}-\frac{1}{(ip+ik)^2+q^2} \right] &= -\frac{\theta(p-\mu)}{4pk} \frac{4pk}{(p^2+k^2-q^2)^2-4p^2k^2}\\
        &=-\frac{\theta(p-\mu)}{(k^2-p^2-q^2)^2-4p^2 q^2}\\
    \end{split}
\end{equation}
and 
\begin{equation}
    \begin{split}
        -\frac{\theta(p-\mu)}{4pq} \left[\frac{1}{(ip-iq)^2 + k^2}-\frac{1}{(ip+iq)^2+k^2}\right]&= \frac{\theta(p-\mu)}{(k^2-p^2-q^2)^2-4p^2 q^2},
    \end{split}
\end{equation}
which mutually cancel out. However, let us additionally note that even should these terms not cancel one another, they would furthermore vanish in  dimensional regularization. This is easily seen upon denoting $k^2-p^2-q^2 = -2 \vec{p}\cdot\vec{q}=-2pqz $ and then writing
\begin{equation}
   - \int_{p,q} \frac{\theta(p-\mu)}{(k^2-p^2-q^2)^2-4p^2 q^2} = \frac{1}{4}\left[\int_p \frac{\theta(p-\mu)}{p^2 (1-z^2)} \right] \int_q \frac{1}{q^2}.
\end{equation}
Upon applying dimensional regularization, the denominator structure
\begin{equation}
    \frac{1}{pq [(ip-iq)^2+k^2]} =\frac{1}{2p^2 q^2 (1-\textcolor{black}{z})}
\end{equation}
can be explicitly carried out, pertaining to radial integrations with respect to $p, q$ and the angular integration with respect to $z$. Thus, we observe that the (post-temporal) spatial integrands contain no structural differences for two-loop bubbles (containing only loop momenta) even when the previously pathological pole structures are integrated over. This can be easily observed to persist even in the opposite (temporal) order of integration.

\bibliographystyle{utphys}
\addcontentsline{toc}{chapter}{Bibliography}
\bibliography{referencesmain}
\end{document}